\pdfoutput=1

\documentclass[fleqn,usenatbib]{mnras}

\usepackage[T1]{fontenc}

\DeclareRobustCommand{\VAN}[3]{#2}
\let\VANthebibliography\thebibliography
\def\thebibliography{\DeclareRobustCommand{\VAN}[3]{##3}\VANthebibliography}

\DeclareMathAlphabet{\pazocal}{OMS}{zplm}{m}{n}

\newcommand{\M}{\mathrm}

\usepackage{graphicx}	
\usepackage{amssymb}	
\usepackage{multirow}
\usepackage{multicol} 
\usepackage{breqn}      
\usepackage{subfigure}

\title[Time Delay Cosmography of QSO SDSSJ1433]{Time Delay Cosmography: Analysis of Quadruply Lensed QSO SDSSJ1433 from Wendelstein Observatory}

\author[G. Queirolo et al.]{G. Queirolo$^{1,2}$\thanks{E-mail: queirolo@usm.lmu.de },
S. Seitz$^{1,2}$,
A. Riffeser$^{1,2}$, 
M. Kluge$^{2,1}$,
R. Bender  $^{2,1}$,
C. G\"{o}ssl   $^{1}$,
U. Hopp    $^{1,2}$,
\newauthor
C. Ries    $^{1}$,
M. Schmidt $^{1}$,
R. Z\"{o}ller $^{1,2}$
\\
$^{1}$Universit\"{a}ts-Sternwarte,
Fakult\"{a}t f\"{u}r Physik, Ludwig-Maximilians Universit\"{a}t M\"{u}nchen, Scheinerstr. 1, 81679 M\"{u}nchen, Germany \\
$^{2}$Max Planck Institute for Extraterrestrial Physics, Giessenbachstrasse 1, 85748 Garching, Germany
}

\date{Accepted XXX. Received YYY; in original form ZZZ}

\pubyear{2021}

\begin{document}
\label{firstpage}
\pagerange{\pageref{firstpage}--\pageref{lastpage}}
\maketitle

\begin{abstract}
The goal of this work is to obtain a Hubble constant estimate through the study of the quadruply lensed, variable QSO SDSSJ1433+6007. To achieve this we combine multi-filter, archival \textit{HST} data for lens modelling and a dedicated time delay monitoring campaign with the 2.1m Fraunhofer telescope at the \textit{Wendelstein Observatory}. The lens modelling is carried out with the public \texttt{lenstronomy} Python package for each of the filters individually. Through this approach, we find that the data in one of the \textit{HST} filters (F160W) contain a light contaminant, that would, if remained undetected, have severely biased the lensing potentials and thus our cosmological inference. After rejecting these data we obtain a combined posterior for the Fermat potential differences from the lens modelling in the remaining filters (F475X, F814W, F105W and F140W) with a precision of $\sim6\%$. The analysis of the \textit{g'}-band Wendelstein light curve data is carried out with a free-knot spline fitting method implemented in the public Python \texttt{PyCS3} tools. The precision of the time delays between the QSO images has a range between 7.5 and 9.8$\%$ depending on the brightness of the images and their time delay. We then combine the posteriors for the Fermat potential differences and time delays. Assuming a flat $\Lambda$CDM cosmology, we infer a Hubble parameter of $H_0=76.6^{+7.7}_{-7.0}\frac{\M{km}}{\M{Mpc\;s}}$, reaching $9.6\%$ uncertainty for a single system. 
\end{abstract}



\section{Introduction}

The last decades have seen the growing success of the $\Lambda$CDM cosmology model, which envisions a flat, cold dark matter universe with a cosmological constant $\Lambda$ representing dark energy. Such a simple yet powerful model has proven to account for multiple observable properties, such as the temperature anisotropies of the cosmic microwave background \citep[CMB, ][]{aghanim2020planck} and the acceleration of the expansion of the universe \citep{riess1998SNeIa}. However, the higher precision and larger number of measurements have exacerbated, rather than resolved, the tensions within the model. One of the most crucial is the Hubble tension, concerning the Hubble parameter $H_0$, a key parameter that encodes the current expansion rate of the cosmos, and thus its age and size. The variety of methods available to measure $H_0$ is commonly divided between late and early time measurements, referring to the epoch of the $\Lambda$CDM model in which the observed physical probe originated. The tension between the results of such methods is between 4$\sigma$ and 6$\sigma$ \citep{di2021realm}. The same classification can be defined as direct (late time) and indirect (early time) measurements, depending on the analysis required assumptions on the cosmological model. One of the most precise indirect measurements is given by the \textit{Planck} collaboration \citep{collaboration2014planck,ade2016planck} using the Planck satellite data to observe the CMB and map its anisotropies. These small deviations in intensity and polarization from point to point on the sky are used to tightly constrain cosmological parameters assuming a flat $\Lambda$CDM framework. The obtained result corresponds to $H_0=67.4\pm0.5\frac{\M{km}}{\M{Mpc\;s}}$ \citep{aghanim2020planck}. Such a value is several standard deviations lower than any of the late time measurements, one of the most formally precise being $H_0=73.24\pm1.74\frac{\M{km}}{\M{Mpc\;s}}$ from the SH0ES program \citep{Riess_2016} based on the cosmic distance ladder. In this case, several probes (e.g. parallaxes, Cepheid and SNe Ia) are used to measure distances up to the Hubble flow. In this regime, the expansion of the universe dominates the receding velocity of the object and therefore can be correlated with its distance to constrain $H_0$. Another key measurement in the Hubble tension discourse is the late time measurement based on the tip of the red giant branch (TRGB) method \citep{lee1993tip}. Here precise and well-understood stellar physics allows TRGB stars to be used as standard candles, thus allowing for high-precision distance measurements. Applied to cosmology, this method has been used to measure the Hubble parameter obtaining $H_0=69.9\pm0.8(\text{stat})\pm1.7(\text{sys})\frac{\M{km}}{\M{Mpc\;s}}$ \citep{freedman2020calibration}, thus not in tension with the two aforementioned results. 

The $H_0$ tension, if not due to unaccounted systematic measurement errors, would indicate a critical failure of the standard $\Lambda$CDM model and may lead us to a new understanding of physics. For this reason, it is of paramount importance to have unbiased, direct measurements of $H_0$ with uncertainty on the order of $1$\%. In this context, the time delay cosmography (TDC) method \citep{refsdal1964possibility,treu2022strong}
could be highly advantageous. This method is based on strong gravitational lensing of a variable background source. The light of the multiple images reaches the observer at different times, due to the lightpaths different lengths (geometric delay) and the lens' gravitational delay \citep[Shapiro time delay, ][]{shapiro1964fourth}. The relative time delay between different images $\Delta t$ can be measured for variable sources, such as for supernovae or for QSOs. 
 
 This was suggested first by \citet{refsdal1964possibility} as he laid the foundation of this method. Although initially discarded due to technological limitations, TDC is now well within reach of our observational capabilities, as strongly lensed SNae have been recently observed and employed in this framework \citep{kelly2015multiple,grillo2020accuracy,suyu2020holismokes}. Strongly lensed AGNs, quasars or QSOs have been used as well \citep[e.g.][]{suyu2010dissecting}, notably in the H0LiCOW project \citep{suyu2017h0licow,h0licow_XIII}, upon which the TDCOSMO collaboration expanded upon \citep{millon2020tdcosmo,birrer2020tdcosmo}. 
Following such implementation, in this paper we applied TDC to a quadruply lensed QSO, SDSSJ1433+6007 (\ref{sec:glsyst}, henceforth J1433), which lightcurves have been obtained from a 3-year observational campaign carried out at the LMU Wendelstein observatory, using the Wendelstein Wide Field Imager \citep[henceforth \textit{WWFI}, ][]{WWFI_article} of the 2.1m Fraunhofer Wendelstein optical telescope \citep[henceforth \textit{WST}, ][]{wwfi_2014}. 
The aims of this study are to:
\begin{itemize}
    \item Contribute to the study of the Hubble tension: While the accuracy and precision reachable for a single system might be limited compared to the result of collaborations such as TDCOSMO \citep{birrer2020tdcosmo}, this study has the peculiarity of being an independent implementation of TDC for QSOs, thus serving as a fundamental verification for possible biases. Moreover, the system J1433 has not yet been studied within the framework of TDC and therefore is a valuable addition to the limited but growing number of lensed QSO used for cosmology.
    \item Add \textit{WST} as follow-up facility: 
    This study reinforces the argument that 2-meter ground-based telescopes are well suited for high cadence, high Signal-to-Noise-Ratio (SNR), and multi-year observational campaigns for time-domain surveys. This will be of crucial importance to follow up the wealth of strong lenses expected to be discovered with the advent of the ``Big Data Era'' of astrophysics, as large-scale surveys such as Euclid \citep{holloway_23}, the Roman Space Telescope \citep{weiner2020predictions}, the Chinese Space Station Telescope \citep{gong2019cosmology}, and the Vera Rubin Observatory \citep{Oguri_2010,smith2019discovery} will increase the number of suitable systems by three orders of magnitudes \citep{collett2015population}. This project thus adds \textit{WST} to the growing list of reliable ground-based telescopes for time-domain surveys employed for cosmological studies.

\end{itemize} 
Tied to the first point, it is here worth noting that this study differed from the methodology of the H0LiCOW project and of most modern lens modelling methods, as the model of the lens mass is not obtained by a multifilter analysis of the high-resolution exposures available, but it is instead obtained by analysing each of those independently, and the $\Delta \phi$ is later constrained by combining the posterior of the individual model. This approach allowed for an in-depth study of possible biases and pitfalls that such modelling can present. The extensive investment of researcher time proved to be necessary in order to achieve the level of precision required for cosmological inference (as shown in Section \ref{subsec:discard_f160}).

When required, the default cosmology is assumed to be a flat $\Lambda$CDM with $\Omega_{m,0}=0.3$, and $H_0=70\frac{\M{km}}{\M{Mpc\;s}}$, taking care not to be biased by such an assumption. 
All results are obtained by marginalising over every other parameter and taking the 16th, the 50th, and the 84th percentile, representing the median and the $1\,\sigma$ deviation.
 The lens-system-specific Python scripts implemented are available in Github\footnote{\url{https://github.com/GiacomoQueirolo/SDSSJ1433_lens_modelling}}.
 \section{Gravitationally Lensed System}\label{sec:glsyst}
\begin{figure}
	\includegraphics[width=\columnwidth]{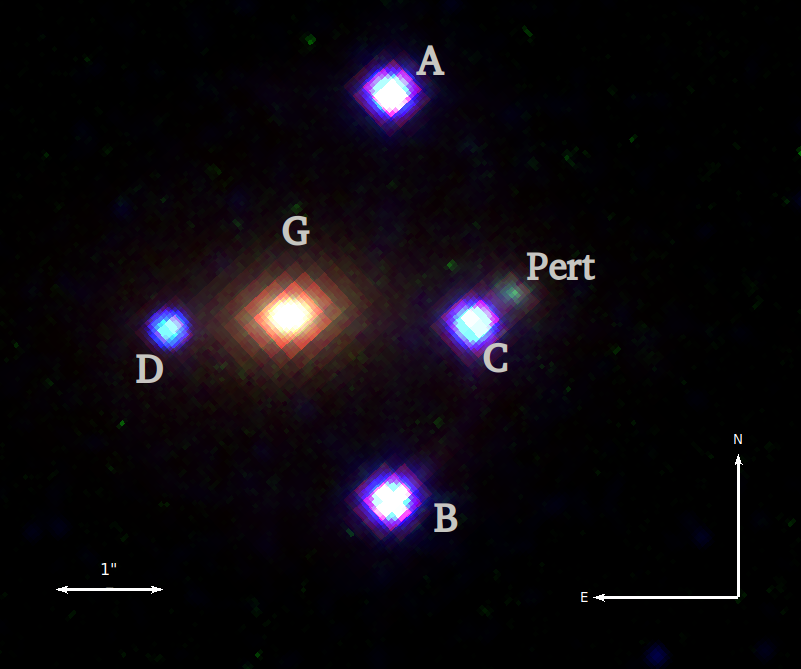} 
    \caption{Colour image from \textit{HST} using all available filters of the system SDSSJ1433+6007. The blue point sources are the four lensed images of the QSO and are indicated with capital letters from A to D (following \citet{agnello_SDSSJ1433}). The central red galaxy is the main lens, indicated with G and ``Pert' indicates a smaller galaxy, northwest of image C, acting as a perturber for the main lens  (likely its satellite galaxy). The North and East directions and the 1-arcsecond scale are also illustrated.}
    \label{fig:J1433_color}
\end{figure}

In this paper, we study the quadruply-lensed QSO SDSSJ1433+6007 (Figure \ref{fig:J1433_color}), discovered by  \citet{agnello_SDSSJ1433}, henceforth A18. Previous analyses of this system were carried out in A18, in \citet{shajib_SDSSJ1433} 
and in \citet{schmidt22_STRIDES},
which differed in data quality (A18), modelling procedure and scope.
The system is composed of a large elliptical foreground galaxy (\textbf{G}) at redshift $z_{\text{lens}}=0.407$ (A18), which acts as the main gravitational lens (henceforth ``main lens'') and a less luminous companion, which we assume to be located at the same redshift and will refer to as satellite galaxy which we treat as a ``perturber'' (\textbf{Pert.}) of the lens model.  The lensed object is a bright QSO located at redshift $z_{\text{source}}=2.737$ (A18) and it appears quadruply lensed (images A, B, C and D). 
All components are shown in the colour Figure \ref{fig:J1433_color}. The host galaxy of the QSO,  due to cosmological redshift, is only visible in the infrared, forming partial arcs or arclets at and around the positions of the images (visible in the individual infrared filters, Figure \ref{fig:all_HST_fig}). \

This system was selected from the pool of 220 lensed QSO's \citep{GLQ_site} known in 2019 based on the following criteria:
\begin{itemize}
    \item More than two observed lensed images
    \item Large separation of the images ($\Delta\theta\gtrsim1$ '')
    \item High luminosity of the images in visible light ($g' \lesssim 23$ mag)
    \item High declination ($dec\gtrsim10^{\circ}$), implying visibility from \textit{WST} for most of the year
    \item Limits on the expected time delay ($1$ day$\lesssim\Delta t\lesssim100$ days)
    \item Simplicity of the system (e.g. avoiding galaxy clusters or high-density environments)
    \item Presence of high-resolution, public imaging data (e.g. \textit{HST} archival data)
    \item Known spectroscopic redshift of source and lens galaxy
\end{itemize}
The first five points were intended to allow and facilitate the observations for the lightcurves (see Section \ref{sec:lc_analysis}) and, correspondingly, their analysis. More than two images meant that, in case one image could not be used (for example due to its excessively low luminosity), the analysis could still succeed, albeit with reduced constraints. The large separation of images, their high luminosity and the high declination all contributed to facilitating the observations given the location of the telescope, estimated limiting magnitude and average seeing. The limits of the time delays were chosen by taking into account the length of the observational campaign to be on the order of one year and the expected sampling of the observations to be on the order of a few days.
The latter two points were chosen to simplify and thus reduce the uncertainty of the lens mass model (see Section \ref{sec:lens_mod}) and the cosmological constraints (see Section \ref{sec:h0}).
The two separate analyses, the lens modelling and the time delay measurement, require two complementary sets of data. For the first, high-resolution \textit{HST} images are used (Section \ref{subsec:HST_data}) to model the lens mass distribution and obtain the Fermat potential (Section \ref{sec:lens_mod}). For the second, high cadence, high SNR lightcurves were obtained from ground-based observations (Section \ref{subsec:creation_lc}) to measure the time delay between the images (Section \ref{sec:lc_analysis}). In Section \ref{sec:flux_ratio} we measure the flux ratio obtained from the two analysis and their discrepancy.  In the final Section \ref{sec:h0} the results are combined to constrain the Hubble parameter. We discuss the obtained results in Section \ref{sec:conclusion}.

\section{Lens Modelling} \label{sec:lens_mod}

For the mass modelling of the lens system we use the public  \texttt{lenstronomy}\footnote{\url{https://lenstronomy.readthedocs.io}} software \citep{BIRRER2018,Birrer_2015} version 1.11.2.
In this Section, we describe the preprocessing of the data (Section \ref{subsec:HST_data}), namely the data reduction, the PSF modelling (Section \ref{subsubsec:psf}) and masking (Section \ref{subsubsec:masking}). We then introduce the choices for mass and light profiles employed in the modelling (Section \ref{subsec:profiles}) and the constraints added from the observed light distribution (Section \ref{subsec:custom_LogL}). These elements are used in the fitting of the model (Section \ref{subsec:lens_modelling}), from which the posteriors are obtained and combined (Section \ref{subsec:compare_&_combine}).
The results of the lens model are also presented and compared to a previous lens model from the literature in (Section \ref{subsec:compare_with_litterature}).

For our analysis, we decided to carry out the lens modelling in all \textit{HST} filter's exposures (hereafter simply filters) separately, and therefore obtain an independent posterior for the lens model parameter for every single filter.

Our models in the different filters share the same mass profile and relative parametric prior, but there are also procedural differences which we will describe below. This approach presents multiple challenges, but it allows us to recognise some critical distinct features in the data and to tackle them individually (e.g. see Section \ref{subsec:discard_f160}). A combined analysis might produce faster results but could be affected by larger uncertainties or unknown and overlooked biases.
Each modelling result is individually assessed based on the presence of structures in the residual image (``goodness'' of the residual) and the convergence of the non-linear solver runs. 
We then compare the posterior of each filter and evaluate the tension between them. If present, we investigate its origin and resolve the tension. Finally the posteriors for the 

Fermat potentials at the image positions for all filters are combined to yield a single result (see Section \ref{subsec:compare_&_combine}) which is then used
to constrain $H_0$ in Section \ref{sec:h0}.
The software \texttt{lenstronomy} operates on angles rather than on pixel positions, and all input and output are therefore defined following this convention. 
All common features between the different models are hereafter described with respect to one of the filters, F814W. Details on the filters are discussed in Section \ref{subsec:HST_data}. Eventual differences from this common approach are then highlighted in the appropriate Section. 

\subsection{\textit{HST} Data for Lens Modelling} 
\label{subsec:HST_data}
\begin{figure}
    \includegraphics[width=\columnwidth]{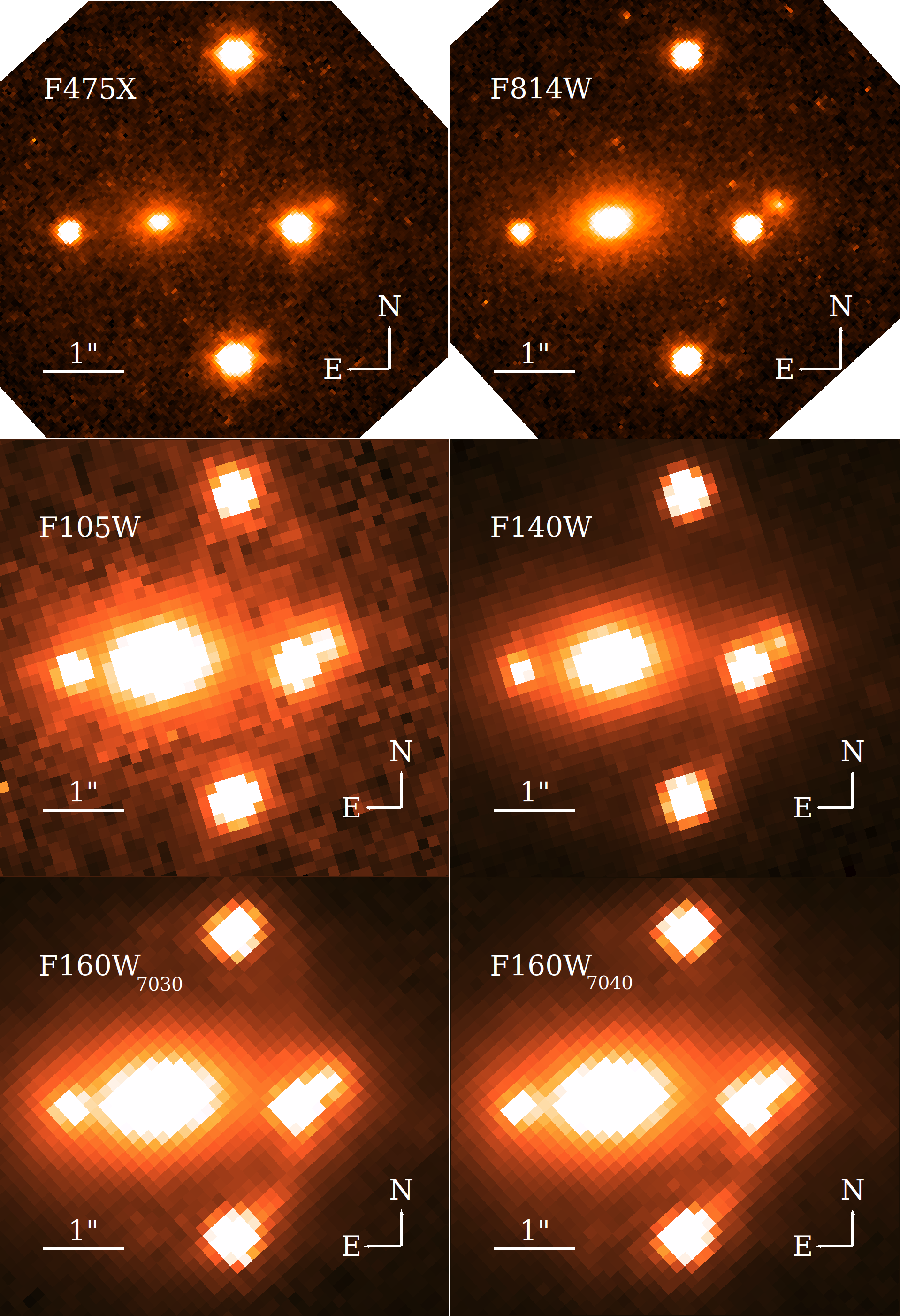}
    \caption{Available \textit{HST} data in the two optical and three near-infrared filters, with the two exposures of F160W. In F140W and F160W we can observe the arclets produced by the multiply imaged host galaxy. For other details refer to Figure \ref{fig:J1433_color}. }\label{fig:all_HST_fig}
\end{figure}
The lens modelling was based on six archival \textit{HST} exposures \citep{HST_archive}, reduced and calibrated, in multiple filters (Figure \ref{fig:all_HST_fig}).
The exposures were taken in two optical filters (F475X and F814W) and three near-infrared filters (F105W, F140W, and F160W) with the Wide Field Camera 3 (WFC3).
The observations were carried out between May 2018 (F475X, F814W, and F160W) and  February 2019 (F105W and F140W). Two consecutive exposures were taken for filter F160W 
(indicated by F160W$_{\M{7030}}$ and F160W$_{\M{7040}}$ when referred to them individually), one after the other and differing only for their exposure time; both were analysed separately. The pixel scale varies between the higher resolution of the optical filters of $\sim 0.040 \frac{"}{\M{pix}}$ and the infrared filters' lower resolution of $\sim 0.128 \frac{\M{"}}{\M{pix}}$ 
(see Table \ref{tab:HST_obs}).
\begin{table}
    \centering
	\begin{tabular}{lcccr} 
		\hline
		Filter & Central  & Exposure & Exposure & Resolution \\
		 & Wavelength [\AA] & Date [yr] & Time [s] & [$\frac{\M{"}}{\M{pix}}$] \\
\hline 
F475X & 49360. & 2018-05-04 & 1504  & 0.040  \\
F814W &  8048.1 & 2018-05-04 & 1428 & 0.040  \\

F105W & 10551.0 & 2019-02-14 & 124.23 & 0.128 \\
F140W & 13922.9 & 2019-02-14 & 446.93 & 0.128 \\
F160W$_{\text{7030}}$ & 15369.2 & 2018-05-04 & 998.47  & 0.128\\
F160W$_{\text{7040}}$ & 15369.2 & 2018-05-04 & 1198.46 & 0.128 \\ 
\hline
	\end{tabular}	\caption{Specifics for the \textit{HST} exposures.}
	\label{tab:HST_obs}

\end{table}
The error frames of the optical filters are not provided in the archive, while they are present in an early reduction step for the infrared filters, but affected by the distortion of the optics, for which the corresponding science images are already corrected. Two different approaches are followed in order to obtain reliable error frames. For the optical filters, the error frames are obtained starting from the raw exposures (also available in the \textit{HST} archive) following the standard Gaussian error propagation:
\begin{equation} \label{eq:err_frame}
   \text{err}\left[\frac{e-}{sec}\right] =
   \frac{\sqrt{\text{(image + sky)}[\frac{e-}{sec}]  \times \text{TEM}[sec] + \text{RN}^2 [e-] } }{\text{TEM }[sec] },
\end{equation}

where RN is the Read-Out-Noise and TEM is the time exposure map. The sky value as well as the science image and error frames are in units of electrons per second. 
The sky is estimated by $\sigma$-clipping the raw exposure,  measuring the median value of the remaining pixels and normalising by the exposure time. Equation \ref{eq:err_frame} is evaluated for each of the four amplifier regions of the CCD separately. The error frames are computed for each single exposure, and these are then propagated to form the final error frame.
As a consistency check, we verified that the scatter of the sky is compatible with the median of the corresponding pixels in the error frame.
We obtained the near-infrared error frames simply by correcting for the distortion of the optics. 
The same procedure previously described as a consistency test for the optical frames showed that the error frames in the infrared filters were affected by undersubtracted ``sky'' brightness. 
To account for this we subtract the median of the ``sky'' ($\text{sky}_{\M{\M{new}}}$) from the science frame and update the error frame following $\text{err}_{\M{new}}=\sqrt{\text{TEM}^2\cdot \text{err}^2 + \text{TEM}\cdot {\text{sky}_{\M{new}}}}/\text{TEM} $ accounting for the exposure map.
This will mostly affect the quality of the point spread function (PSF), as it will be discussed in Section \ref{subsubsec:psf}.

The images are further cropped into a square of $\sim6$'' of edge around the main lens galaxy, in order to limit the modelling to the region of interest.

\subsubsection{PSF Modelling} \label{subsubsec:psf}
Along with the science image and its error frame, a pixelized PSF model (and corresponding error frame) can be given as an input to \texttt{lenstronomy}.

This is then used as a PSF kernel to convolve the modelled light profiles in order to compare them to the input image and compute the likelihood.
It is known that lens model results are strongly dependent on the precision of the PSF models \citep[e.g. ][]{shajib2022tdcosmo} 
as this correlates with the precision of the position of the images and thus of the model in general, as well as the final value of the Fermat potential at the positions of the images. 
To obtain a reliable PSF model we followed the same procedure for all \textit{HST} filters and ground-based observations (\ref{subsec:creation_lc}) using the software \texttt{psf} \citep{Rieffeser_psf}. 
The PSF model is produced by supersampling selected point sources in a flux-conserving way and then stacking them. This significantly increases the resolution of the PSF model and therefore the precision of the image positions, but is mostly effective when multiple sources are considered. The best results for the optical images are obtained by taking the two brightest QSO images, A and B (Figure \ref{fig:J1433_color}), as references for the PSF modelling after the lens light subtraction (described in detail in \ref{subsec:light_prof&llm}), and considering a supersampling of a factor 5. The other images C and D are both dimmer and affected by the perturber light or the main lens light, respectively, and thus are discarded as candidate sources for the PSF modelling.
This approach is possible for both optical filters.

For F105W the PSF can be modelled on the QSO's images A and B once the lens light is locally subtracted. 

Thus the same isophotal technique explained in Section \ref{subsec:light_prof&llm} is applied to subtract at the first order the lens light at the QSO's position. 
Finally, we subtract a constant value from the PSF kernel to correct for undersubtracted lens light, allowing the wings to converge to zero; thus its numerical value is obtained by averaging at the kernel's edge.
For both F140W and F160W, the light of the QSO's images is too severely blended with the arclets of the lensed host galaxy light, the main lens and the perturber light, and therefore can not be used for the modelling. Instead, multiple nearby bright stars are taken as reference  (two for F140W and three for both exposures of F160W, the number depending on the presence and brightness thereof) allowing for a supersampled PSF model. This choice is in general sub-optimal, as the shape of the PSF is colour-dependent and the QSO images are significantly bluer than the stars taken into consideration. This effect is negligible in the near-infrared filters, for which the low pixel resolution renders such PSF differences less severe.
A last step is needed, which is the masking of small light contaminants, i.e. visually detected perturbations of the PSF wings, which might arise due to nearby small secondary objects or straylight. Assuming at the first order the PSF to be circularly symmetrical (especially on the outer wings), we substitute the affected pixels with pixels at the circularly symmetrical opposite of the PSF kernel. The same substitution is done for the error frame. This correction affects only a few pixels and is limited to the kernel's wings
We test the accuracy of our PSF models by comparing their encircled energy (EE) with the one reported in the literature \citep{STSCI_EE} and we find them in agreement, as shown in Figure (\ref{fig:psf_EE_f814}).
Nevertheless, the main test for the goodness of the PSF model was the lens modelling itself. The choice for the models was therefore iteratively optimised with this reference, in particular with respect to the residuals at the image positions. 

\begin{figure}
\includegraphics[width=\columnwidth]{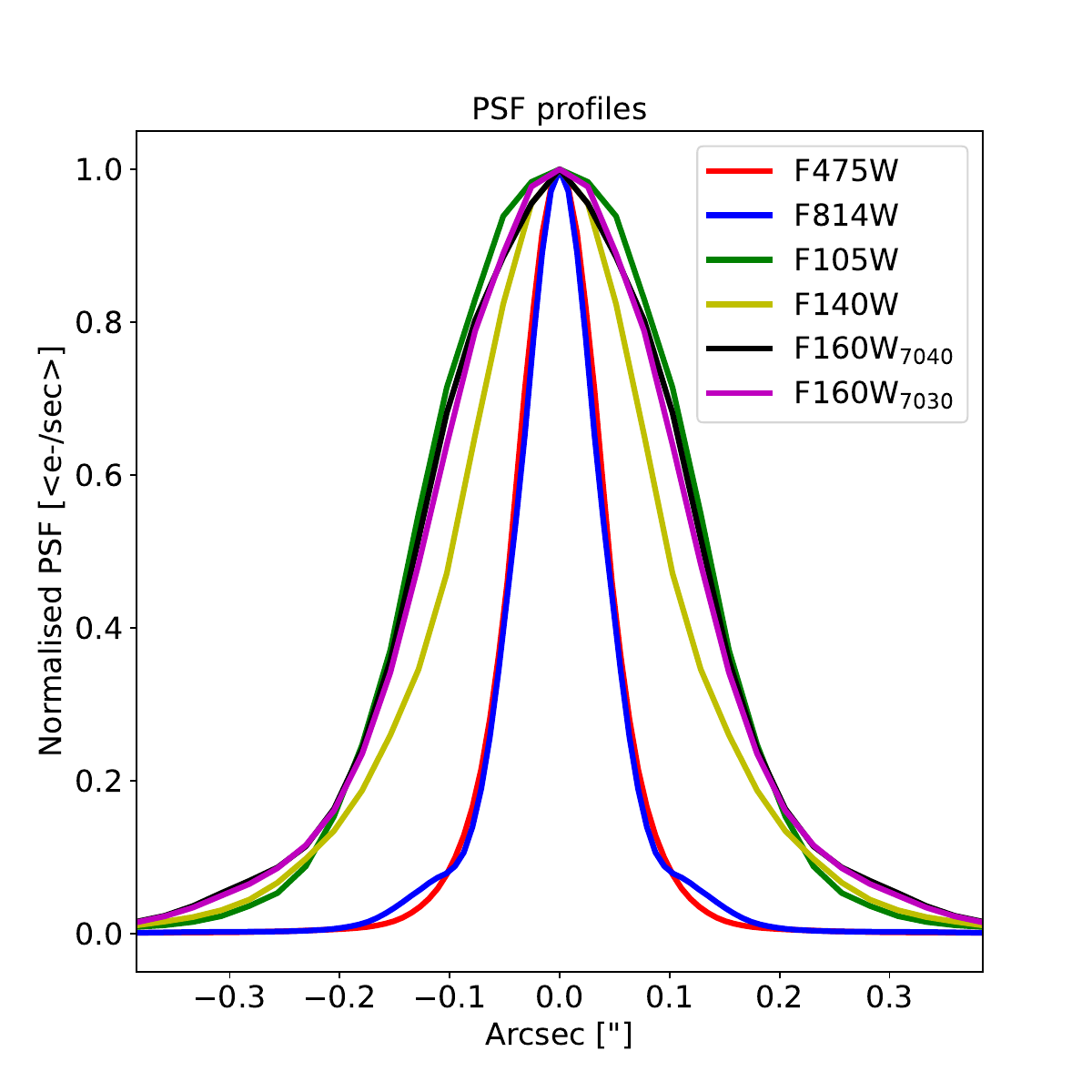} 
\caption{PSF profiles of the different filters, normalised to their maximum value.}
\end{figure}
\begin{figure}
\includegraphics[width=\columnwidth]{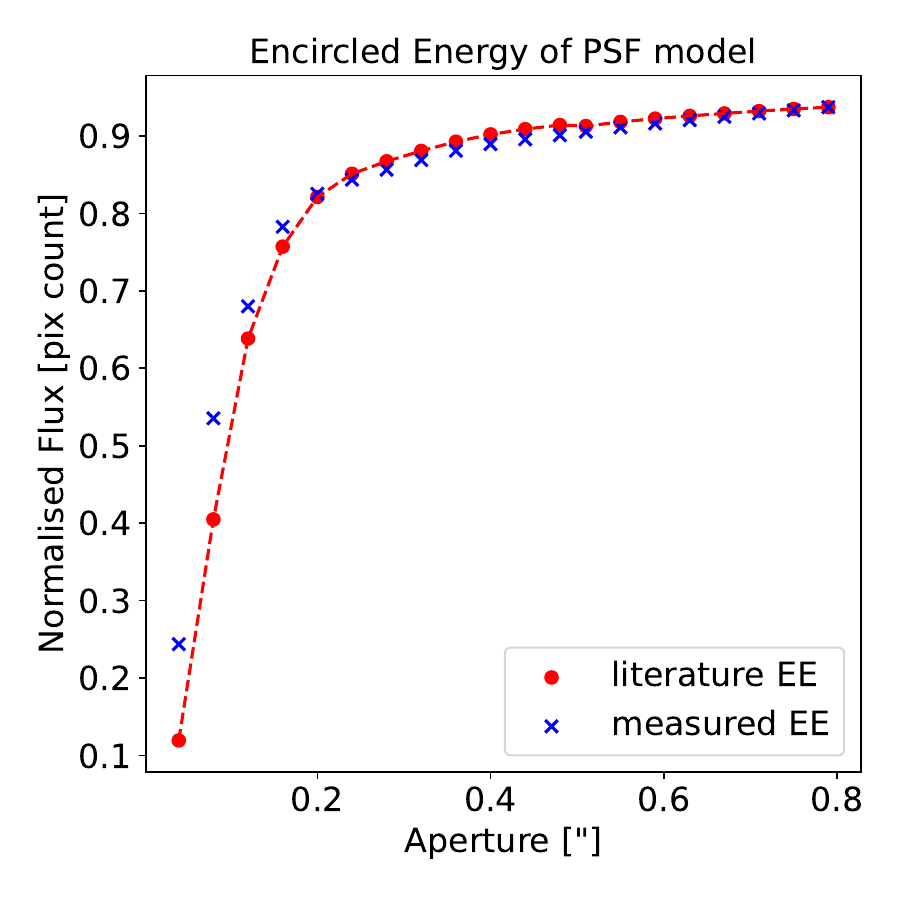} 
\caption{Encircled energy (EE) of the PSF for filter F814W. The ``literature'' values of the EE are taken from \citet{STSCI_EE}. In the Appendix are reported the equivalent plots for the other filters (see Figure \ref{fig:app_EE}).}
    \label{fig:psf_EE_f814}
\end{figure}

\begin{table}
\centering
\resizebox{.8\columnwidth}{!}{
\begin{tabular}{l c r}
\hline
    Filter & FWHM ["] & $\M{R}_{\M{EE}}$ ["] \\
    \hline
        F475X & 0.095 &  0.067\\
        F814W & 0.079 & 0.075\\
        F105W & 0.257 & 0.144\\
        F140W & 0.205 &  0.158 \\
        F160W$_{\M{7030}}$ & 0.257 &0.16\\
        F160W$_{\M{7040}}$ & 0.257 &0.16\\
        \hline
\end{tabular} } 
\caption{Table with FWHM of the various PSF models for the different filters and their radius at which the EE=50$\%$, refereed to as EE radius ($\M{R}_{\M{EE}}$). It can be seen that the FWHM does not always increase with the central wavelength $\lambda$ as expected from a diffraction-limited telescope as \textit{HST}, following $\theta\propto \lambda/\M{d}$ (where d is the telescope aperture, which is constant for all filters, and considering the small angle approximation). This is however very likely due to the distorted shape of the PSF at the wings, possibly due to the presence of diffraction spikes or tracking errors. Instead, $\M{R}_{\M{EE}}$, which is on average significantly smaller than the FWHM as it is dominated by the core of the PSF kernel, recovers the expected dependency on $\lambda$.} 
\label{tab:f814w_psf_fwhm}
\end{table}

\subsubsection{Masking} \label{subsubsec:masking}
To avoid introducing external signals from nearby sources that do not belong to the lens system, \texttt{lenstronomy} allows the user to input a mask image.

In this study, the centroid of the lens system is computed from the initial position of the lensed images, measured with \textbf{\texttt{SExtractor}} \citep[ as later described in \ref{subsubsec:priors}]{sextractor},
and everything further than 3'' from that point is masked for all filters. 
Moreover, the centre of the main lens light is masked, depending on the filter, between 3 and 4 pixels (infrared and optical, respectively), in order to account for residuals at the core from the lens light model. Neither of these masks affects significantly the final model, but they do lower the reduced $\chi^2$.  
Nearby light features that are not believed to belong to the lens system are masked as well; for most filters, these are outside the 3'' radius of the overall masking.

\subsection{Profiles}
\label{subsec:profiles}
\texttt{lenstronomy} models the lens system by using parametric profiles for the mass and light components. This Section describes the approach employed to model each of these.

\subsubsection{Mass Profiles}
\label{subsec:mass_prof}

As in A18 and \citet{shajib_SDSSJ1433}, the mass profile of the main galaxy lens, a massive elliptical, is modelled by an elliptical power law (PEMD following the \texttt{lenstronomy} convention). 

The perturber on the west side of the system is likely to be a satellite galaxy of the main lens. This was not visible in SDSS observations, on which the previous model of A18 was based, and therefore was not modelled. In agreement with the automated modelling choices of \citet{shajib_SDSSJ1433}, we assume it to be at the same redshift of the lens and we model it with a ``singular isothermal sphere'' (SIS).

Finally,  we consider an external shear to account for distortions along the line of sight that were not explicitly modelled. 

These modelling choices were the same for all analysed filters. 
\subsubsection{Light Profiles and Lens Light Model} \label{subsec:light_prof&llm}

For the optical filters, the main lens light profile is modelled a priori and subtracted from the data. Then only the lens light-subtracted image is input to \texttt{lenstronomy}. This step reduces the number of parameters and results in faster and better constraints of the remaining parameters. 

Moreover, as described below  (\ref{subsec:custom_LogL}),  this procedure provided us with lens light parameters that are used as prior values for the lens mass model. 

To obtain this light profile model, the QSO's images have to be subtracted, in particular their extended wings as their light would bias the luminosity of the galaxy. Thus they are first subtracted using a PSF model re-scaled to match the image brightness and resampled to match the position of the QSO with sub-pixel precision. This model is obtained following the same approach explained in Section \ref{subsubsec:psf}, this time taking as reference isolated field stars, as this allows the production of larger PSF kernels which fully contain the extended wings light. The residual of the PSF subtraction in the core of the image is masked, as well as nearby sources. Similarly to the masking process of the PSF model for small contaminants described in Section \ref{subsubsec:psf}, now assuming the lens light profile to be elliptically symmetrical, the masked pixels are substituted by pixels at the elliptically symmetrical opposite with respect to the centre of the light distribution.
Finally, the remaining main lens light is iteratively fitted using elliptical isophotes at increasing radii, up to a semi-major axis of $a_{\text{fix}}^{1/4} =1.013{\M{arcsec}}^{1/4}$, after which the shape parameters are fixed. This is set as the obtained parametric model is stable enough to fit the extended light beyond $a_{\text{fix}}$ while avoiding being biased by unmasked neighbours.

\begin{figure*}
	\includegraphics[width=.8\textwidth]{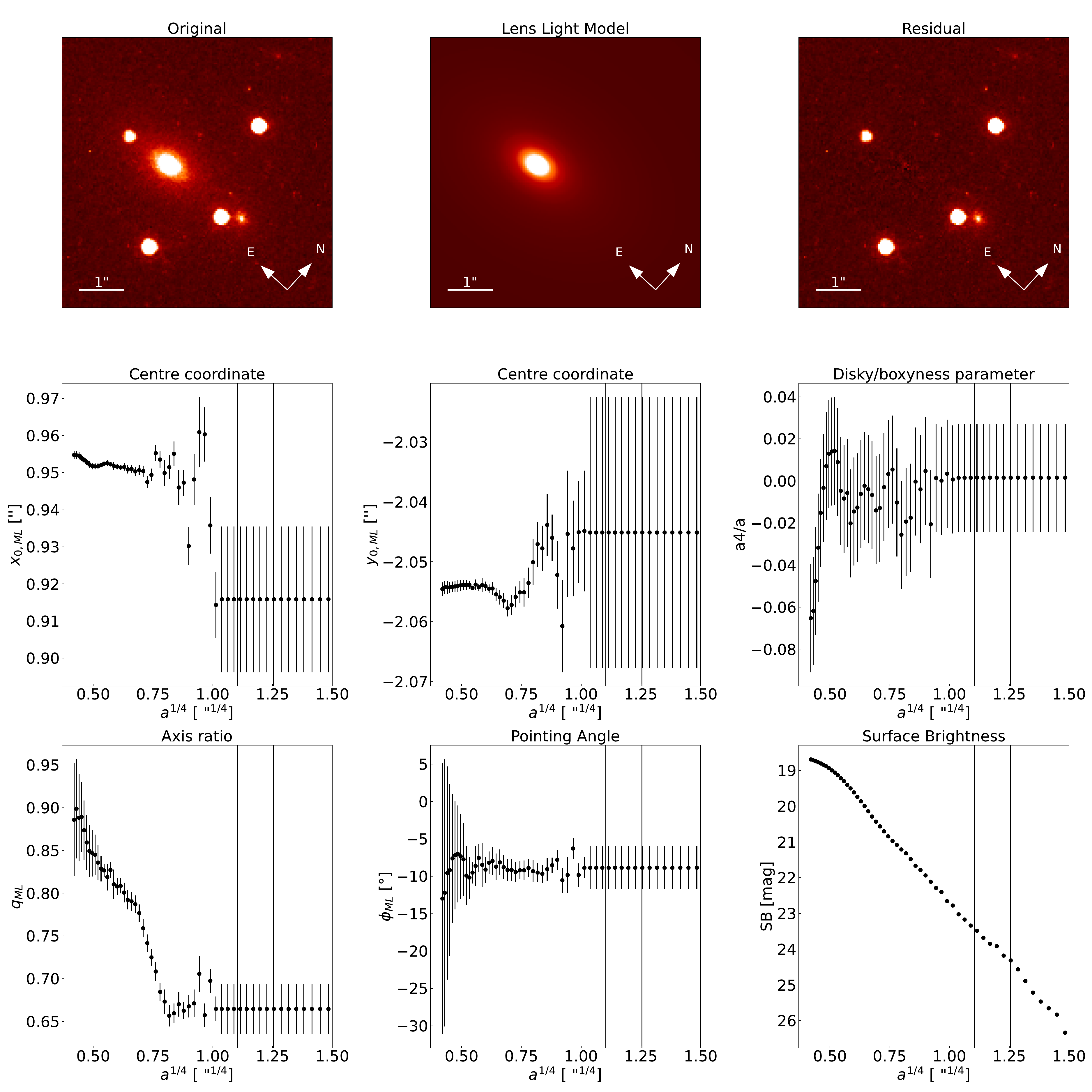} 
    \caption{Result for the main lens light modelling and subtraction for F814W. On the first row, the original science frame is shown, alongside the reconstructed model of the lens light and the resulting subtracted image. Note the small residuals in the centre, which are later masked during the lens modelling (\ref{subsubsec:masking}). On the second and third rows, the parameters of the isophotes are reported relative to the isophotes major axis $a^{1/4}$.
    The two black vertical lines indicate the approximated range of position of the QSO's images. Note that for radii $a^{1/4}$ larger than $a^{1/4}_{\text{fix}}=1.013{\M{arcsec}}^{1/4}$ the parameters are fixed and that the disky/boxyness parameter $a4$ is normalised by the major axis $a$. The centre coordinates $x_{0,ML}$,$y_{0,ML}$ are in arcseconds with respect to the position of image A.}
    \label{fig:isophote_output} 
\end{figure*}

We produced these models following the isophote fitting procedure presented in \citet{klugeisophotespy} and \citet{Kluge2023rhea}. The model's output based on F814W can be seen in Figure \ref{fig:isophote_output}.

We find the galaxy to be more spherical in the centre with an increasing ellipticity towards the outer regions, as shown in Figure (\ref{fig:isophote_output}).

The lensed images are in a range ($1.1$ arcsec$^{1/4}\lessapprox a^{1/4} \lessapprox 1.25$ arcsec$^{1/4}$) where the lens light parameters are already fixed. 

This is the most crucial region for the mass modelling as it contains most of the information used to constrain it, i.e. the lensed arclets of the extended host galaxy and QSO's images.

This procedure is especially effective for the optical filters due to higher pixel resolution, lower main lens light luminosity and absence of arclets of the host galaxy. These features concurred to make the lens light easily recognizable from the neighbouring lights. For the near-infrared images, instead, the blending is too severe and this analysis is not sufficiently precise to subtract the lens light a priori, which we instead fit during the lens modelling run as a parametric model (see Section \ref{subsec:profiles}).  Nevertheless, this procedure can produce a reliable parametric prior for the lens light parameters, although with larger uncertainties. For this reason, this analysis was carried out for the infrared filters as well.  We then considered the averaged values for the centre of the lens light ($x_{c,LL}$,$y_{c,LL}$ in Figure \ref{fig:llm_study}) and ellipticity parameter ($q_{LL}$ and $\phi_{LL}$) between all filters' isophotal result as prior values for the main lens mass modelling. 
The details are described in subsection \ref{subsec:custom_LogL}.

Moreover, as described in Section \ref{subsubsec:psf}, this subtraction is effective at the first order for F105W to subtract the lens light at the QSO's image positions. In this filter, the main lens light is not strongly blended with the QSO's images and the lensed arclets of the host galaxy are not visible. Thus the isophotal lens light model is subtracted in order to model the PSF using the QSO images A and B as reference.

We fit the main lens light for all near-infrared filters during the lens modelling procedure as a ``Sérsic Ellipse''. This allows us to maintain the flexibility of the model to accurately decompose the components of the light profiles.

For similar reasons, we do not subtract the light of the perturber a priori, as its light is blended with the light of the QSO's image D for all filters. We model it instead as a circularly symmetric ``Sérsic'' profile during the modelling run, as it proves to be well-fit even without considering an elliptical component.

For F140W and F160W, the lensed host galaxy is bright enough to be visible and can be modelled explicitly. The parametric model chosen is another circular ``Sérsic'' profile defined in the source plane. A ``Sérsic Ellipse'' was initially considered, but it proved to be too little information in the lensed arclets for an accurate model and it resulted in a overfit of the data.

For the other filters, the host galaxy light is too dim and is therefore not modelled. 
This can be explained by considering the host galaxy redshift $z_{\text{source}}=2.737$  and due to the Balmer break it is expected to be brightest at $\lambda \gtrapprox 1360$nm. 
It is unclear whether a very faint arc can be seen in F814W due to its high resolution and low ``sky'' brightness. In this case, since the result is not affected when an extended source profile is considered, we ignore it in order to not overfit the noise.

The QSO's images are then modelled as a point source in the source plane. Their amplitudes are free to vary and do not depend on the magnification ratio predicted by the lens modelling (see Section \ref{sec:flux_ratio}). 

Thus the model is not biased by intrinsic variability, time-delay and microlensing effects.
Finally, a uniform background ``sky'' brightness is considered to account for any undersubtracted background light, possibly a residual of the previous lens light subtraction. This final degree of freedom does not affect noticeably the result and can be discarded in further analyses without significant loss of accuracy.  

\subsubsection{Joint Parameters} \label{subsubsec:joint_params}

\texttt{lenstronomy} allows the joining of the numerical value of specific parameters between different profiles in order to better constrain the model. This feature is used in the following cases:
\begin{itemize}
    \item the coordinates of the centre of the host galaxy and the QSO have to coincide in the source plane
    \item the coordinates of the centre of the perturber light profile and its mass profile have to be the same

\end{itemize}
Other possible ``joint parameters'' are available, for example, fixing the mass and light profile for the main lens to be coincident. However, given that the light profiles differ between the filters, such constraints bias the models due to variations in the light distribution between filters. Instead, in order to link the mass profile to the observed light profile, log-likelihood terms are added as described in Section \ref{subsec:custom_LogL}.

\subsubsection{Priors}
\label{subsubsec:priors}
We here present the parameters of the chosen profiles used for the model. For the mass profile, these are the Einstein radius  $\theta^{\M{ML}}_{\M{E}}$, the exponent of the power law $\gamma^{\M{ML}}$, the two polar components of the ellipticity $e^{\M{ML}}_1$ and  $e^{\M{ML}}_2$, the centre coordinates $x^{\M{ML}}$ and  $y^{\M{ML}}$ for the main lens ($\M{ML}$) and, similarly, $\theta^{\M{P}}_{\M{E}}$,   $x^{\M{P}}$,  $y^{\M{P}}$ for the perturber ($P$). Finally, the two polar components of the external shear are indicated by $\gamma^{\M{Shear}}$ and $\psi^{\M{Shear}}$.

Additionally, there are eight positional free parameters for the QSO's image positions in the image plane,  $x_i^{\M{QSO}}$ and $y_i^{\M{QSO}}$ for $i=\{A,B,C,D\}$, along with their independent (see \ref{subsubsec:joint_params}) intensities $I_{0,i}^{\M{QSO}}$. All such parameters are common to all filters. Furthermore, each specific filter also produced between 4 to 10 free parameters for the luminous profiles depending on the subtraction of the main lens light (\ref{subsec:light_prof&llm} ) and the visibility of the extended host galaxy: for the main lens, perturber and host galaxy Sérsic profiles there are the half-light radii $R^{\M{ML}}$, $R^{\M{P}}$, $R^{\M{HG}}$, the Sérsic indexes $n^{\M{ML}}$, $n^{\M{P}}$, $n^{\M{HG}}$, and the central intensities  $I_0^{\M{ML}}$, $I_0^{\M{P}}$, and $I_0^{\M{HG}}$, respectively. The Sérsic profile of the main lens, when modelled, is elliptical, thus there are two more elliptical polar components  $e^{\M{ML, S\acute er}}_{1}$ and  $e^{\M{ML,  S\acute er}}_{2}$. Finally, the uniform background ($\M{BG}$) has one component, a constant intensity $I^{\M{Bkg}}$. A recapitulation of the parameters used in the different models is shown in Table \ref{tab:prior_params}.

\begin{table*}
\resizebox{.7\textwidth}{!}{
\begin{tabular}{lccccccc}
          Model&Parameter&  Variable&  F475X&  F814W&  F105W&  F140W&  F160W\\ \hline \hline
          \multirow{6}{*}{$\M{PEMD}^{\M{ML}}$} 
          & X Center & $x^{\M{ML}}$ & $\checkmark$ & $\checkmark$ & $\checkmark$ & $\checkmark$ & $\checkmark$ \\
          & Y Center & $y^{\M{ML}}$ & $\checkmark$ & $\checkmark$ & $\checkmark$ & $\checkmark$ & $\checkmark$ \\[1mm]
          & Einstein radius & $\theta^{\M{ML}}_{\M{E}}$& $\checkmark$ & $\checkmark$ & $\checkmark$ & $\checkmark$ & $\checkmark$ \\[1mm]
          & Power & $\gamma^{\M{ML}}$& $\checkmark$ & $\checkmark$ & $\checkmark$ & $\checkmark$ & $\checkmark$ \\[1mm]
          & Polar Ellipticity (1) & $e^{\M{ML}}_1$& $\checkmark$ & $\checkmark$ & $\checkmark$ & $\checkmark$ & $\checkmark$ \\[1mm]
          & Polar Ellipticity (2) & $e^{\M{ML}}_2$& $\checkmark$ & $\checkmark$ & $\checkmark$ & $\checkmark$ & $\checkmark$ \\[1mm]
          \hline
          \multirow{3}{*}{$\M{SIS}^{\M{P}}$} 
          & X Center & $x^{\M{P}}$ & $\checkmark$ & $\checkmark$ & $\checkmark$ & $\checkmark$ & $\checkmark$ \\[1mm]
          & Y Center & $y^{\M{P}}$ & $\checkmark$ & $\checkmark$ & $\checkmark$ & $\checkmark$ & $\checkmark$ \\[1mm]
          & Einstein radius & $\theta^{\M{P}}_{\M{E}}$& $\checkmark$ & $\checkmark$ & $\checkmark$ & $\checkmark$ & $\checkmark$ \\[1mm]
          \hline
          \multirow{2}{*}{External Shear}
          & Shear Power & $\gamma^{\M{Shear}}$ & $\checkmark$ & $\checkmark$ & $\checkmark$ & $\checkmark$ & $\checkmark$ \\[1mm]
          & Shear Angle & $\psi^{\M{Shear}}$ & $\checkmark$ & $\checkmark$ & $\checkmark$ & $\checkmark$ & $\checkmark$ \\[1mm]
          \hline
          \hline
          \multirow{8}{*}{QSO} 
          & X Coord. image A & $x_{\M{A}}^{\M{QSO}}$ & $\checkmark$ & $\checkmark$ & $\checkmark$ & $\checkmark$ & $\checkmark$ \\[1mm]
          & Y Coord. image A & $y_{\M{A}}^{\M{QSO}}$ & $\checkmark$ & $\checkmark$ & $\checkmark$ & $\checkmark$ & $\checkmark$ \\[1mm]
          & X Coord. image B & $x_{\M{B}}^{\M{QSO}}$ & $\checkmark$ & $\checkmark$ & $\checkmark$ & $\checkmark$ & $\checkmark$ \\[1mm]
          & Y Coord. image B & $y_{\M{B}}^{\M{QSO}}$ & $\checkmark$ & $\checkmark$ & $\checkmark$ & $\checkmark$ & $\checkmark$ \\[1mm]
          & X Coord. image C & $x_{\M{C}}^{\M{QSO}}$ & $\checkmark$ & $\checkmark$ & $\checkmark$ & $\checkmark$ & $\checkmark$ \\[1mm]
          & Y Coord. image C & $y_{\M{C}}^{\M{QSO}}$ & $\checkmark$ & $\checkmark$ & $\checkmark$ & $\checkmark$ & $\checkmark$ \\[1mm]
          & X Coord. image D & $x_{\M{D}}^{\M{QSO}}$ & $\checkmark$ & $\checkmark$ & $\checkmark$ & $\checkmark$ & $\checkmark$ \\[1mm]
          & Y Coord. image D & $y_{\M{D}}^{\M{QSO}}$ & $\checkmark$ & $\checkmark$ & $\checkmark$ & $\checkmark$ & $\checkmark$ \\[1mm]
          \hline
          \multirow{4}{*}{ QSO Intensities}
          & A & $I_{0,\M{A}}^{\M{QSO}}$ & $\checkmark$ & $\checkmark$ & $\checkmark$ & $\checkmark$ & $\checkmark$ \\[1mm]
          & B & $I_{0,\M{B}}^{\M{QSO}}$ & $\checkmark$ & $\checkmark$ & $\checkmark$ & $\checkmark$ & $\checkmark$ \\[1mm]
          & C & $I_{0,\M{C}}^{\M{QSO}}$ & $\checkmark$ & $\checkmark$ & $\checkmark$ & $\checkmark$ & $\checkmark$ \\[1mm]
          & D & $I_{0,\M{D}}^{\M{QSO}}$ & $\checkmark$ & $\checkmark$ & $\checkmark$ & $\checkmark$ & $\checkmark$ \\[1mm]
          \hline
          \multirow{3}{*}{ Sérsic$^{\M{P}}$}
          & Half-light Radius &  $R^{\M{P}}$& $\checkmark$ & $\checkmark$ & $\checkmark$ & $\checkmark$ & $\checkmark$ \\[1mm]
          & Sérsic index &  $n^{\M{P}}$& $\checkmark$ & $\checkmark$ & $\checkmark$ & $\checkmark$ & $\checkmark$ \\[1mm]
          & Intensity &  $I_0^{\M{P}}$& $\checkmark$ & $\checkmark$ & $\checkmark$ & $\checkmark$ & $\checkmark$ \\[1mm]
          \hline
          \hline
          \multirow{5}{*}{ Sérsic$^{\M{HG}}$}
          & Half-light Radius &  $R^{\M{HG}}$& - & - & - & $\checkmark$ & $\checkmark$ \\[1mm]
          & Sérsic index &  $n^{\M{HG}}$& - & - &- & $\checkmark$ & $\checkmark$ \\[1mm]
          & Intensity &  $I_0^{\M{HG}}$& - & - & - & $\checkmark$ & $\checkmark$ \\[1mm]
          \hline
          \multirow{7}{*}{ Sérsic$^{\M{ML}}$}
          & X Center & $x^{\M{ML, S\acute er}}$ & - & - & $\checkmark$ & $\checkmark$ & $\checkmark$ \\[1mm]
          & Y Center & $y^{\M{ML, S\acute er}}$ & - & - & $\checkmark$ & $\checkmark$ & $\checkmark$ \\[1mm]
          & Polar Ellipticity (1) & $e^{\M{ML, S\acute er}}_1$& - & - & $\checkmark$ & $\checkmark$ & $\checkmark$ \\[1mm]
          & Polar Ellipticity (2) & $e^{\M{ML, S\acute er}}_2$& - & - & $\checkmark$ & $\checkmark$ & $\checkmark$ \\[1mm]
          & Half-light Radius &  $R^{\M{ML}}$& - & - & $\checkmark$ & $\checkmark$ & $\checkmark$ \\[1mm]
          & Sérsic index &  $n^{\M{ML}}$& - & - & $\checkmark$ & $\checkmark$ & $\checkmark$ \\[1mm]
          & Intensity &  $I_0^{\M{ML}}$& - & - & $\checkmark$ & $\checkmark$ & $\checkmark$ \\[1mm]
          \hline
          
          Background&Intensity&$I_0^{\M{Bkg}}$& $\checkmark$ & $\checkmark$ & $\checkmark$ & $\checkmark$ & $\checkmark$ \\[1mm]
          \hline\hline
          \multicolumn{3}{|c|}{Total \textit{N} of Parameters} & 27 & 27 & 34 & 37 & 37 \\
          \multicolumn{3}{|c|}{Total \textit{N} of non-linear Parameters} & 21 & 21 & 27 & 29 & 29 \\
          \hline
    \end{tabular} }
    \caption{Parameter used for the different filters. The double lines separate the list in mass parameters, light parameters correlated to the mass profiles and light parameters independent from the mass profiles. All intensities ($I_0^{\M{ML}}$, $I_0^{\M{P}}$,  $I_0^{\M{HG}}$, $I_0^{\M{Bkg}}$ and $I_{0,i}^{\M{QSO}}$) are linear parameters and are not sampled explicitly by the non-linear solver (see Section \ref{subsec:lens_modelling}). The centre of the Sérsic profile of the host galaxy is not a free parameter, as it is defined to be identical to the position of the QSO in the source plane (Section \ref{subsubsec:joint_params}).}
    \label{tab:prior_params}
\end{table*}
Each of these parameters is introduced with the same identical uniform prior between their two limits.

All parameters' priors are by default set as uniform in \texttt{lenstronomy}, based on the minimum and maximum given by the user. Furthermore, the non-linear solvers (see \ref{subsec:lens_modelling})  are initialised around a starting value with a given radius, both of them also inputted by the user.
Such values for most of these parameters are based on a prior knowledge of the system, obtained from the previous mass or light models, as well as physical constraints. 
Some iterative adaptation of the prior space was also implemented during the modelling to reduce the parametric space, while at the same time avoiding overfitting.
The only parametric priors that differed from this approach are the positions of the observed luminous components: QSO's lensed images, main lens centre, and perturber centre. These parameters' values are obtained by analysing each frame with \textbf{\texttt{SExtractor}} \citep{sextractor}, and are not iteratively adapted as their parametric freedom is deemed fairly well constrained. There is furthermore no significant tension between the central coordinates of the main lens obtained here and those obtained from the isophotal light modelling described in Section \ref{subsec:light_prof&llm}, therefore such two priors can be considered to be equivalent.

Note that the uniform prior in the lens parameters does not correspond to a uniform prior in the Fermat potential parameter, as this is a non-linear combination of these parameters. This will be later discussed in \ref{subsec:compare_&_combine}. 

\begin{figure}
    \centering
    \includegraphics[width=0.5\textwidth]{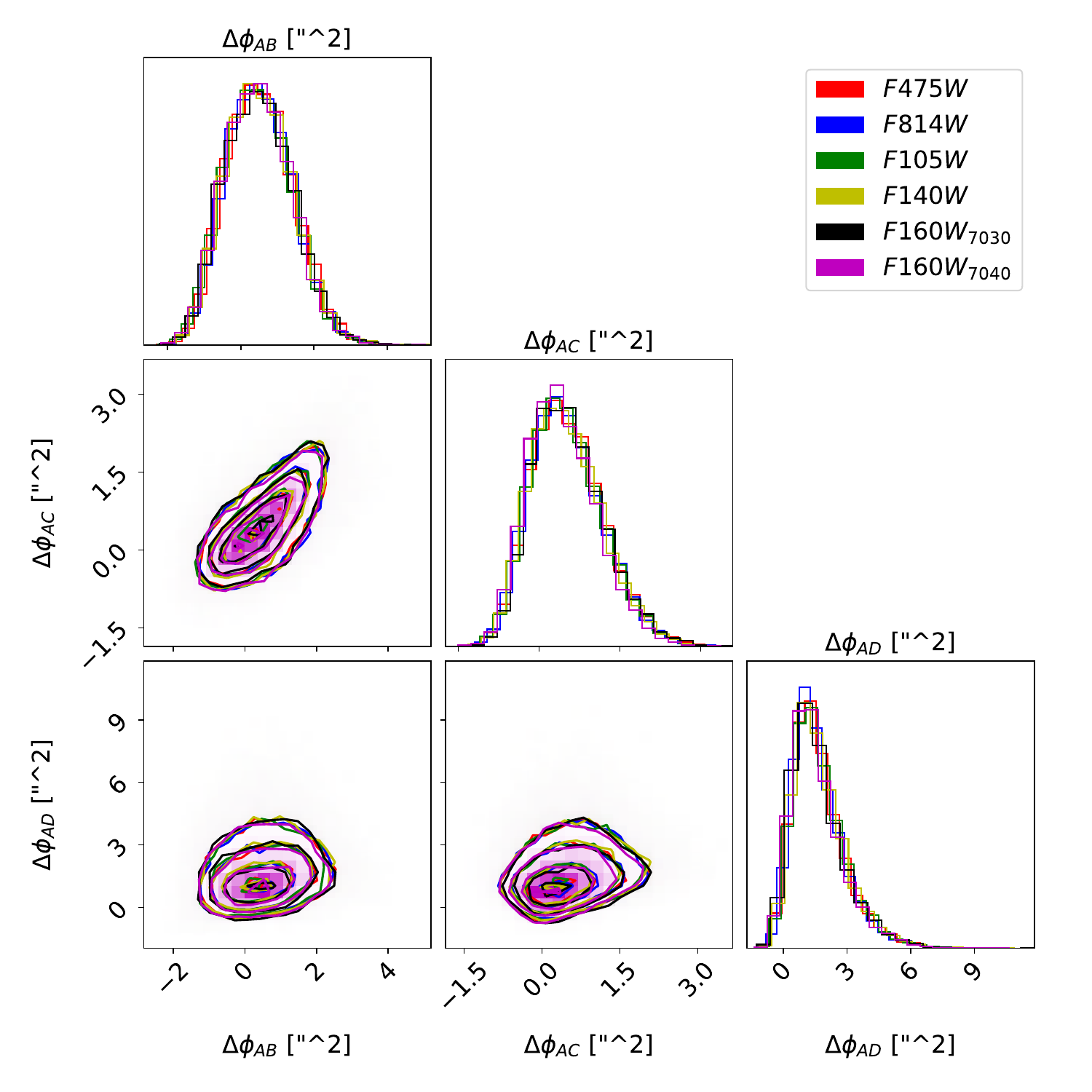}
    \caption{Superposition of the prior distribution for the Fermat potential differences at the image positions $\Delta \phi_{ij}$  for the different filters. We propagate them from the lens model's priors for each filter.
    It can be seen that the small differences in prior do not correspond to a significant difference in the prior of $\Delta \phi$.}
    \label{fig:prior_Df_sup}
\end{figure}

Considering the full priors for the different filters' models, it can be seen that they will not present the same dimensionality, as different luminous profiles are considered with their relative parameters. These are nonetheless only nuisance parameters and are marginalised over after the lens modelling. For this reason, each filter's model could have independent priors for these parameters. In principle, this, as well as the fact that the initial image positions are independently measured, could also affect the prior of the Fermat potential, as it depends on the lens parameters as well as on the position of the images. 
Fortunately, it can be shown numerically that these effects are negligible and the priors are fully compatible, as shown in Figure (\ref{fig:prior_Df_sup}).  We obtain these priors by numerically propagating a sampled lens prior for each filter to $\Delta\phi$. The prior of the Fermat potential difference is therefore considered to be the same for all filters. Finally, the obtained prior in $\Delta\phi$ covers a significantly larger parametric volume compared to the posterior obtained, on the order of one magnitude larger, as can be seen from comparing it to the posterior later reported in Figure \ref{fig:comb_df_prob}.

\subsection{Likelihood Terms}\label{subsec:custom_LogL}

\begin{figure}
    \includegraphics[width=\columnwidth]{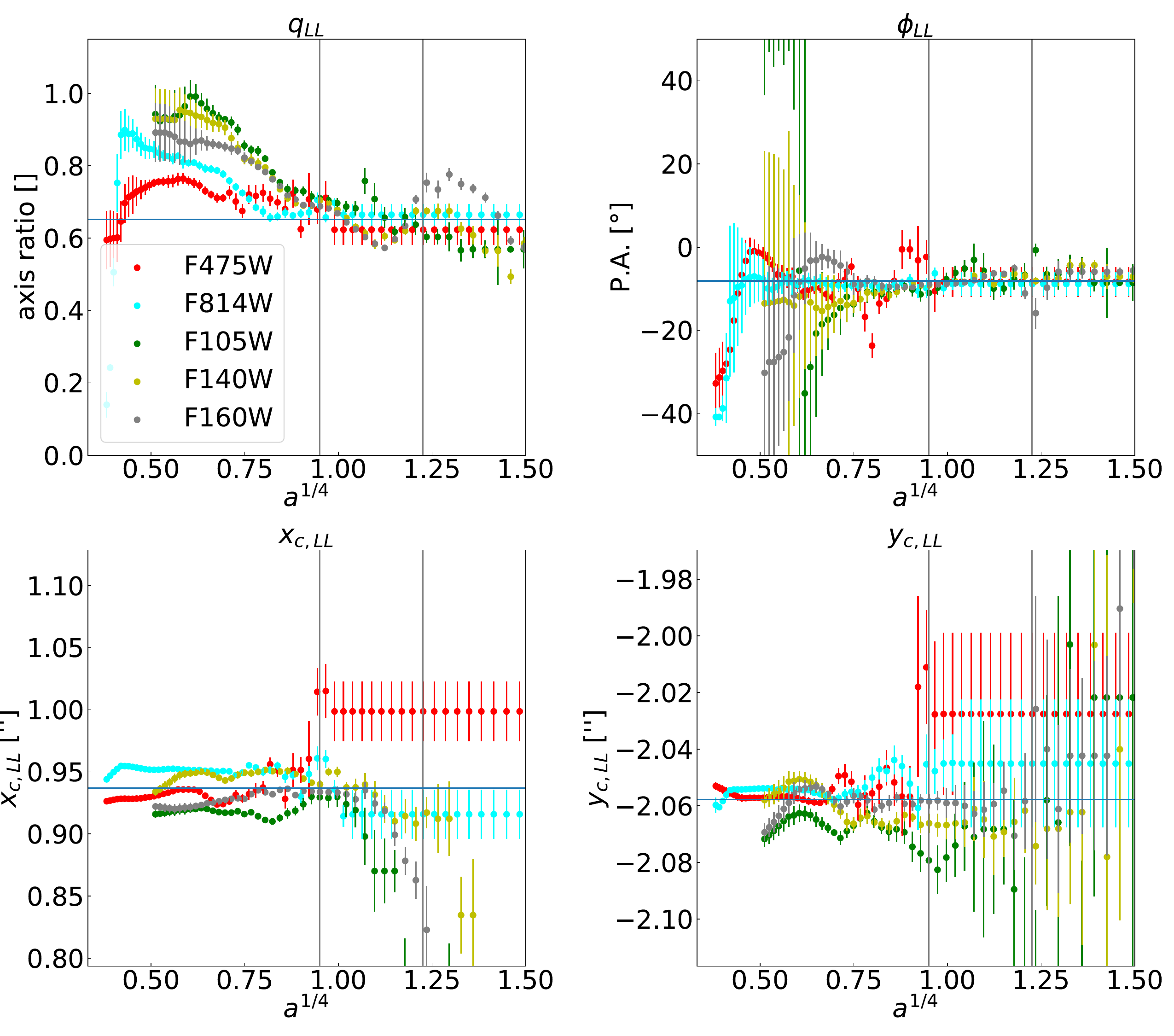}
    
    \caption{Comparison between lens light parameters between the different isophotal fits. The various colours indicate the filter and $a^{1/4}$ is the semi-major axis from the centre of the isophotes. The black vertical lines represent the approximate range in which the QSO's images are located with respect to the lens centre, whereas the blue horizontal line is the mean taken within this area. These means are taken as prior for the mass model of the main lens, as explained in Section \ref{subsec:custom_LogL}, and their values are shown in Table \ref{tab:llm_study}. The differences in starting value for $a^{1/4}$ are due to the different pixel resolutions of the optical and the near-infrared filters.} 
    \label{fig:llm_study}
\end{figure}

In \texttt{lenstronomy}, the likelihood is taken as log-likelihood and is by default computed from a $\chi^2$ function depending on the data, the model, and the error frame obtained for a given set of parameters. The term ``\texttt{check_matched_source_position}'' was considered in this analysis, adding a punishing factor for models in which the backward ray-traced images did not match a single source position in the source plane. 
Of particular significance, similar additional punishing terms ``\texttt{custom_logL_addition}'' were introduced in order to constrain the model given the observed lens light. Specifically, we considered the centre and ellipticity parameters of the lens light as priors for the corresponding lens parameters. 
These were obtained by analysing the isophotal fitting described in \ref{subsec:light_prof&llm} from all filters. As previously described, this was useful to subtract the lens light only for the optical filters, but it produced a reliable parametric model for all exposures. Thus we could obtain the centroid and ellipticity of the main lens light. In order to take advantage of all information from all filters and, at the same time, maintain the same prior and constraints for all models, we considered the averages over the fits in all filters as the same priors for all lens mass models. The coordinates of the main lens light (MLL) centroids were then $x_{c,LL}$, $y_{c,LL}$, and the two ellipticity parameters were the axis ratio $q_{LL}$ and the pointing angle $\phi_{LL}$. Given the 
dependence of such parameters on the radius, we took the average within the region limited between the closest and furthest QSO's image from the centre. This roughly corresponds to $0.94 <a^{1/4}<1.23$ in units of arcsec$^{1/4}$, indicated by the two vertical grey lines in Figure \ref{fig:llm_study}. A further motivation for this is that the inner region can have local variation, for example for $q_{LL}$, which indicates that the lens is rounder at the centre. Furthermore, at $a^{1/4}>a_{\text{fix}}^{1/4}$ the isophotes parameters are fixed. Moreover, the lens analysis is most sensitive to the profile within and at the images' positions, while is less affected by external effects.

\begin{table}\centering
\resizebox{.6\columnwidth}{!}{
    \begin{tabular}{c|c}
    \hline
$\langle q_{\M{LL}}\rangle$ [\, ]& 0.65 $\pm$ 0.04 \\
$\langle \phi_{\M{LL}}\rangle$ [$^\circ$] & -8.1 $\pm$ 1.5 \\
$\langle x_{\M{LL}}\rangle$ ['']& 0.94 $\pm$ 0.01 \\
$\langle y_{\M{LL}}\rangle$ ['']& -2.06 $\pm$ 0.01  \\
\hline
    \end{tabular}}
    \caption{The prior values obtained for the axis ratio $q_{\M{LL}}$, the pointing angle $\phi_{\M{LL}}$ and the central coordinates $x_{\M{LL}}$ and $y_{\M{LL}}$ of the main lens luminous profile. Those values are obtained from the isophotal analysis of the lens light as shown in Figure \ref{fig:llm_study} and described in Section \ref{subsec:light_prof&llm}. Note that the coordinates are defined with respect to image A.}
    \label{tab:llm_study}
\end{table}
These measured averages $\langle q_{\M{LL}} \rangle,\langle \phi_{\M{LL}} \rangle,\langle x_{\M{LL}}\rangle$ and $\langle y_{\M{LL}}\rangle$  were input as prior into all models and entered the calculation of the log-likelihood, for each of the steps of the non-linear solvers, as ``\texttt{custom_logL_addition}'' terms. Their specific values are reported in Table \ref{tab:llm_study}. 

For the axis ratio $q$, it's assumed that the mass profile should not be more elliptical than the light profile. Hence, similar to the approach of \cite{schmidt22_STRIDES}, the punishing term is computed as follows:
\begin{equation}   \label{eq:custom_logL_qll} 
\log L_{\M{q}}   = 
\begin{cases}
 0 & \M{if } q\leq (\langle q_{\M{LL}} \rangle -0.1) \\
 - \frac{[q- (\langle q_{\M{LL}}\rangle -0.1) ]^2}{2 \sigma_{q_{\M{LL}}}^2} &  \M{else}
\end{cases}
\end{equation} 
 $\sigma_{q_{\M{LL}}}$ is obtained from the propagated uncertainty on $\langle q_{\M{LL}} \rangle$ scaled by a factor of 3.
For the pointing angle $\phi$ the punishing term is a standard  $\chi^2$ function, assuming a $\sigma_{\phi_{\M{LL}}}=4.5^{\circ}$, whereas for $x_{\M{c}}$, $y_{\M{c}}$ the Euclidean distance between the mass and prior light centroid $d_{\M{c,LL}}(x_{\M{c}},y_{\M{c}},x_{\M{c,LL}},y_{\M{c,LL}})$ is computed and enters the $\chi^2$ function with $\sigma_{d_{\M{c,ML}}}=0.4''$. Hence:

\begin{align} 
\log L_{\phi}&=-\frac{(\phi-\phi_{\M{LL}} )^2}{2 \sigma_{\phi_{\M{LL}}}^2} \\
\log L_{d}&= -\frac{\left(d_{\M{LL}}\right)^2}{2 \sigma_{d_{\M{LL}}}^2} =-\frac{\left(\sqrt{(x_c^2-\langle x_{\M{LL}} \rangle )^2 + (y_c-\langle y_{\M{LL}} \rangle )^2} \right)^2}{2 \sigma_{d_{\M{LL}}}^2}  
\end{align}
We argue that these constraints are less stringent compared to those implemented in \citet{schmidt22_STRIDES}, in particular for the axis ratio $q$. This is due to the low number count of the SLACS sample upon which these correlations are drawn, composed of 63 lenses.  

\subsection{Modelling Run} \label{subsec:lens_modelling}

The modelling process was composed of two steps; firstly a Particle Swarm Optimisation \citep[PSO, ][]{pso} was run in order to locate the optima of the parametric space. We chose to initialise the PSO run with 300 particles and 800 steps. 

Following the implementation of \texttt{lenstronomy}, the PSO convergence criteria are based on the proximity of the particles in the parameter space to the point found to correspond to the maximum likelihood during the optimisation. 
The point where the PSO converges was used to produce the model shown in the plots. From this position, we ran a Monte Carlo Markov Chain (MCMC) implemented with \texttt{emcee} \citep{emcee}. We used a burn-in of over 2000 steps followed by 8000 steps, and a ``walker ratio'' of 10, i.e. the ratio of walkers to the number of non-linear parameters. Given that this varies between the different models (due to the different light profiles considered, see Section \ref{subsec:light_prof&llm} and Table \ref{tab:prior_params}, the number of walkers spanned between 210 and 290. The MCMC sampled the posterior allowing us to obtain the covariance of the model parameters around the optimum.
\begin{figure*}
    \includegraphics[width=.8\textwidth]{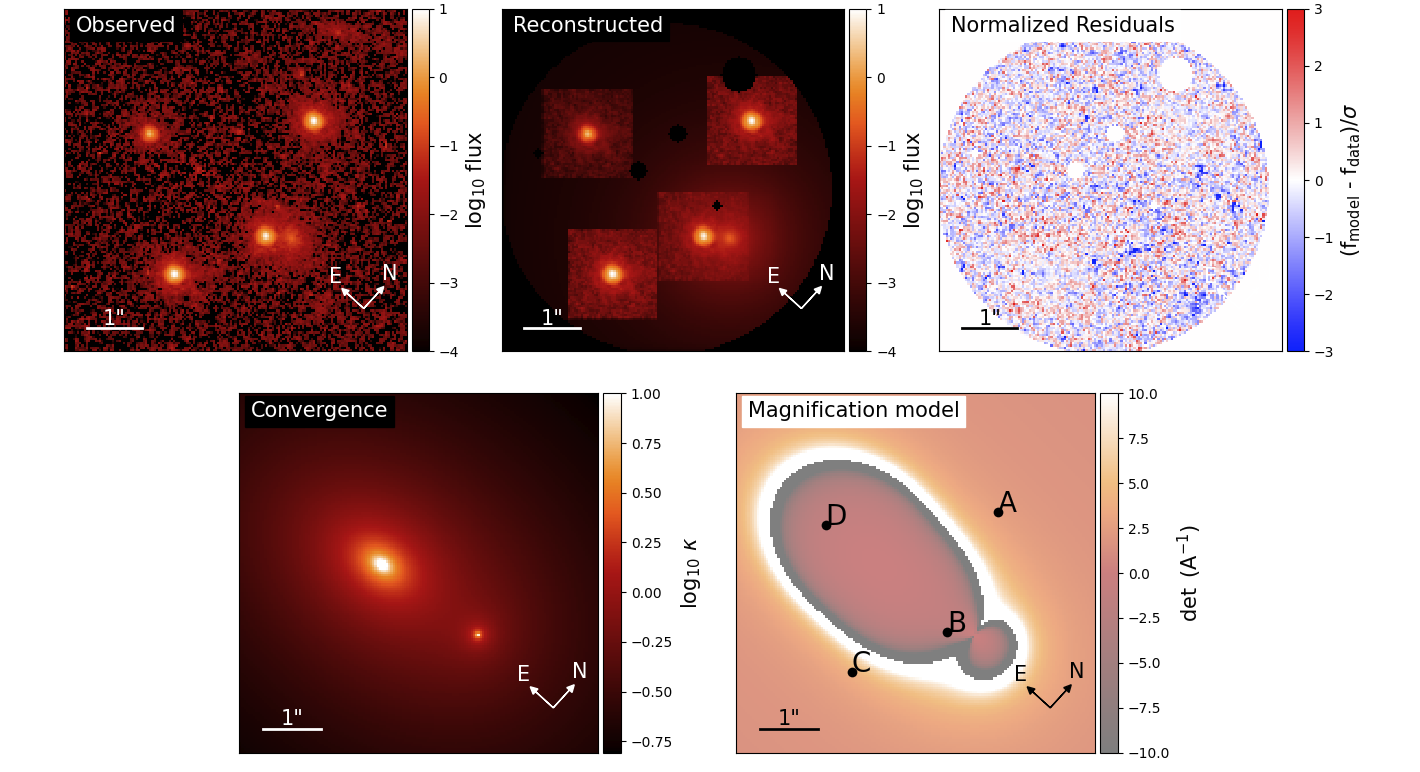}
    \label{fig:lenstronomy_f814w}
    \caption{The resulting model for F814W: \textit{Top}: ``Observed'' is the input image, note that here the main galaxy light has been previously modelled and subtracted (\ref{subsec:light_prof&llm}), leaving a negligible residual in the centre. ``Reconstructed'' is the reconstructed light profile, considering also the masking; note the visible square of the PSF at the images' positions and the masking of the main lens light residual. ``Normalised Residual'' is the result of the subtraction of the previous two images normalised by the error frame; the masked pixels are ignored. \textit{Bottom}: ``Convergence'' is the convergence map in logarithmic scale. ``Magnification model'' shows the magnification map and the images' positions.}
\end{figure*}

Once completed, the output of the program reconstructs a modelled image of the lens system which can be compared with the data, as well as a normalised residual map, a convergence map, a magnification map, and the reconstructed unlensed map of the background source plane with the position of the QSO and the reconstructed host galaxy (when visible) - for reference, consider Figure (\ref{fig:all_HST_fig}).
We inspected the models obtained from each single filter to confirm the reliability of the results. Firstly, no strong residual had to be present in the normalised residual map - this may be the case when an imperfect PSF model is considered. Secondly, the MCMC chain had to be converged. No universal definition for such convergence exists and we instead relied upon several indicators. Firstly, the convergence can be visually verified by plotting the ``MCMC behaviour'', i.e. the average position of the walkers during the run for each parameter. If a strong trend can be seen outside the model is considered not converged. Secondly, we observed the corner plot of the MCMC chain which shows the direct parameter correlations; distributions that showed multimodality are deemed not converged.
We reconsidered the non-converged model by either running the MCMC (and/or PSO) longer or testing new initial parameters, as well as their boundaries.  
 Not all parameters had the same weight, as this analysis is focused on lens mass parameters; therefore non-convergence of light parameters (most notably for the source $R^{\M{HG}}$ and $n^{\M{HG}}$
) were not highlighted as an issue. 
This is not meant to be a quantitative or precise approach, but only a very inexpensive test (in terms of time and complexity). A quantitative convergence criterion is then added, following \citet{ertl23} (eq. 7). 
We consider the medians of the log-likelihood of the first and last $\sim 70,000$ points after burn-in and if the difference is smaller than 2, we deem the chain to be fully converged. If this convergence criterion was met, the comparison with other models is a further test of the goodness of the single result.

Once these tests were passed successfully, the posterior for the Fermat potential $\phi$, $P(\phi|\mathcal{D})$, could be calculated from the MCMC chain of each model. This depends on the images' positions and is obtained by computing it for each single MCMC step. 
The differences were then considered with respect to image A as in $\Delta \phi_{\M{A}j} = \phi_j - \phi_{\M{A}}$, where {$j$=B, C, D}. This produced posteriors in a three-dimensional parametric space, $P(\Delta \phi|\mathcal{D})$. 
At this point, we compared the Fermat posterior to the other filter's model posteriors. This step is of crucial importance since the results must not show tension between each other, if the modelling and data are well understood, as they are independent representations of the same phenomenon. Any such tension would indicate a bias or a lack of convergence in one or more of the models and require an in-depth analysis to be solved. The most common reasons for this are wrong parametric constraints, sub-optimal PSF model, non-converged PSO or MCMC and finally an under-constrained modelling approach. While the first three can be easily solved, the last is the most delicate problem and requires careful correction. In particular, this was the hint that the initial model required the additional log-likelihood terms discussed in Section \ref{subsec:custom_LogL} and the main issue that forced F160W to be discarded from the dataset (see Section \ref{subsec:discard_f160}).

Once all the results are under control and achieving minimal tension in the posteriors, it was possible to continue with the combination thereof discussed in Section \ref{subsec:compare_&_combine}.

\subsection{Discarding F160W}
\label{subsec:discard_f160}
\begin{figure}
    \includegraphics[width=\columnwidth]{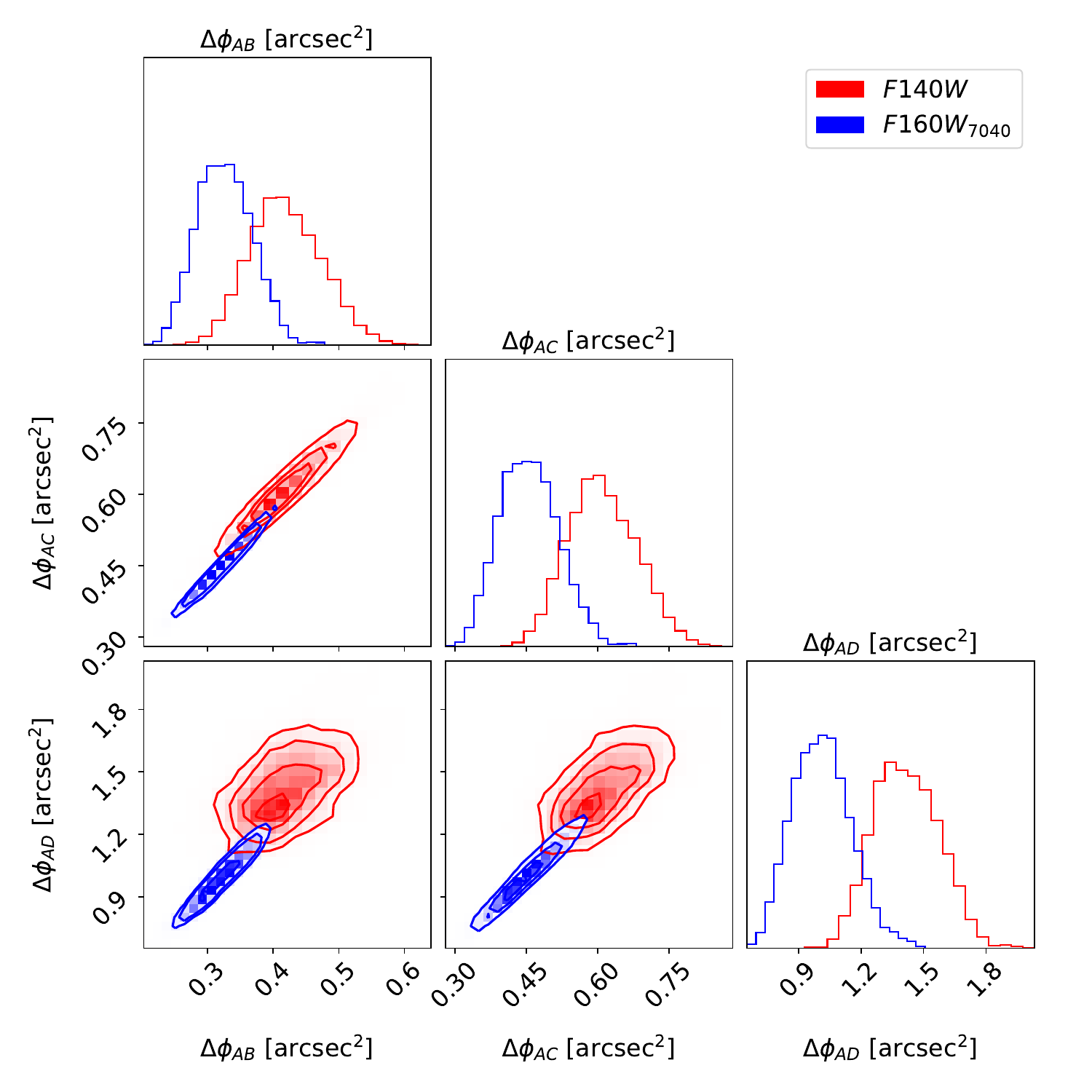}
    \caption{Comparison between the MCMC posterior of the lens model for F140W and  F160W$_{\text{7040}}$ in Fermat potential differences. The tension between the distribution is 1.7$\sigma$.}
    \label{fig:Df_f160w_wo_mask}
\end{figure}

When comparing the F160W models with the other filters' results, we found a strong tension between their $\Delta \phi$ posteriors. We carefully checked that such tension was not due to imperfect PSF modelling or wrong lens modelling (e.g. lack of convergence or ill-defined parameters freedom). Furthermore, the tension was present not only with respect to the optical filters but also with the posterior obtained from F140W. This made it unlikely that the tension could be due to modelling choices, as both filters had a similar modelling set. Being the second most red exposure, they shared features such as the presence and modelling of the extended source light, the brightness of the main lens and the pixel resolution.    
This is unexpected as the two filters should have differed at most for the brightness of the light sources, and this would not affect significantly the end result of the modelling. Instead, a tension of 1.7$\sigma$ was found in their $\Delta \phi$ posteriors, between both F160W$_{\text{7030}}$  and F160W$_{\text{7040}}$ and F140W (see Figure \ref{fig:Df_f160w_wo_mask}). 

In order to explain this tension, we compared the modelled light distributions of F140W to the observed image in F160W. Taking the parametric results of the model of F140W, a mock image was created convolving F140W's light profiles (i.e. lenses lights, lensed host galaxy and QSO images) with the PSF of F160W$_{\M{7040}}$ and leaving the amplitude for each luminous profile free to vary. These were then fitted in order to minimise the residual between this mock model and the exposure of F160W$_{\M{7040}}$. As seen in the residual shown in Figure \ref{fig:f160w_resid_modf140w}, where the standard masks are applied, residual structures remain visible.
The residuals at the images' positions or at the edge of the mask for the lens light are explainable due to random scatter in luminosity, but an unexplained point-like source appears on the top left region of the image, at $\sim 1''$ from the position of image A. This is located near the observed arclets generated from the source light and is not observable in any other filter, while it appears again in the same analysis applied on F160W$_{\M{7030}}$, although noisier due to the lower exposure time. The resolution is too low to define it clearly, but it has a radius of approximately 3 pixels (equivalent to $\sim0.37$'') within which the average SNR is 1.3. We refer to this light as ``\textit{contaminant}''.
Other features might be considered, such as a ``dim'' arclet East of image D or another smaller ``point source" near image B, but both have very low SNR and were deemed not to be significant. Similar results are obtained for F160W$_{\M{7030}}$.

Various hypotheses were considered to explain this finding.
Given the fact that the optical and the F160W exposures were taken the same day, while the two other infrared filters were taken 9 months later, it may have been a serendipitous observation of some other foreground moving object, such as galactic red and dim star, which must have been below the noise in the optical and have moved out of the area of interest by the time the two other exposures were taken. Finally, it could indicate a more complex structure of the host galaxy of the lensed QSO, i.e. features that would not be described appropriately by a single Sèrsic profile, such as the presence of eccentricity, of substructures or a secondary nearby source, for example. Only in the latter case, this light could be used to further constrain the mass model, with the cost of adding complexity and parametric freedom to the model.
Given the uncertainty regarding its nature, the low resolution and SNR of the image and the risk of overfitting if taken into account in the modelling, we opted for the masking of the affected pixels. 

\begin{figure}
    \includegraphics[width=\columnwidth]{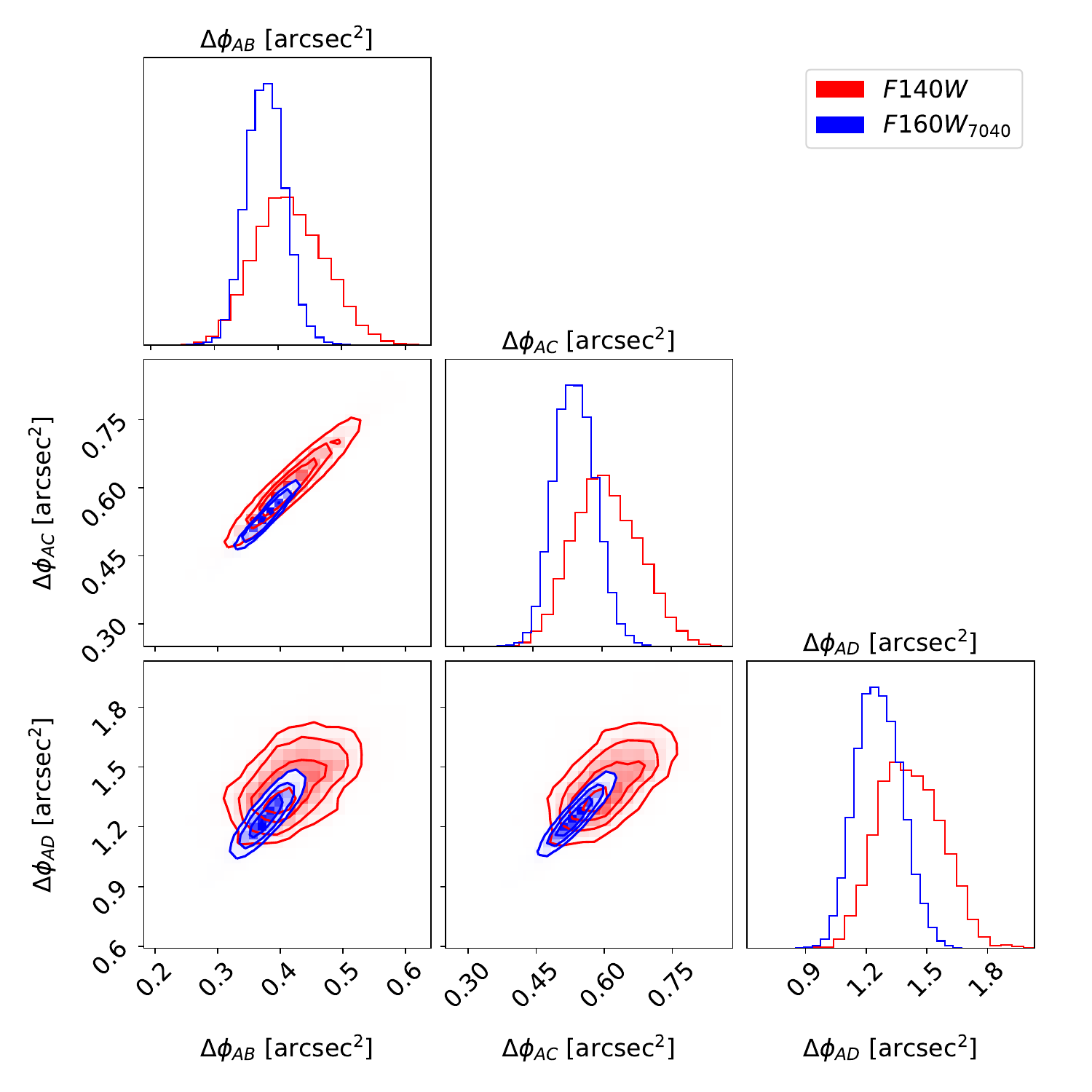}
    \caption{Comparison between the MCMC posterior of the lens model for F140W and  F160W$_{\M{7040}}$ in Fermat potential differences once the contaminant had been masked (see text \ref{subsec:discard_f160}).  Compared to the model where the mask was not implemented (see Figure \ref{fig:Df_f160w_wo_mask}), the tension is  reduced to less than $0.8\sigma$ for each $\Delta\phi$.}
    \label{fig:Df_f160w_wmask_single}
\end{figure}

\begin{figure}
    \includegraphics[width=\columnwidth]{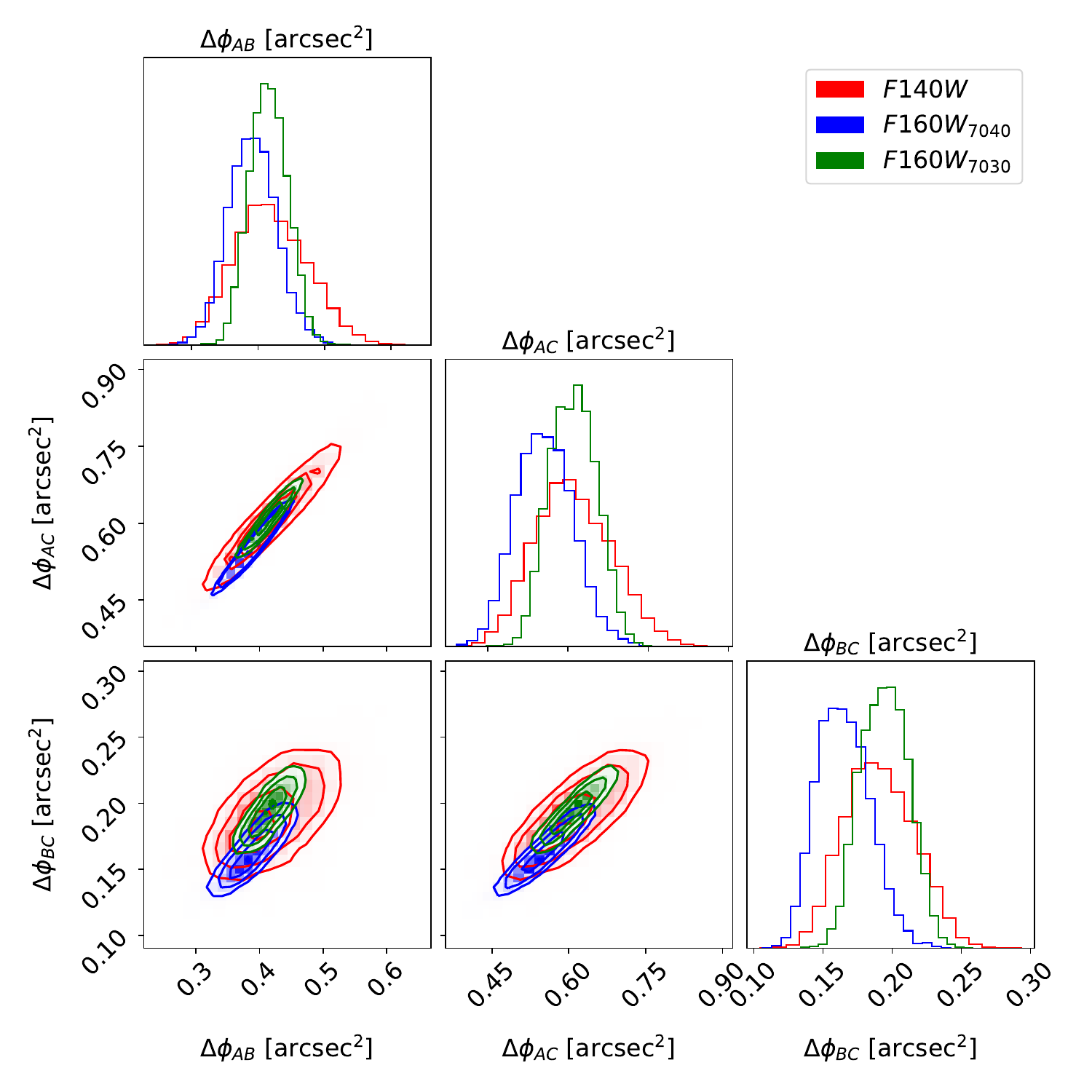}
    \caption{Comparison between the MCMC posterior of the lens model for F140W, F160W$_{\M{7040}}$ and F160W$_{\M{7030}}$ in Fermat potential differences with the same mask for the contaminant. While the mask reduces their tension with respect to F140W, it produces results which are in tension depending on the exposure considered. The tensions range from $0.8\sigma$ to $1.6\sigma$ }
    \label{fig:Df_f160w_wmask_both}
\end{figure}

\begin{figure}
    \includegraphics[width=\columnwidth]{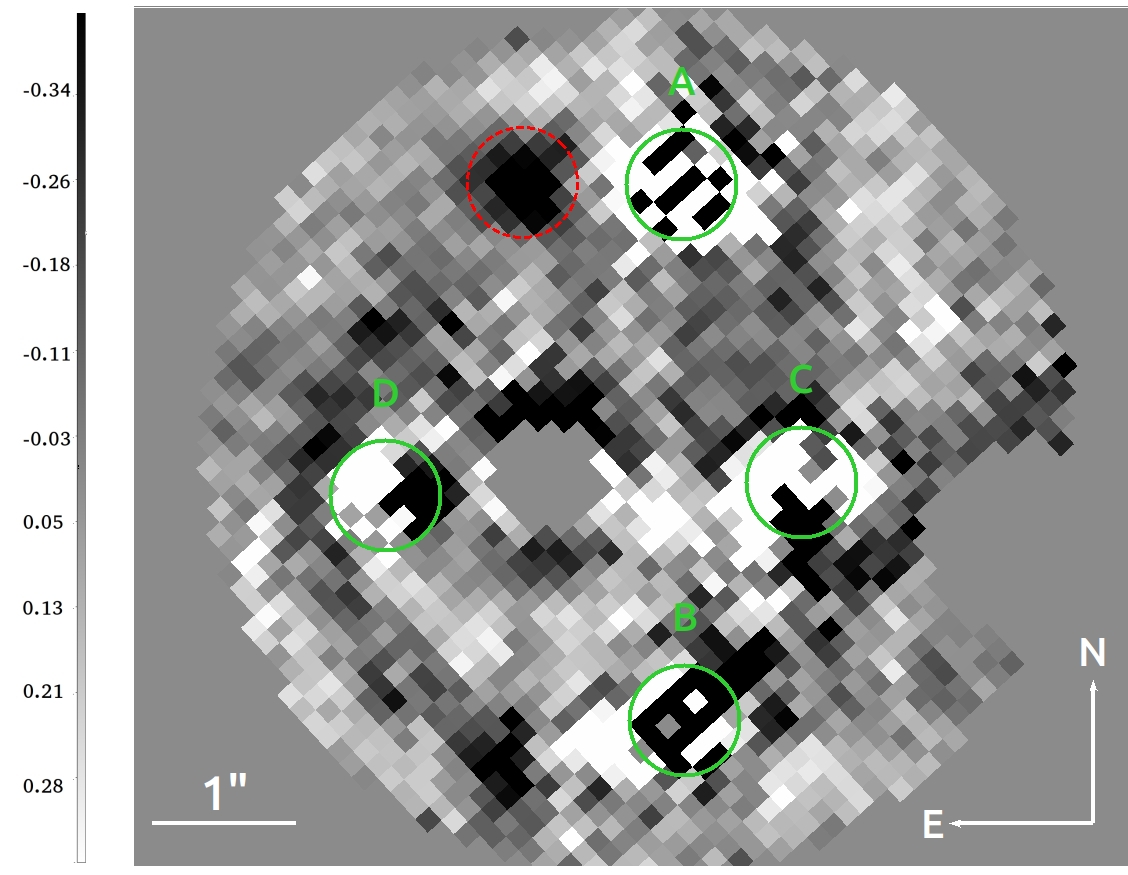}
    \caption{Residual of subtraction of exposure F160W$_{\M{7040}}$ with respect to the reconstructed image obtained from the lens modelling results from filter F140W. The contaminant is shown circled in red, whereas the position of the images is indicated by a green circle and the corresponding letter.}
    \label{fig:f160w_resid_modf140w}
\end{figure}

We implemented the same mask for both F160W exposures (F160W$_{\M{7030}}$ and F160W$_{\M{7040}}$) and repeated the modelling.
The two obtained $\Delta \phi$ posteriors present a reduced tension with the other filters' posteriors (e.g. the posterior of F160W$_{\M{7040}}$ compared to the one of F140W, Figure \ref{fig:Df_f160w_wmask_single})  on the order of $0.6\sigma$. However, they show a tension on the order of $1.5\sigma$ within themselves, i.e. a tension between the results of the modelling of the same filter F160W for the two different exposures 7030 and 7040, as seen in Figure \ref{fig:Df_f160w_wmask_both}.
We conclude that this filter's most informative region, where the lensed arcs are located, is also affected by unknown contaminants. Those cannot be completely and safely masked without heavily biasing its result, as such contaminants are strongly blended with crucial modelling features, i.e. the lensed arcs and QSO images. These effects appear predominantly in this specific filter due to a combination of higher brightness of the redshifted host galaxy at the observed wavelength, lower pixel resolution and higher seeing. 
For this reason, we decided to discard both exposures of F160W for the lens modelling, which requires a high level of confidence to be used for cosmological inference.

\subsection{Combined Posterior} \label{subsec:compare_&_combine}
We now come to the combination of posterior $P(\boldsymbol{\Delta \phi}| D_i)$ of the Fermat potential differences $\boldsymbol{\Delta\phi}$, where $\boldsymbol{\Delta \phi}$ indicates the three-dimensional vector over all couple of images differences, i.e. $\boldsymbol{\Delta \phi}=[\Delta \phi_{AB},\Delta \phi_{AC},\Delta \phi_{AD}]$. We have furthermore introduced $D_i$, where $D_i\in \boldsymbol{D}=(D_1, D_2,..., D_N)$ and
$\boldsymbol{D}$ contains the $N$ dataset relative to the exposures in the different filters, which are conditionally independent, given that they are different observations of the same phenomenon.

We combine $P(\boldsymbol{\Delta \phi}| D_i)$ to infer the combined posterior $P(\boldsymbol{\Delta \phi}| \boldsymbol{D})$.  
Considering the independence of the dataset and the fact that the prior $P(\boldsymbol{\Delta \phi})$ is identical for all $\boldsymbol{\Delta \phi}$ as shown in (\ref{subsubsec:priors}), the combined posterior can be computed using Bayes' theorem:

\begin{align}
P(\boldsymbol{\Delta \phi}|D_i) &\sim P(D_i|\boldsymbol{\Delta \phi})\cdot P(\boldsymbol{\Delta \phi}) 
  \\[1pt]
P(\boldsymbol{\Delta \phi}| \boldsymbol{D}) &\sim \Pi_{i}^N P(D_i|\boldsymbol{\Delta \phi}) \cdot P(\boldsymbol{\Delta \phi}) 
\\[1pt]
P(\boldsymbol{\Delta \phi}| \boldsymbol{D}) &= a \frac{\Pi_{i}^N P(\boldsymbol{\Delta \phi}|D_i)}{P(\boldsymbol{\Delta \phi})^{N-1}} \label{eq:combined_posterior}
\end{align}

$P(D_i|\boldsymbol{\Delta \phi})$ being the likelihood and $a$ being a normalisation constant which takes into consideration the evidence.

Equation \ref{eq:combined_posterior} can be divided into two calculations, firstly of the posterior $P(\boldsymbol{ \Delta \phi}| D_i)$  for each exposure $i$, and secondly of the prior $P(\boldsymbol{ \Delta \phi})$. 
$P(\boldsymbol{ \Delta \phi}| D_i)$ is obtained by computing a normalised histogram in the three dimensions [AB,AC,AD], having the same bin grid for every filter, based on the Fermat posteriors MCMC from Section \ref{subsec:lens_modelling}. 
We numerically derive a prior distribution for $\boldsymbol{ \Delta \phi}$ from the prior of the lens model parameters and the modelling choices. We estimate its density using the  Kernel Density Estimator (KDE) implemented with \texttt{sklearn.neighbors.KernelDensity} \citep{scikit-learn}.

\begin{figure}
    \includegraphics[width=\columnwidth]{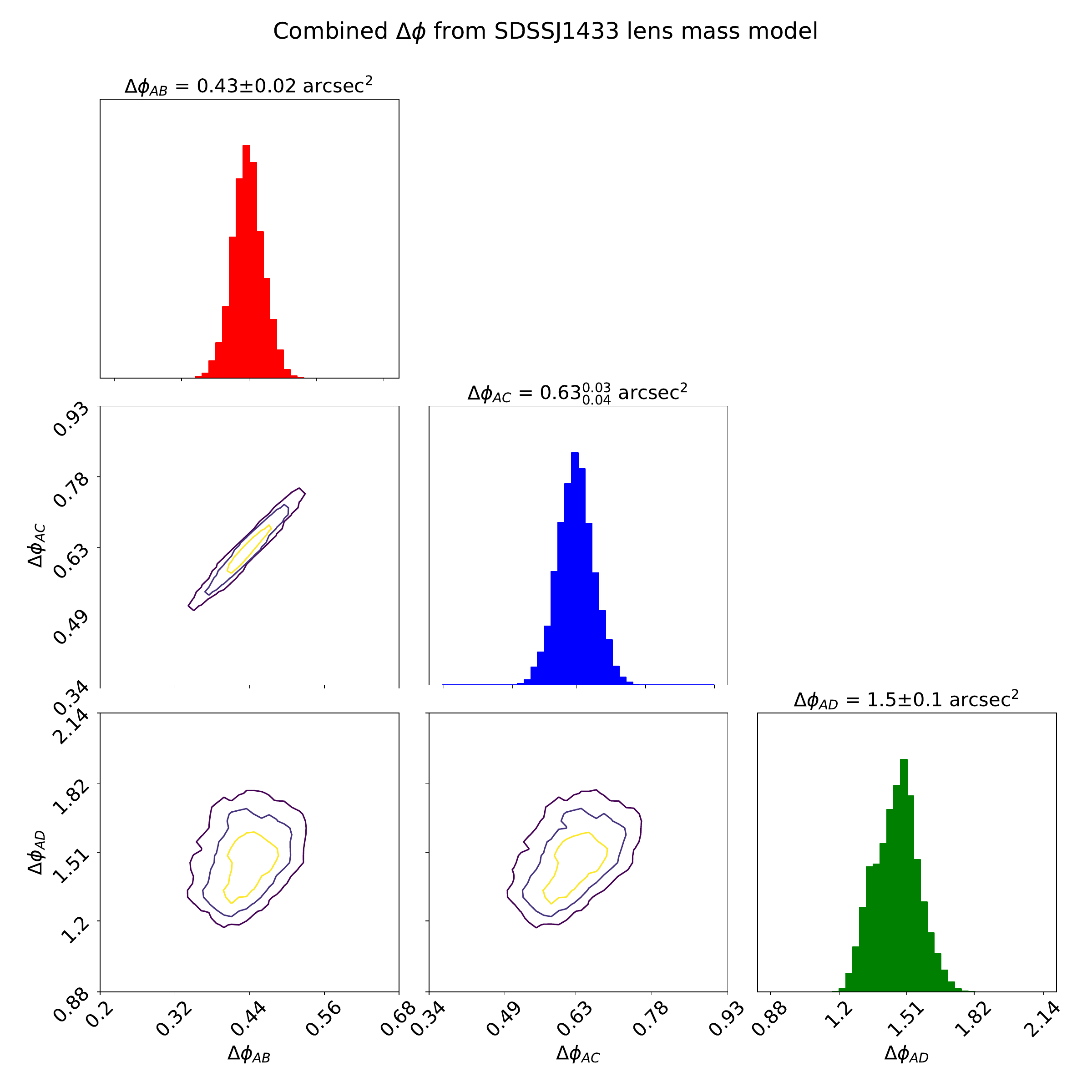}
    \caption{Resulting combined posterior of the differences of Fermat potential $\Delta \phi$ at the positions of the images A, B, C, and D from the modelling \ref{subsec:lens_modelling} following \ref{subsec:compare_&_combine}. The contour levels indicate the 68, 95 and 99.7$\%$ confidence level. }
    \label{fig:comb_df_prob}
\end{figure}

The normalisation factor is then the variable $a$ of equation \ref{eq:combined_posterior}, which is defined such that the integral of the posterior is unity.
 Once the prior and the posterior have been estimated, having taken care that the same binning was considered in all calculations, the resulting $P(\boldsymbol{ \Delta \phi}|D_1,...,D_N) $ is computed following equation \ref{eq:combined_posterior} for each bin. The result is then a histogram in the $\boldsymbol{\Delta\phi}$ parametric space from which the median and the 1$\sigma$ uncertainties are computed (see Table \ref{tab:comb_df_res}). The resulting combined posterior is shown in Figure \ref{fig:comb_df_prob} along with the corresponding corner plots and numerical values.

\begin{table} \centering 
\resizebox{.9\columnwidth}{!}{
\begin{tabular}{ccc}
\hline
  $\Delta \phi_{AB}$ [arcsec$^2$] & $\Delta \phi_{AC}$ [arcsec$^2$] & $\Delta \phi_{AD}$ [arcsec$^2$]\\
\hline
  0.43$\pm$0.02 & 0.63$_{-0.04}^{+0.03}$ & 1.5$\pm$0.1  \\
\hline
\end{tabular} }
\caption{Combined posterior for the difference of Fermat potential at the image positions.}
\label{tab:comb_df_res}
\end{table}

\subsection{Resulting Model and Comparison with Previous Results} \label{subsec:compare_with_litterature}

Regarding the mass model, we present our result for the filter F814W and compare it to results from the literature obtained for the modelling of J1433.

\subsubsection{Total Mass and Mass to Light Ratio}

We restrict the mass profile results to within the range of [$0.2''\lesssim a \lesssim 16''$]  
to allow for a meaningful comparison with the light profile and hence the Mass-to-Ligh ratio (M/L) profile. The minimum radius is defined by the radius of the smaller isophotes, roughly 11 pixels for F814W, corresponding to $\sim 0.2''$ and $\sim 1.1$ kpc on the lens plane. 
Given that the only observables over which the mass model is based are located around $\theta_{\M{E}}$ or within such radius, any mass information at radii larger than $\sim 2\theta_{\M{E}}$ is extrapolated. For this reason, the plot is limited to $a\leq10\theta_{\M{E}} \approx 16''$, corresponding to $a\sim82.6$ kpc. 
Then $M(\theta_{\M{E}})$, the total mass within the Einstein radius $\theta_{\M{E}}$, is computed from the posterior distribution of the model and results to be $(3.1 \pm0.4)\cdot 10^{11} M_\odot $. Approximating the model to a singular isothermal sphere, it is possible to obtain an analytical estimate of the velocity dispersion using the equation $\sigma_{\M{v}}= c\sqrt{\theta_{\M{E}}\frac{D_{\M{s}}}{4\pi D_{\M{ds}}}}\approx 275\frac{\M{km}}{\M{s}}$ (e.g. \citet{meylan2006gravitational}, eq. 52).

Further on, using the obtained light profile (see Section \ref{subsec:light_prof&llm}), the M/L is estimated by comparing the projected luminosity with the total mass (i.e. baryonic and dark matter) within a given major axis. The luminosity is corrected for the distance modulus of $5\cdot \log_{10}(D_{\M{L}}(z_{\text{lens}})/10 pc)=41.73$, where $D_{\M{L}}$ is the luminosity distance), cosmic dimming of $2.5\cdot \log_{10}((1+z_{\text{lens}})^4)=1.48$, galactic extinction  and  $K$ correction. The last two depend on the filter used and were obtained from the literature. We obtain $K($F814W$)=-0.14$ and $K($F475X$)=0.09$  using the tool available at the site \url{http://kcor.sai.msu.ru/} based on \citet{2010K_corr} and \citet{2012K_corr}.  We obtain the galactic extinction from \citet{NED_ext}, which results to be $A_{\M{F814W}}=0.014$ mag and $A_{\M{F475X}}=0.029$ mag.  

\begin{figure}
    \includegraphics[width=\columnwidth]{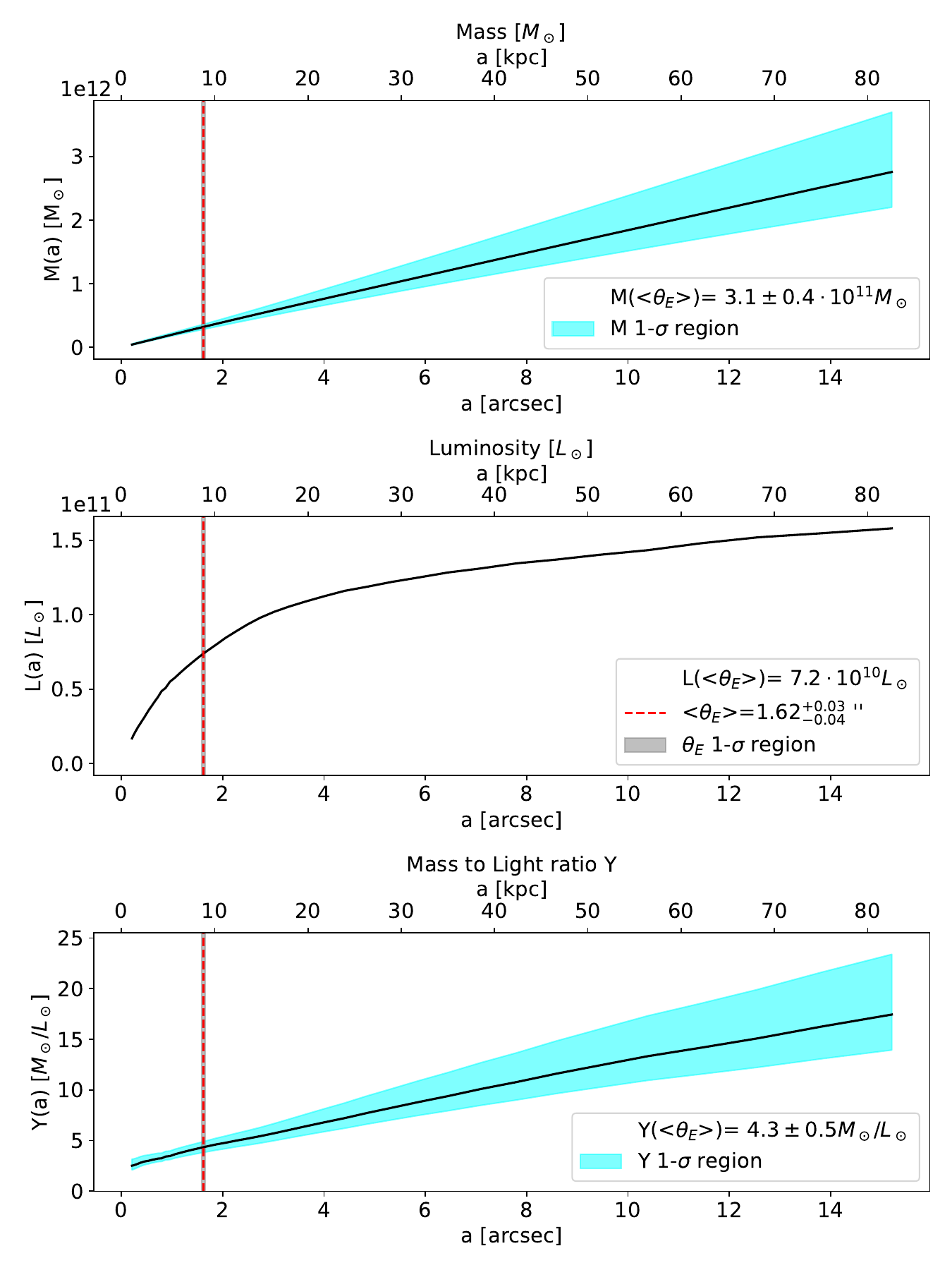}
    \caption{Cumulative mass from lens mass modelling (\ref{sec:lens_mod}),  enclosed luminosity from isophotal fitting (\ref{subsec:light_prof&llm}), and corresponding Mass to Light ratio $\Upsilon$  for filter F814W. All these are plotted with respect to the semi-major axis of the isophotes. The cyan-shaded region indicates the 1-$\sigma$ region obtained from the posterior of the lens modelling. The luminosity has negligible uncertainty, which is not taken into account. The vertical red dashed line indicates $\theta_{\M{E}}$ obtained from the mass model, along with its corresponding grey 1$\sigma$ region. The luminosity error, as measured from the error frame, is here negligible.}
    \label{fig:MtL}
\end{figure}

As seen in Figure \ref{fig:MtL}, $\Upsilon$ grows approximately linearly with $r$ as a result of the shape of the mass profile.  We report its value at the Einstein radius to be $\Upsilon(\theta_{\M{E}})=4.3\pm 0.5 M_\odot/L_\odot$ . 
\begin{table}
\begin{tabular}{llccr}
\hline
& &This paper (F814W) & S22 & Tension \\ \hline
$\theta_{\M{E}}^{\M{ML}}$ & ["] & $ 1.62^{+0.03}_{-0.04} $ & $1.581_{-0.002}^{+0.003}$ & 1.16\\[1.5mm]
$\gamma^{\M{ML} }$ & ["] & $ 2.05^{+0.09}_{-0.1} $ & $1.92\pm0.03$ & 1.28\\[1.5mm]
$q^{\M{ML}}$ & [] & $ 0.67\pm0.07 $ & $0.96\pm0.01$ & 4.15\\[1.5mm]
$\phi^{\M{ML}}$ & [$^\circ$] & $ -7.0^{+3.9}_{-4.1} $ & $-28_{-2.6}^{+4.5}$ & 3.93\\[1.5mm]
x$^{\M{ML}}$ & ["] & $ 0.93^{+0.04}_{-0.05} $ & $0.931\pm 0.006$ & 0.03\\[1.5mm]
y$^{\M{ML}}$ & ["] & $ -2.069\pm0.005 $ & $-2.038\pm 0.006$ & 4.0\\[1.5mm]
$\theta_{\M{E}}^{\M{Pert}}$ & ["] & $ 0.2\pm0.05 $ &  & \\[1.5mm]
$\gamma^{\M{Shear}}$ & [] & $ 0.11\pm0.03 $ & $0.127\pm0.004$ & 0.71\\[1.5mm]
$\psi^{\M{Shear}}$ & [$^\circ$] & $ -77.0^{+2.8}_{-3.8} $ & $-82\pm0.4$ & 1.63\\[1.5mm]
\hline
\end{tabular} 

\caption{Comparison between mass-profile results for F814W and S22 (adapted for the frame of reference). The tension is discussed in \ref{subsubsec:comp_w_lit}. Note that S22 does not report the value for the $\theta_{\M{E}}^{\M{P}}$ explicitly.}\label{tab:comp_s22}
\end{table}

\subsubsection{Main Lens Colour} \label{subsubsec:main_lens_colour}
\begin{figure}
        \includegraphics[width=\columnwidth]{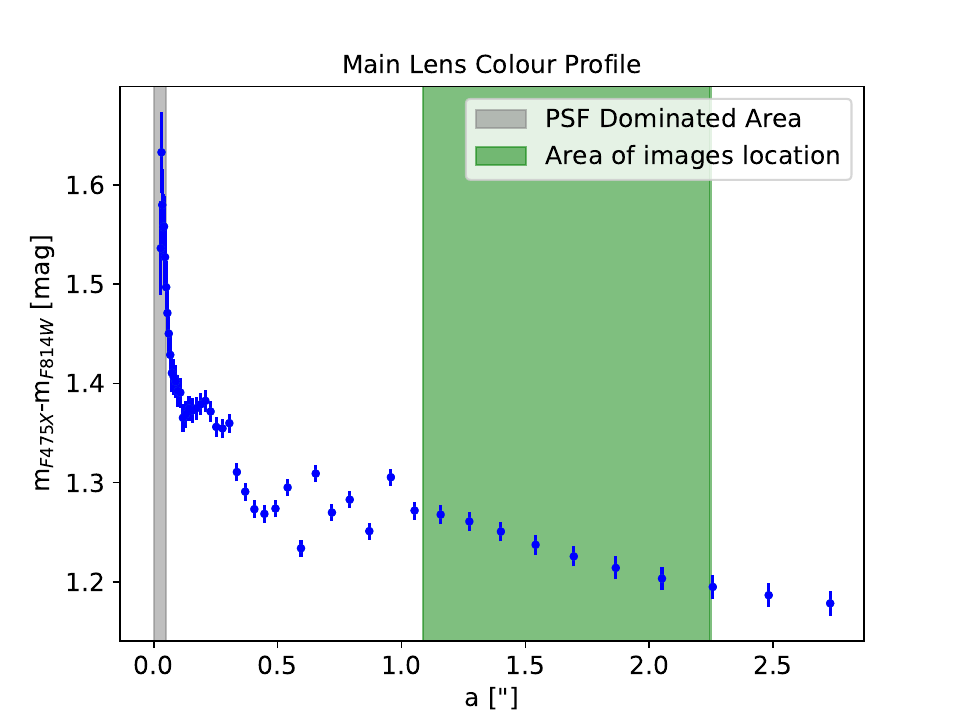}
    \caption{The integrated colour profile obtained from isophotal light models of the main lens in filter F475X and F814W with respect to the isophotes' major axis $a$. This is the integrated colour, i.e. the colour given by the enclosed luminosity within the isophote, not the colour at the given isophote. The green vertical region indicates the location of the QSO images, whereas the grey region approximately corresponds to the FWHM of the PSF of F475X, i.e. $\sim0.05''$, indicating where the modelling is most affected by PSF.}
    \label{fig:colour}
\end{figure}

Given the isophotal light models obtained for both optical filters, F475X and F814W, the integrated colour of the main lens can be measured. We take the elliptical apertures using the python package \texttt{photutils} \citep{photutils} matching the isophotes and compute the enclosed luminosity of the models, as well as the corresponding error from the error frames. The enclosed luminosity difference in the two filters is the integrated colour for a given major axis value $a$. In Figure \ref{fig:colour} the integrated colour is shown with respect to the major axis $a$ of the isophotes. This plot is cut after $a>3''$ from the centre of the lens, as it reaches the limiting surface brightness.
This is an issue in this analysis as the uncertainties of the two models are combined. 
We furthermore ignore the grey area, i.e. $a<0.05''$, where the differences in the PSFs most strongly affect the model. 
The negative trend with radius, i.e. the colour becoming ``bluer'', is expected from elliptical galaxies \citep{saglia2000evolution}.

The resulting integrated colour within the isophote with major axis $0.05''<a<3''$  is $\Delta \text{mag}_{F475X-F814W} = 1.18\pm$0.01 mag. The colour information, while important in its own right, is beyond the scope of this paper. Thus the error is approximately obtained from photon noise and ignores any uncertainty due to other isophotes parameters (centre coordinates, ellipticity and boxiness). Such uncertainties should be nonetheless limited within the radius we are here considering. Thus the reported error, while underestimating the full uncertainty, can be considered as a good approximation. 

\subsubsection{Comparison with Literature}\label{subsubsec:comp_w_lit}

\begin{figure}
    \centering
    \includegraphics[width=\columnwidth]{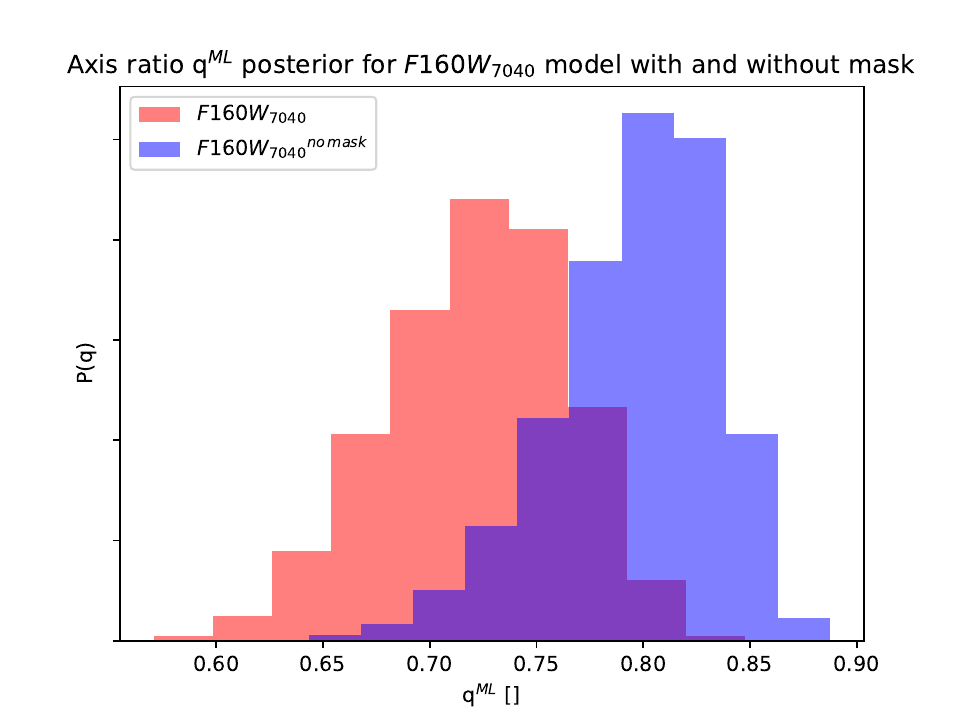}
    \caption{Comparison of axis ratio of the mass profile of the main lens $q^{\M{ML}}$ for two models of F160W$_{\text{7040}}$: with and without masking the contaminant ("no mask"). Note the higher mean value and negative skewness for the second case.}
    \label{fig:f160w_qwwomask}
\end{figure}

We compare our results to \citet{schmidt22_STRIDES} (hereafter S22). This study shared the same \textit{HST} data (although limited to F475X, F814W and F160W), the same choice of mass profiles and the same lens modelling software. 
The scope of S22 was different from the one here presented, as it aimed to introduce an automated lensing pipeline and had specific requirements for which model was to be considered ready for cosmological inferences. Furthermore, the S22 modelling approach differed from ours as it was run using the three exposures in parallel.  In Table \ref{tab:comp_s22} the resulting lens parameters are compared. Most parameters are in agreement between the analysis of S22 and ours, apart from the ellipticity parameter $q^{\M{ML}}$ and $\phi^{\M{ML}}$ which present a high level of tension between 4.15 and 3.93, respectively. Note that in S22 the pointing angle $\phi^{\M{ML}}$ presents a difference larger than 18$^{\circ}$ from its luminous counterpart. In our modelling, this is unlikely to happen as such large differences are suppressed (see Section \ref{subsec:custom_LogL}). This can be due to the fact that the axis ratio $q^{\M{ML}}$ appears very large in S22, indicating an almost spherical distribution. This renders $\phi^{\M{ML}}$ degenerate with respect to the model, hence a tension in such a parameter is not significant. On the contrary, the tension in $q^{\M{ML}}$ will strongly affect the mass distribution and therefore the cosmological inference based on it. 
This tension might very well be due to the presence of F160W considered in the S22 model. As it can be seen in Figure C5 in their Appendix, this filter leaves a significant residual at the same position as found for the ``contaminant'' object described in Section \ref{subsec:discard_f160}. Our models of F160W without masking the ``contaminant'' are found to have a higher axis ratio: $q^{\M{ML}}=0.80^{+0.03}_{-0.04}$, where the posterior distribution showed a negative skew (see Figure \ref{fig:f160w_qwwomask}), indicating that the model could not converge to higher value due to the punishing term in the Likelihood shown in equation \ref{eq:custom_logL_qll}.
This suggests that the higher axis ratio is correlated to the model fitting the light of the ``contaminant''.  This further convinced us that without a sound approach to the modelling of F160W, such a dataset had to be discarded (Section \ref{subsec:discard_f160}). 

We further compared the differences in Fermat potential at the images' position $\Delta\phi$ (see S22, Table 8).  
This comparison is done with respect to our combined Fermat difference posterior of (\ref{subsec:compare_&_combine}, Table \ref{tab:comb_df_res}).  The reason is that we wanted to compare our final result, based on multifilter modelling, with a similarly informed model from the literature. Note that in our result F160W has been discarded.

The results are shown in Table (\ref{tab:comparison}). They are in some tension (in all cases higher than 1.5), which can be explained by the difference in ellipticity previously described. 
\begin{table}
    \centering
\resizebox{\columnwidth}{!}{
 \begin{tabular}{|lccr|}
\hline
& This paper (all filters) & S22 &Tension \\ \hline

$\Delta\phi_{AB}$ [$\text{arcsec}^2$] & 0.43$\pm$0.02&$ 0.36\pm0.03 $&1.75 \\[1mm]
$\Delta\phi_{AC}$ [$\text{arcsec}^2$] & 0.63$_{-0.04}^{+0.03}$&$ 0.49\pm0.03 $&2.72 \\[1mm]
$\Delta\phi_{AD}$ [$\text{arcsec}^2$] & 1.5$\pm$0.1&$ 1.03\pm0.03 $&3.83 \\
\hline
\end{tabular}} 	
 	
\caption{Comparison of results for the combined Fermat potential difference between all filters with respect to S22.} 
	\label{tab:comparison}
\end{table} 
Overall, the results differ due to the modelling approach and most importantly due to the difference in the data employed in the modelling. In particular, we point out that modelling in parallel without a sound approach for the detection and treatment of unknown systematic, such as the ``contaminant'' of F160W,  can result in severely biased results. This is therefore strongly advised against in the framework of cosmological inference. We also point out that S22 refers to the model of J1433 to be ``far from cosmography grade'' based on a metric of the stability of the model. 
\section{Lightcurves Analysis and Time Delay Estimate} \label{sec:lc_analysis} 
To constrain the time delay between the images their luminosity is observed over time, creating the ``lightcurves'' (LCs) of the images. These are analysed by correlating the observed variability between the couples given a time and magnitude shift. 
This type of analysis carries several problems when applied to real observations. A full overview can be found in \citet{cosmog_pycs_XI}.  
The main issue encountered in this analysis is the low brightness of image D ($\sim 22.8$ mag in $g'$ band, as seen in \ref{fig:lc_init}), which, coupled with the limited overlap with the other lightcurves due to the large expected time delay of this image ($\sim 100$ days), resulted in a slightly less constrained time delay measurement.
Other possible issues (e.g. lack of variability, microlensing amplitude, limited overlap of the lightcurves) were solved or diminished by the total length of the observation campaign (1223 days), its high observation frequency ($\sim$ 1 observation/ 4 night) and overall high quality of data ($\langle \M{Seeing}\rangle =1.16''$ and $\langle \M{Transparency}\rangle =86.72\%$, see Figure \ref{fig:seeing_and_trsnp_distr}). In this chapter, we discuss the observational campaign, how the lightcurves are extracted from the data (\ref{subsec:creation_lc}), and how these are analysed to obtain time delay estimates (\ref{subsec:td_analysis}). The time delay error estimation is further discussed in (\ref{subsec:dt_error}). Finally, we discuss the combination of different time delay estimates into a common result (\ref{subsec:comb_dt}).

\subsection{\textit{WST} Data for Lightcurves Compilation}\label{subsec:WST_data}
\begin{figure}

\includegraphics[width=\columnwidth]{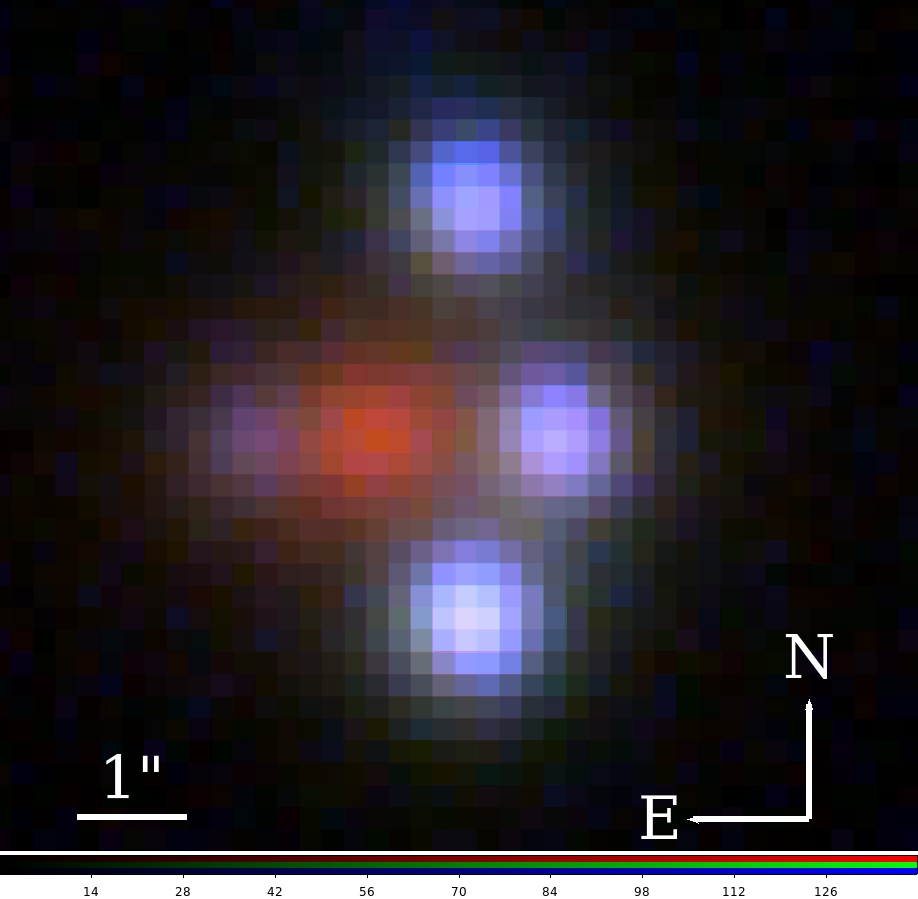}
    \caption{Colour image from \textit{WST} in three bands: $u'$, $g'$ and $r'$, corresponding to blue, green and red in the image, respectively. Note the redder colour of the central lens light, while the perturber galaxy is completely blended with image C. These stacks were obtained from \citet{zoeller23}.}
    \label{fig:Color_WST_ugr}
\end{figure}
\label{subsec:creation_lc}
The observation campaign was carried out in two bands, $g'$ and $i'$, 
from 7/02/2020 till 15/06/2023 at the \textit{Wendelstein Observatory} using the \textit{WWFI}. The pixel scale of the camera is $0.2''/$pixel. The $i'$ band was eventually discarded for the analysis after less than 2 years of observation as its SNR was too low to be informative. 
Each observation was then reduced following the data reduction pipeline detailed in \citet{Kluge_diss} and \citet{kluge2020structure}. The pipeline corrects for bias, aligns the CCD, divides for flatfield, masks bad pixels, multiplies for the gain and computes a first approximated astrometry. The error frames are generated by the propagation of the statistical uncertainties. 
This first basic reduction is concluded by masking the charge persistence for each frame per day sequentially and by creating a star catalogue for each night. The satellites are manually masked, and then the single exposures are resampled in order to be stacked, obtaining a final single image per night. Similarly, the coadded error frames are obtained by resampling and propagating the individual error images of the single exposures.

\begin{figure}
    \includegraphics[width=\columnwidth]{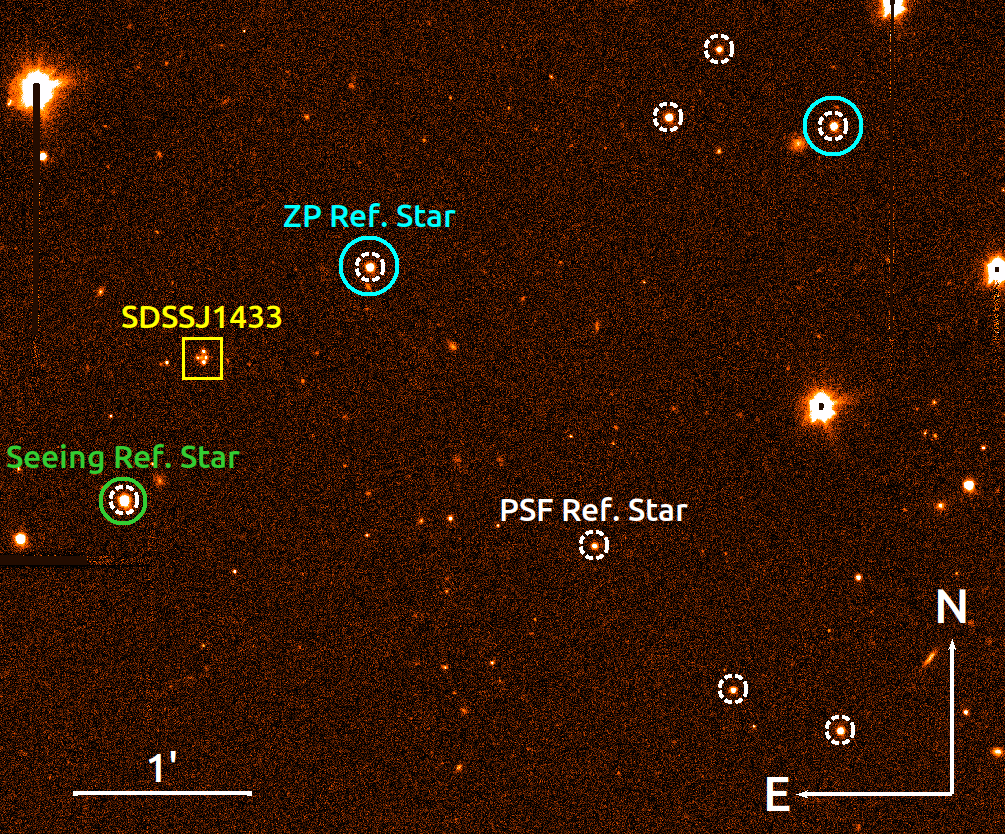}
    \caption{Positions of PSF reference stars (shown in white), ZP reference stars (shown in cyan) for J1433 (in the yellow box) in the reference image, taken 31/07/2020. The star circled in green is the one considered when estimating the seeing. For their coordinates, see Table \ref{tab:ref_star_SDSSJ1433}}
    \label{fig:ref_stars_SDSSJ1433}
\end{figure}

For each night, the PSF full-width half maximum (FWHM) is automatically measured in a non-parametric way using the SB profile of field stars in order to estimate the seeing. The photometric zeropoint (ZP) is determined by comparing the aperture photometry of field stars with the Pan-STARRS1 DR2 catalogue \citep{panstarrs}.
We increase the ZP precision by recalibrating it with PSF photometry. First, we estimate the FWHM of the PSF for the single night (i.e. the seeing) from a ``Seeing reference star''. We then create a supersampled PSF model based on a set of bright stars near the lens system, the ``PSF reference stars'', following the approach outlined in Section \ref{subsubsec:psf}. The positions of these stars are reported in Table (\ref{tab:ref_star_SDSSJ1433}) and are shown in Figure (\ref{fig:ref_stars_SDSSJ1433}). Finally, we use the PSF model to estimate the brightness of the two ``ZP reference stars'', chosen for their stable luminosity, in order to account for ZP corrections.
\begin{table}
    \centering
    \begin{tabular}{|c|c|c|}
    \hline
    R.A. (J2000) & Dec. (J2000)& \\ \hline
14:33:26.46  &  +60:06:25.27  & Seeing Reference Star \\
14:33:15.12  &  +60:07:46.45  & ZP Reference Star 1 (mag$_{g'}=18.064$)\\
14:32:53.65  &  +60:08:35.26 & ZP Reference Star 2 (mag$_{g'}=18.306$)\\
14:33:01.28  &  +60:08:38.29 &  \\
14:32:58.96  &  +60:09:01.82 &  \\
14:33:04.72  &  +60:06:10.16 &  \\
14:32:58.31  &  +60:05:20.49 &  \\
14:32:53.35  &  +60:05:06.44 &  \\
\hline 
    \end{tabular}
    \caption{Sexagesimal sky coordinates of the stars used as references in the PSF modelling for J1433. The first one is the star considered when estimating the seeing. The second and third are the reference stars used for the ZP calibration (see text \ref{subsec:WST_data}) with the corresponding magnitude measured for the reference night (31/07/2020, see Figure \ref{fig:ref_stars_SDSSJ1433}) and taken as the reference magnitude for all observations. }
    \label{tab:ref_star_SDSSJ1433}
\end{table}

\begin{figure}
  \includegraphics[width=\columnwidth]{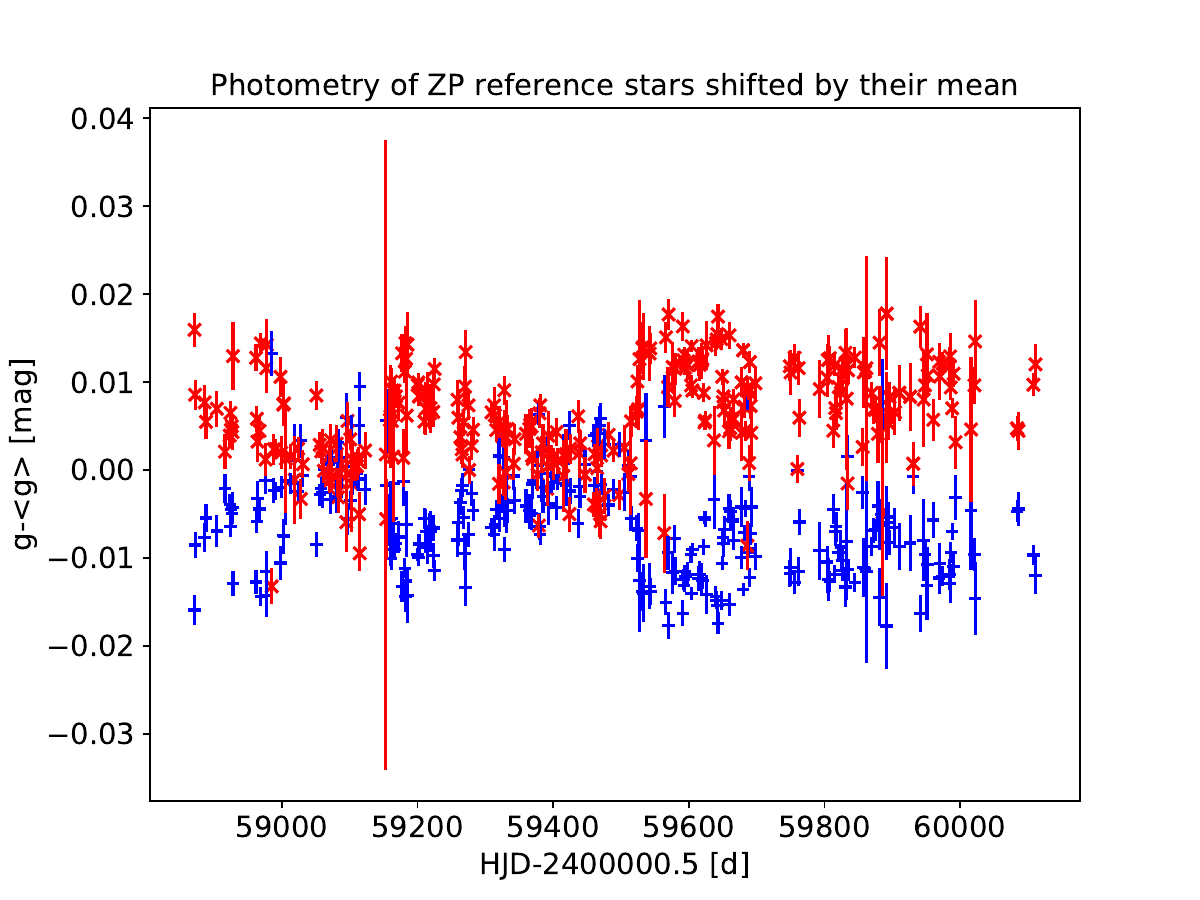}
 \caption{Compared photometry of the ZP reference stars (see Table \ref{tab:ref_star_SDSSJ1433} and Figure \ref{fig:ref_stars_SDSSJ1433}) corrected for their reference magnitude and ZP nightly variation (see equation \ref{eq:zp_stars}). The remaining low variability indicates the stability of such stars and their limited scatter ($\sim 0.006$ mag) does not contribute significantly to the lightcurve error budget.}
 \label{fig:phot_ref_star_g}
\end{figure}

These ``ZP reference stars'' are chosen by considering a large sample of nearby stars and measuring their photometry over time, discarding the ones that appeared variable with respect to the others. We were left with two stable stars.  
This can be seen in Figure \ref{fig:phot_ref_star_g} 
where their lightcurve in the $g'$ band is plotted after the subtraction of the expected magnitude and corrected for the nightly ZP recalibration. This corresponds to the equation :
\begin{equation}
\Delta \text{mag}_{i,S_j} = \text{mag}_{i,S_j} - \text{mag}_{\text{ref},S_j} - \delta\text{ZP}_i
    \label{eq:zp_stars}
\end{equation}
where $\Delta \text{mag}_{i,S_j}$ is the resulting datapoint for ZP star $S_j$ , $\text{mag}_{i,S_j}$ is the measured magnitude of $S_j$ for the $i$-th night,  $\text{mag}_{\text{ref},S_j}$ is the reference magnitude of the $S_j$ star  and $\delta$ZP$_i$ is the ZP correction for the corresponding $i$-th night obtained. Said ZP correction is obtained by averaging the difference of the observed magnitude for the ZP stars with their reference magnitude for each $i$-th night:  $\delta$ZP$_i =\langle mag_{S_j,i}-\text{mag}_{\text{ref},S_j} \rangle _j$. 
Their reference magnitudes $\text{mag}_{\text{ref},S_j}$ are the magnitudes observed from the reference night (31/07/2020, see Figure \ref{fig:ref_stars_SDSSJ1433}) and are  reported in Table (\ref{tab:ref_star_SDSSJ1433}).
As a validation measure, this process is applied to the images obtained in the $i'$ band, and the same set of stars results to be stable in that filter as well.
The standard deviation of $\Delta \text{mag}_{i,S}$ for both ZP reference stars is $\sim0.006$ mag, thus negligibly affecting the lightcurves error budget. 
Once we obtained a precise estimate of the ZP, we employed the \textit{HST} images previously introduced for the lens modelling to subtract all objects besides the QSO's images. F475X, whose bandwidth and centring are most similar to the $g'$ band, was adapted to the \textit{WST} images. Firstly, an HST-PSF model is created following the same approach as in Section \ref{subsubsec:psf}  by taking multiple bright but unsaturated, nearby stars in the field.  
We used this PSF to subtract the QSO's light. Note that the PSF here described differs from the one obtained previously (see Section \ref{subsubsec:psf}) for its purpose. Here the objective is to correctly subtract the outer wings of the QSO, whereas in Section \ref{sec:lens_mod}, the PSF has to fit precisely its position, hence its core. 
The possible residuals in the centre of the QSO's image were interpolated using their neighbouring pixels. 
After this, the pixel resolution of the whole HST image is degraded to the \textit{WST} resolution of $0.2\frac{\M{"}}{\M{pix}}$.  
At this point, the image is shifted, rotated, cropped, and finally resampled in order to coincide in the pixel grid with the reference \textit{WST} image.  
The obtained image, which will be henceforth referred to as \textit{HSTtoWST}, is now used for all observations. This image has now the PSF resolution of \textit{HST}, while regridded to the pixel grid of \textit{WST}.

\begin{figure}
    \includegraphics[width=\columnwidth]{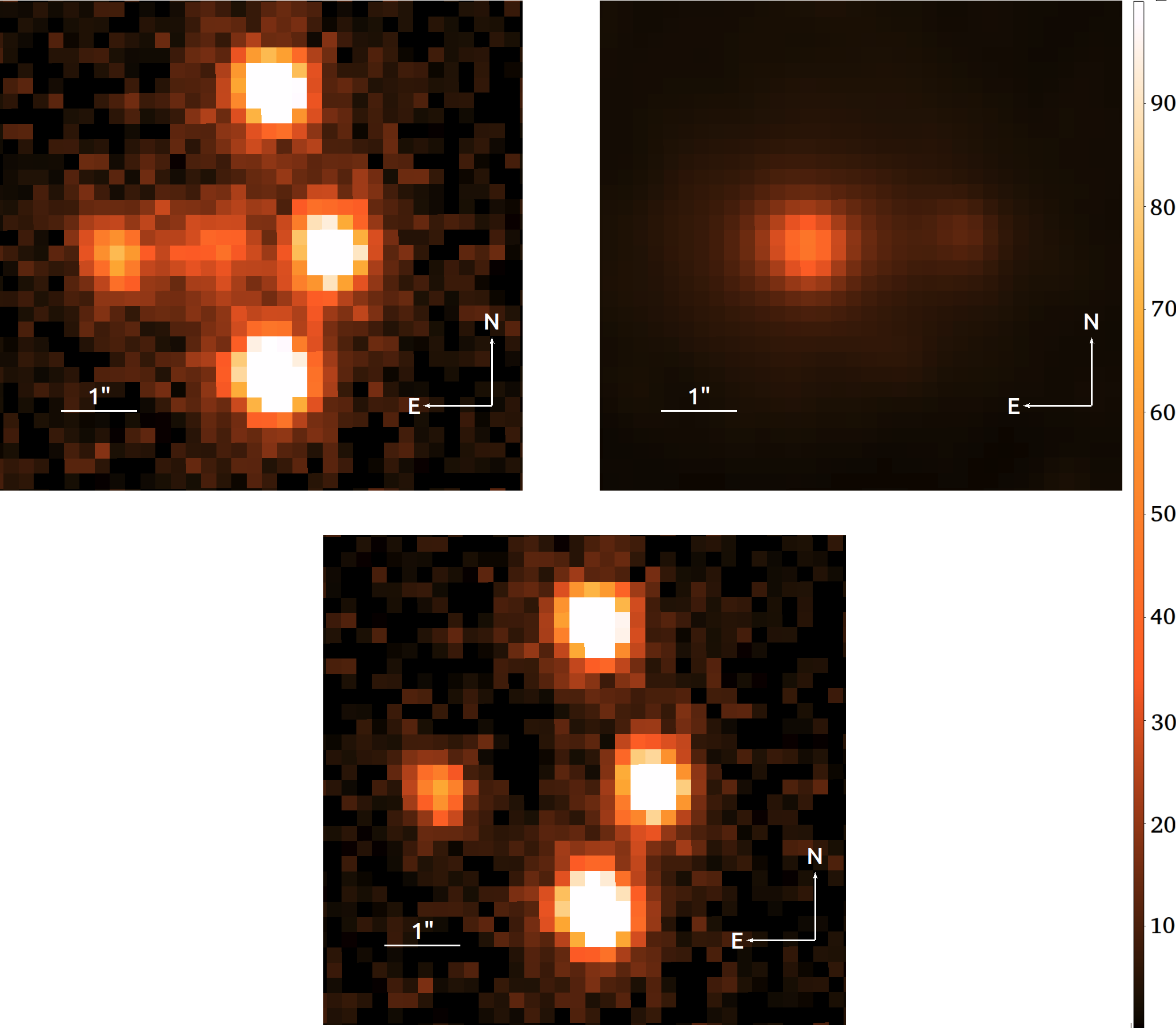}
    \caption{Subtraction of the lens light (both main lens and perturber) for the reference observation (31/07/2020). Top left: reduced \textit{WST} observation, top right: model of lens light obtained from \textit{HST} observation in F475X brought to \textit{WST} resolution (i.e. ``HSTtoWST'' image for the night, see text \ref{subsec:creation_lc}), bottom: resulting subtracted image.}
    \label{fig:HSTtoWST_sub}
\end{figure}

Following this, for each night we convolve \textit{HSTtoWST} such that the PSF are identical to the \textit{WST} observation. 
We then iteratively refine the astrometric solution and the convolution kernel, as the estimation of the convolution kernel is sensible to precise astrometry.
The obtained image is then corrected by the ZP of the \textit{WST} image and scaled by an additional empirical scaling factor to account for differences in \textit{HST}'s ZP. This free factor is obtained by minimising the residual, and we find it to be 1.11. This image is then subtracted from the single night exposure, leaving only the QSO images light without any additional source. An example of the result is shown in Figure \ref{fig:HSTtoWST_sub} for the reference night (31/07/2020, see Figure \ref{fig:ref_stars_SDSSJ1433}). 

Now the single night photometry of the QSO's images can be measured using PSF photometry based on the model obtained from the PSF reference stars (\ref{fig:ref_stars_SDSSJ1433}). In order to do so, the PSF model, which was previously used to refine the ZP using the ``ZP reference stars'', is scaled to match the flux of the QSO images, while optimising for a common shift of their positions to account for the remaining imprecision of the astrometry, on the order of $0.3$" (or 1.5 pixels in the \textit{WST} resolution). Finally, these values are converted to magnitudes.
\begin{figure}
	\includegraphics[width=1.\columnwidth]{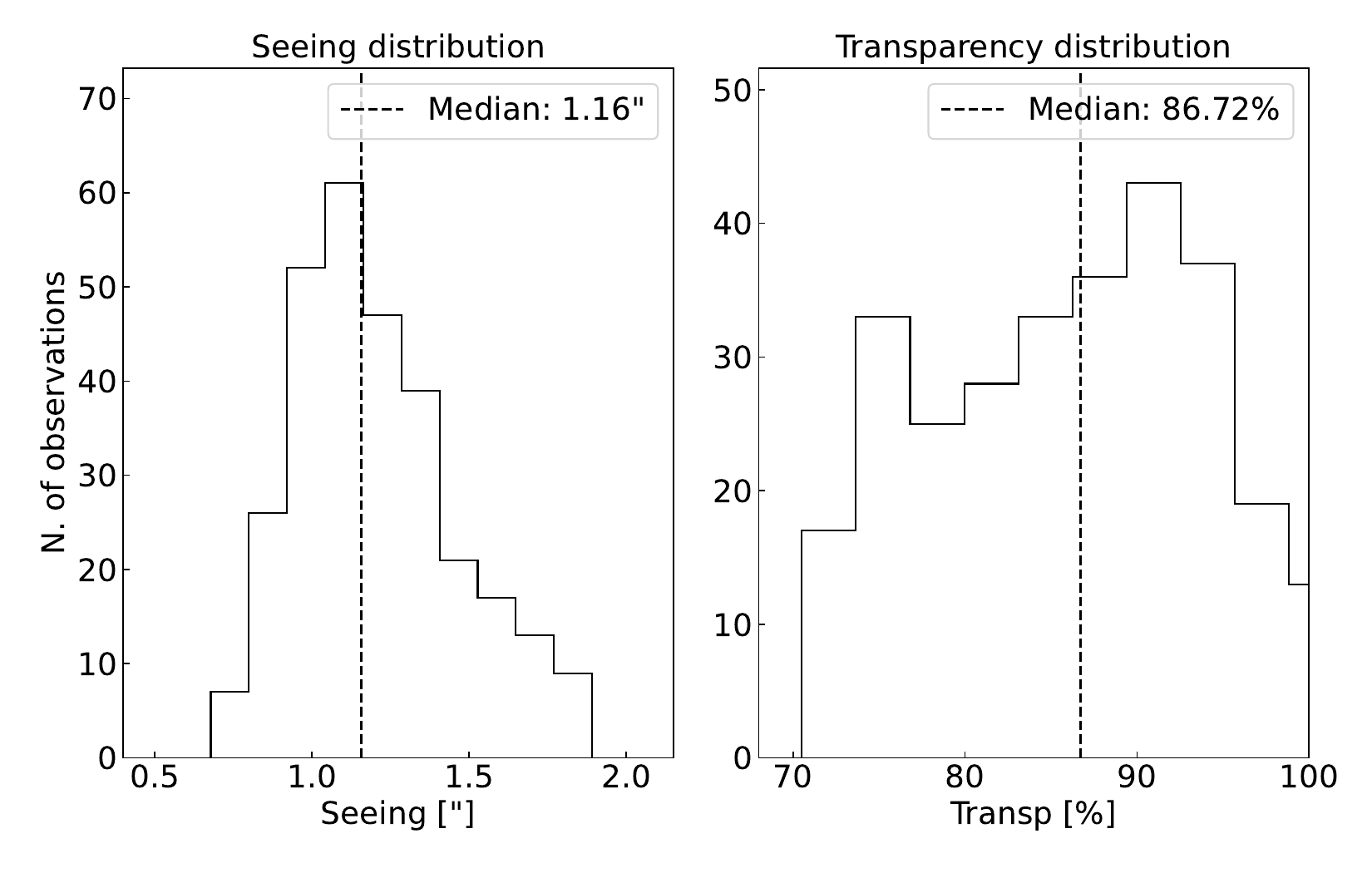}
    \caption{Distribution of seeing and transparency for the observations used in the lightcurve analysis (\ref{subsec:td_analysis}) after the cutoff (see text \ref{subsec:WST_data}). The medians are 1.16'' for the seeing and 86.72\% for the transparency.}
    \label{fig:seeing_and_trsnp_distr} 
\end{figure}
To increase the precision of the measurement, we implement a threshold on the data quality for each night. We thus exclude from the lightcurve analysis any night whose sky transparency was lower than $70\%$ relative to the best night and whose seeing is larger than 1.9''. The seeing and transparency for the remaining datapoints are shown in Figure \ref{fig:seeing_and_trsnp_distr}. No further cut of the sample is required, and the resulting dataset has an average seeing of 1.16'' and 86.6\% transparency.
The final result is a lightcurve for each QSO image with 297 datapoints over three years and a median magnitude precision of  $0.015\leq \sigma_{\text{mag}}\leq 0.055$ 
depending on the brightness of the QSO images. The resulting lightcurves can be seen in Figure \ref{fig:lc_init}.
\begin{figure*}
	\includegraphics[width=.8\textwidth]{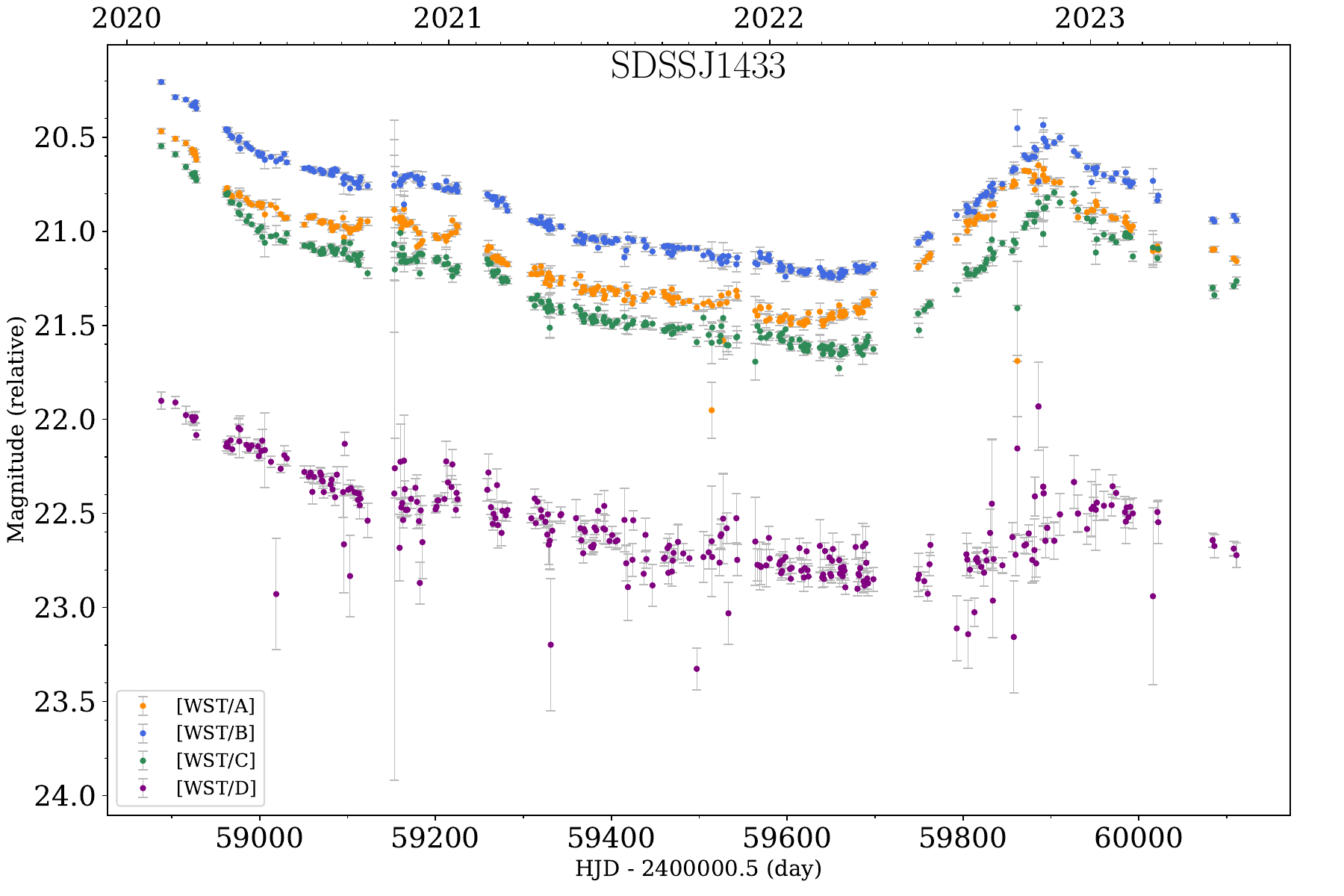}
    \caption{Lightcurves for the QSO's images observed from the WWFI at the 2.1 Meter Telescope of the Wendelstein Observatory in the $g'$ band.}
    \label{fig:lc_init} 
\end{figure*}

\begin{table}\centering
\resizebox{.9\columnwidth}{!}{
    \begin{tabular}{rl}
    \hline
         Period of Observation & 7 February 2020- 15 June 2023\\
         Total observed Nights & 432\\ 
         N. Datapoints & 297\\
         Sampling & 4.1 days \\
         Pixel scale & 0.2 ''/pixel \\
         Median Seeing &1.16 '' \\
         Median Transparency & 86.72 \% \\
         Median $\sigma_{\text{mag}}$ [mag] &0.015 - 0.055 \\
         Exposure Time (bright time) & 12$\times$240s\\
         Exposure Time (dark time) & 6$\times$240s\\
         Relative Photometric Error &  1.6\% - 5.7\% \\ \hline 
    \end{tabular}}
    \caption{Details of \textit{WST} observations. The final number of datapoints is $69\%$ of the total number of observed nights due to quality constraints (see Section \ref{subsec:WST_data}). The sampling is the mean number of days between the observations and is computed with respect to the number of datapoints used. The median $\sigma_{\text{mag}}$ indicates the range of median uncertainty on the magnitudes of the lightcurves. Thus it depends on the luminosity of the images. Here the two extremes are shown for images B and D, respectively the brightest and dimmest. The number of exposures varies between dark and bright times by a factor of two.}
    \label{tab:wst_obs}
\end{table}
\subsection{Analysis}\label{subsec:td_analysis}
We follow the framework of the COSMOGRAIL collaboration \footnote{\url{https://www.epfl.ch/labs/lastro/scientific-activities/cosmograil/}} for the time delay analysis. We carry it out with the publicly available Python package \texttt{PyCS3} (Python Curve Shifting) presented in \citet{cosmog_pycs_XI}, using the free-knots spline fitting technique. This method assumes that the intrinsic lightcurve, i.e. the unlesed QSO lightcurve, can be represented by a spline, a piecewise polynomial function. The locations where the polynomial pieces connect, called ``knots'', are then also free parameters, in addition to their polynomial coefficients.
\texttt{PyCS3} describes all QSO images' lighcurves with one intrinsic spline, adjusting the shifts in order to minimise the $\chi^2$. To account for the presence of extrinsic variability, i.e. microlensing, an additional spline or polynomial is considered for each individual QSO images' lightcurve, whose parameters are adjusted simultaneously with the ones of the intrinsic curve. This ``microlensing fit'' or extrinsic fit is then adapted in order to account for the distortions of a single lightcurve with respect to the intrinsic spline and is therefore a relative correction between lightcurves. In fact, given two lightcurves, it is undefined which one is affected by microlensing and how, but only their relative distortion can be described. This degeneracy has no effect on the time delay measurement, while it has to be taken into account when studying the flux ratio (see \ref{sec:flux_ratio}).

The free knot positions avoid the bias introduced by a fixed grid of knots. 
The spline fitting method optimises simultaneously all splines' parameters and both shifts (time and magnitudes).
The optimisation is initialised with equidistant knots and a flat spline, and relies on two main parameters: 
the knot steps $\eta$, the initial time interval in days between knots (inversely proportional to the number of knots),
and the minimum knot distance $\epsilon$. 

The flexibility of the curve is defined by $\eta$: a too small knot step would overfit the data, as it would correspond to too many knots, while a too large one would produce a fit too coarse to appreciate the shortest time variations. The choice of the knot step therefore depends on the amount of freedom needed to fit the intrinsic variability of the lightcurves. Hence a strongly/weakly variable curve would require a small/large knot step. In the case under consideration, it is expected that most scales should be dominated by intrinsic variability, as large scale variation ($\sim 300$ days) common to all lightcurves can be seen by eye while smaller scale ($\sim 10-20 $ days) can be well-fitted. Very short peaks ($\sim 1-2 $ days) or very long trends ($\sim 1000$ days) can only be explained by microlensing. The very case is not explicitly accounted for in this analysis, but this phenomenon affects a small number of datapoints, roughly 2-3 points per peak. This has little effect on the overall result, given the small number of points involved compared to the dataset, and will be accounted for in the error budget, assuming a limited frequency of such events. This appears to be the case in this system, as very few isolated peaks are visible and given the ``goodness'' of the fit later described.
Medium and long-term variability due to microlensing is instead taken into account by the extrinsic fit, which is implemented by correcting every lightcurve by a spline or a polynomial with a low degree of freedom. This phenomenon also appears to have a limited effect in this system, as the lightcurves can be accurately modelled even with minimal microlensing correction, such as using a low-degree polynomial. 
One of the possible reasons for such limited impact of microlensing is the large separation of the images from the centre of the lens (which was one of the selection criteria of such a lensed system).

The time delay estimation follows the method described in \citet{cosmog_pycs_XIX}, but it is limited to the spline fitting method only, as we find good results with this procedure. 

As the whole analysis is dependent on the amount and type of freedom given to both the intrinsic and extrinsic fitting, rather than trying to choose the best set of parameters, the analysis is repeated for all combinations given a set of reasonable parametric choices. Firstly and most importantly the values of $\eta$  are chosen between 30, 35, 40, and 45 days. Secondly, the microlensing correction is defined by the amount of free parameters available for the extrinsic spline. It ranges between a polynomial of degree 0, corresponding to a simple magnitude shift,

a polynomial of degree 1 and a spline of degree 2 with two intervals and a fixed knot. A free knot for the microlensing is seen to overfit the data and is therefore avoided. 
By construction, a magnitude shift is indistinguishable from constant microlensing, and in \texttt{PyCS3} these values are completely degenerate (more about it in \ref{sec:flux_ratio}), hence the polynomial of degree 1 actually corresponds to the case where no microlensing is occurring.  

Moreover, since the external microlensing is only a relative correction, it can be applied either to all four lightcurves or a subset of three of them. The fourth one will be taken as reference one, i.e. as if it was equal to the intrinsic lightcurve. 

However, the two latter described approaches, the polynomial microlensing correction of degree 0 and the correction being applied only to a subset of lightcurves, lead to excessive scatter in the analysis results. This indicates that such methodologies are not stable enough, i.e. do not converge to a single value for the time delay and that microlensing has to be taken into account. 

We thus apply microlensing corrections to all lightcurves, while varying their degree of freedom (e.g. spline or polynomial). This results in a set of 12 analyses, given by 4 knot steps for the intrinsic spline, and 3 possible extrinsic variability corrections each applied to all lightcurves. This extensive parametric search is allowed by its small computational cost and a reliable method for obtaining the combined result, as explained in Section 3.3 of \citet{cosmog_pycs_XIX} and here summarised in \ref{subsec:comb_dt}. 

Previous to the analysis, the lightcurves are shifted in magnitude and time by coarse prior estimates, in order to leave \texttt{PyCS3} a fine adjustment on the order of 10 days. The prior magnitude shift is obtained by measuring the weighted average for the difference in magnitude with respect to image A, while the time delay is computed from the previously measured Fermat potential. In order to convert these values into time delay, a prior value for $H_0$ has to be considered, which could bias the measurement of the time delay. To avoid this, the analysis is repeated multiple times with the initial time delay randomly scattered on the order of 10 days around the expected time delay when considering a dummy value of $H_0=70\frac{\M{km}}{\M{Mpc\;s}}$.  
This scatter corresponds to a prior $H_0$ ranging approximately between 50 and 100 $\frac{\M{km}}{\M{Mpc\;s}}$ depending on the Fermat potential and thus does not bias the final result.

Given such a scatter of initial shifts, for each set of parameters, the analysis is repeated 1000 times and the median of the resulting time delay is taken as a result. The scatter of the resulting distribution, $\sigma_{\M{an}}$, is an indication of the robustness of the analysis but would underestimate the uncertainty by one order of magnitude \citep{cosmograil_XI}. Instead, a different approach is needed for the error estimation, which is explained in the following Section \ref{subsec:dt_error}. This scatter of the distribution is instead considered as a sanity check relative to the set of parameters used in the analysis, and any resulting $\sigma_{\M{an}}>2$ days is flagged and discarded. This allows to cut out all sets of parameters that are too little constraining. Furthermore, $\sigma_{\M{an}}$ is later considered for the error estimation, although its effect on the final error budget is negligible. 

\begin{figure*}
    \includegraphics[width=.8\textwidth]{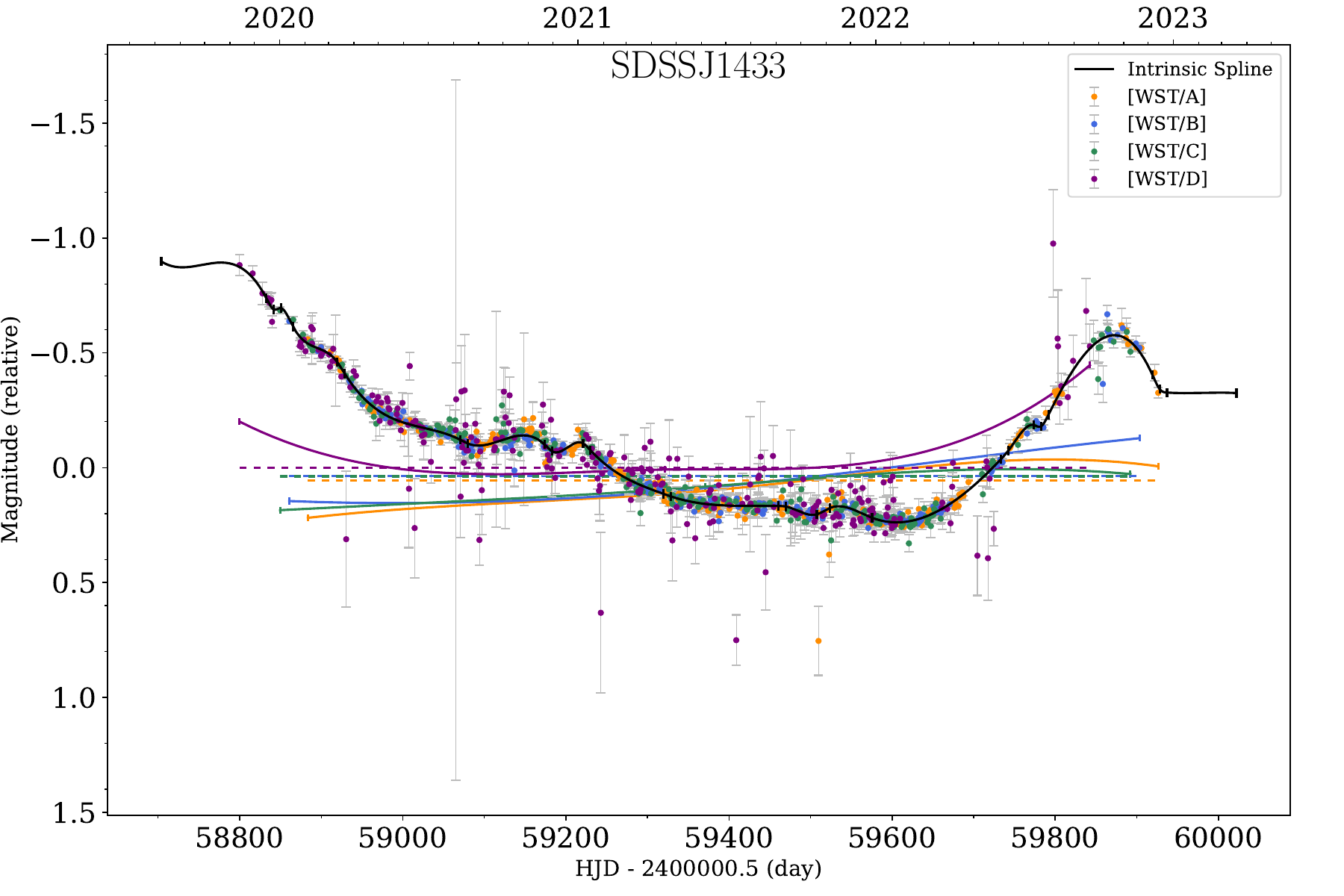}
    \caption{Result from time delay analysis with \texttt{PyCS3}  on the observed lightcurves. The black curve is the common intrinsic spline. The coloured points correspond to the datapoint of each lightcurve shifted by time delay, magnitude shift, and microlensing correction. The latter is plotted in the corresponding colour.
    This example is one of the fits obtained considering an initial knot step of 45 days and a spline microlensing correction}
    \label{fig:dt_analysis_spline}
\end{figure*}

\begin{figure}
    \includegraphics[width=\columnwidth]{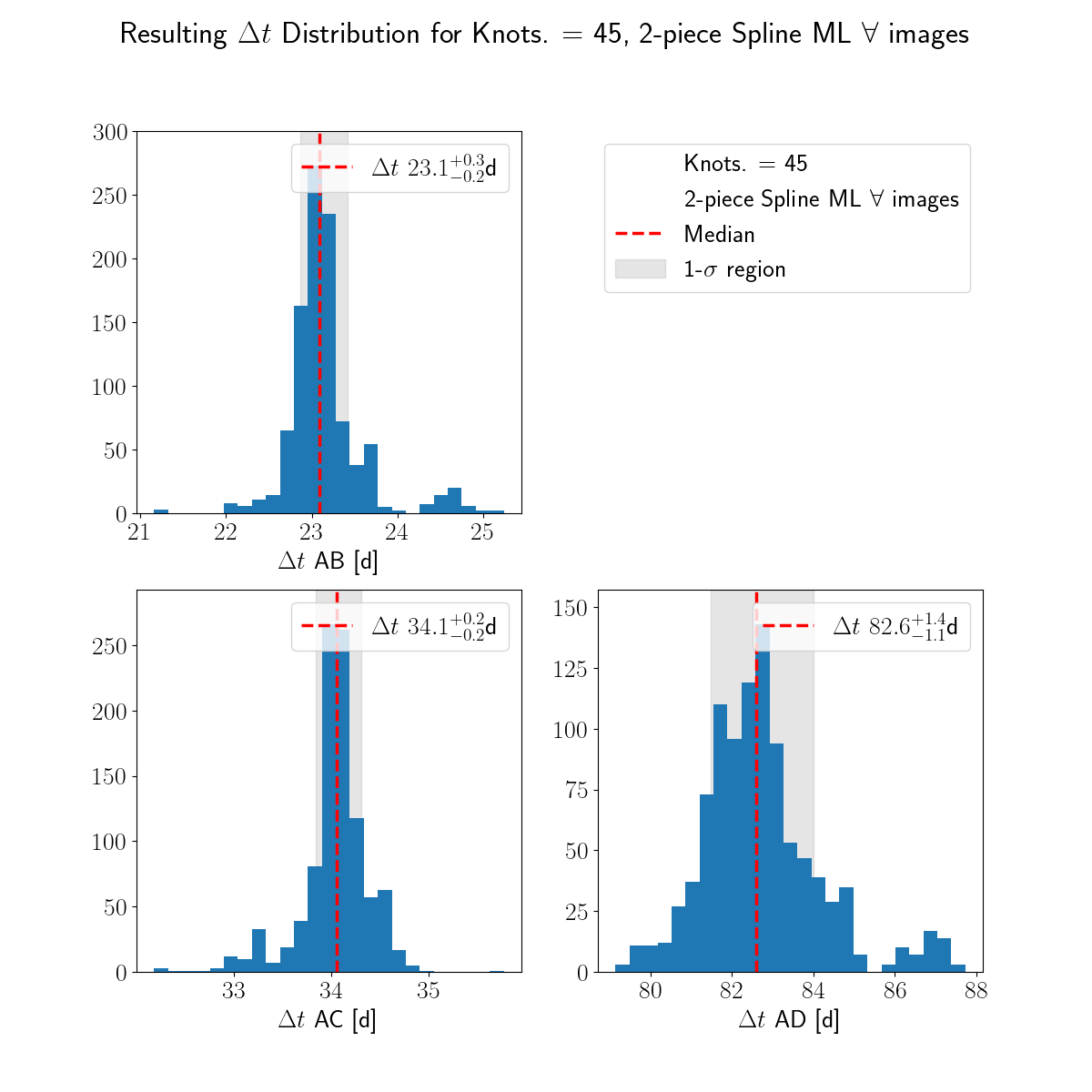}
    \caption{Distribution of time delay results from the 1000 analyses of the observed lightcurves for a set of parameters. $\text{Knots.}$  indicates the knot step, i.e. the initial separation between the position of the knots in days. ``$\text{Splne ML } \forall \text{ images}$' ' refers to the fact that the spline microlensing correction is applied to all images and not a subset of such. }
    \label{fig:dt_res_distrib}
\end{figure}

Given the large number of analyses considered and, for each of them, the large number of instances of the given analysis, it is not possible to report here all the results, but for exemplary purpose, one is shown in Figure (\ref{fig:dt_analysis_spline}), given the parametric set of 45 days as the initial knot step and a spline microlensing correction with one fixed knot.
Once considering the full distribution of results for the time delay analysis we obtain the distribution shown in Figure \ref{fig:dt_res_distrib}.  
Note that the scatter of results, indicating the robustness of the given analysis, is at maximum on the order of a day.
\subsection{Error Estimation} \label{subsec:dt_error}
Following \citet{cosmog_pycs_XI}, the error estimation is based on a ``Monte Carlo'' approach by generating a large dataset of synthetic lightcurves statistically compatible with the observed ones and analysing them as previously done on the real data. Comparing the resulting time delay with the real-time delay of the synthetic lightcurves, the error of the analysis is estimated as $\sigma_{\Delta t} = \Delta t^{\M{real}}_{\M{sim}} - \Delta t^{\M{meas}}_{\M{sim}}$.  

The generative model used for the synthetic lightcurve is described in \citet{cosmog_pycs_XIX}. The following is done in parallel for each set of parameters used in the analysis, such as to obtain independent estimates of the time delay uncertainty for each of them.

Firstly an intrinsic spline is modelled by repeating the initial analysis on the observed lightcurves. This is considered the original intrinsic lightcurve for the simulated dataset. 

It is then shifted by time delays uniformly randomised around the expected delays in a range of $\pm 10$ days and by magnitude shifts similarly randomised around the mean magnitude shift in a range of $\pm 0.5$ mag.  

The obtained lightcurves are then sampled at the same dates of the observations, and the single magnitude datapoint is given a photometric uncertainty equal to the corresponding observed datapoint. Finally, a noise component is added by estimating the power spectrum of the residuals with randomised phases in the Fourier space. This generative procedure is repeated to create a dataset of 800 synthetic lightcurves having compatible ``time delay constraining power'' as the observed dataset but with known time delays. 

The synthetic data are then analysed in the same way as the original (i.e. with the same set of parameters), producing an error distribution, where the median indicates the systematic error $\sigma_{\M{sys}}$ and the scatter the random error $\sigma_{\M{rnd}}$ of the analysis under consideration. These are in fact two components of the total error and it can be seen that they depend on the flexibility of the fitting. If the fit is too flexible and overfitting the data, the resulting distribution would be unbiased but with a large scatter. Conversely, if the fit is given too little freedom to adjust to the data the distribution would present little scatter around a biased result. In order to take into account the scatter of the initial analysis $\sigma_{\M{an.}}$, the latter is added to the systematic error under quadrature, thus $\sigma_{\M{sys}}'=\sqrt{\sigma_{\M{an}}^2 + \sigma_{\M{sys}}^2}$. This only marginally affects the error budget. 

Thus the time delay error for a given analysis is $\sigma^2 = \sigma_{\M{rnd}}^2 + \sigma_{\M{sys}}'{}^{2}$. While this is considered as the time delay uncertainty in this analysis, in Section \ref{sec:h0} the correlation between couples of time delay differences will be taken into consideration explicitly.

\begin{figure}
    \includegraphics[width=\columnwidth]{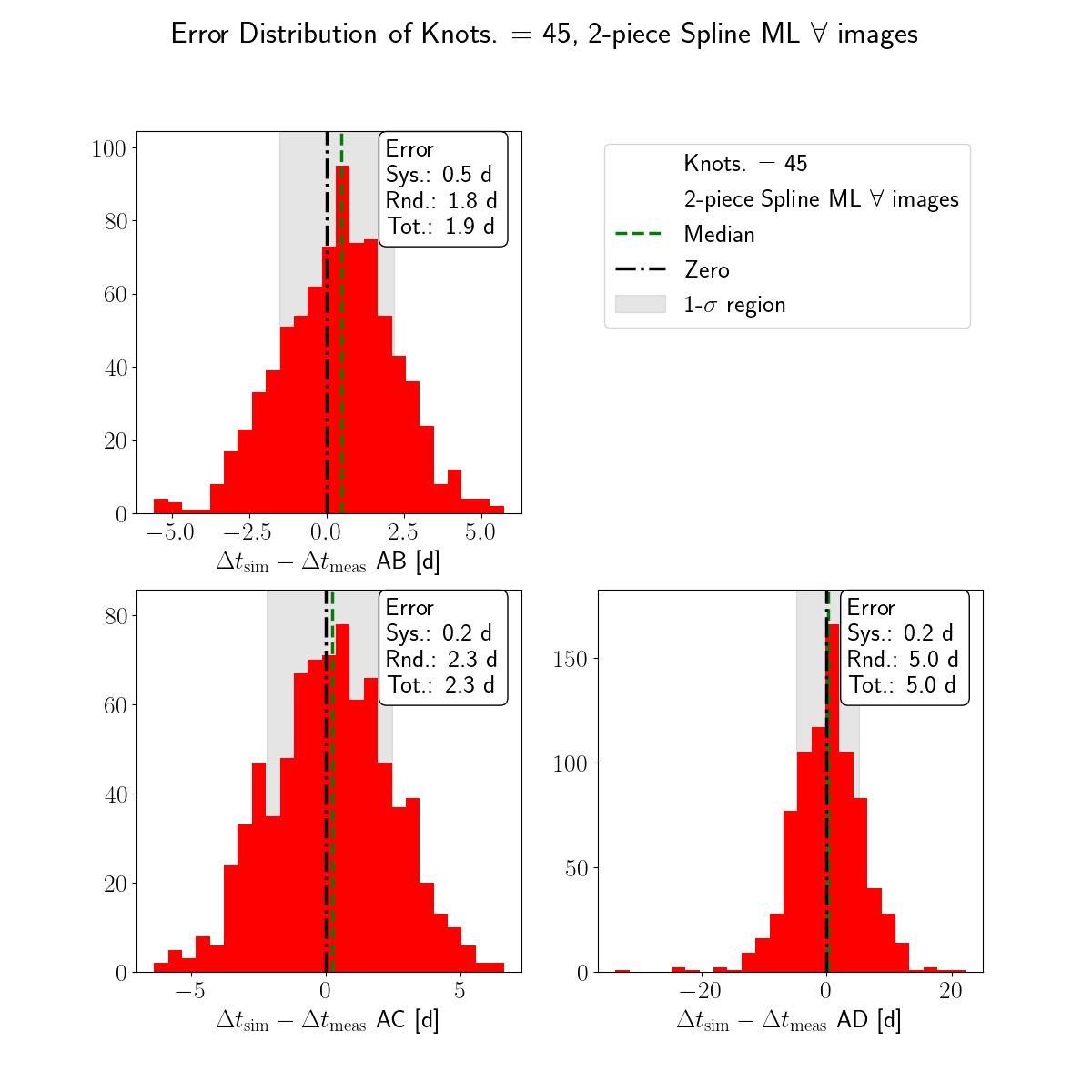}
    \caption{The error distribution of the time delay estimate given a set of parameters. See \ref{fig:dt_res_distrib}  for the description of the parameters used. Note that the error estimate is dominated by the random error, most notably for the time delay AD, due to the low luminosity of image D and thus higher photometric uncertainty. }
    \label{fig:dt_err_distr}
\end{figure}

As in Section \ref{subsec:td_analysis},  we will only show a specific analysis - the same shown in Figure \ref{fig:dt_analysis_spline} and \ref{fig:dt_res_distrib}. Its error distribution is shown in Figure \ref{fig:dt_err_distr}. It can be easily seen that the random error is dominating and that the intrinsic error shown in Figure \ref{fig:dt_err_distr} is negligible.  

\subsection{Combining the Time Delay Estimates}\label{subsec:comb_dt}

Following \citet{cosmog_pycs_XIX}, in particular Section 3.3,  we do not select the most precise estimate nor do we marginalise over all obtained results. We instead opted for a hybrid approach by considering the tension between the results. First, the most precise result is selected. The tension between such a result ${\Delta t_\alpha }^{+\sigma^+_{\alpha}}_{-\sigma^-_{\alpha}}$ and all others ${\Delta t_\beta }^{+\sigma^+_{\beta}}_{-\sigma^-_{\beta}}$ is evaluated as 
\begin{equation}
\label{eq:tau}
    \tau(\Delta t_\alpha,\Delta t_\beta) = \frac{\Delta t_\alpha - \Delta t_\beta}{\sqrt{{\sigma^-_{\alpha}}^2 + {\sigma^+_{\beta}}^2}} 
\end{equation}
given $\Delta t_\alpha>\Delta t_\beta$. Else the signs of the numerator and of the signs labelling the $\sigma$ would be inverted. $\alpha$ and $\beta$ indicate the result for a specific set of analysis parameters, e.g. $\eta$=45 days and 1-degree polynomial microlensing correction, and for a couple of lightcurves, e.g. AB. 
The tension between the results of the two analyses is given by the maximum
tension between all possible lightcurves couples.

If there are results with a tension relative to the initial best result higher than a threshold $\tau_{\text{thresh}} = 0.5$, the most precise of this set of results is selected. This result is then combined with the initial best result. The combined result is now taken as the ``best result'' and the process is repeated until there are no more results with tension $\tau$ relative to the combined one higher than $\tau_{\text{thresh}}$. This approach depends on the value of $\tau_{\text{thresh}}$, which can be adjusted. A larger $\tau_{\text{thresh}}$ would be equivalent to selecting only the optimal result while having a smaller one would combine all the results together.  

\begin{figure}
    \includegraphics[width=\columnwidth]{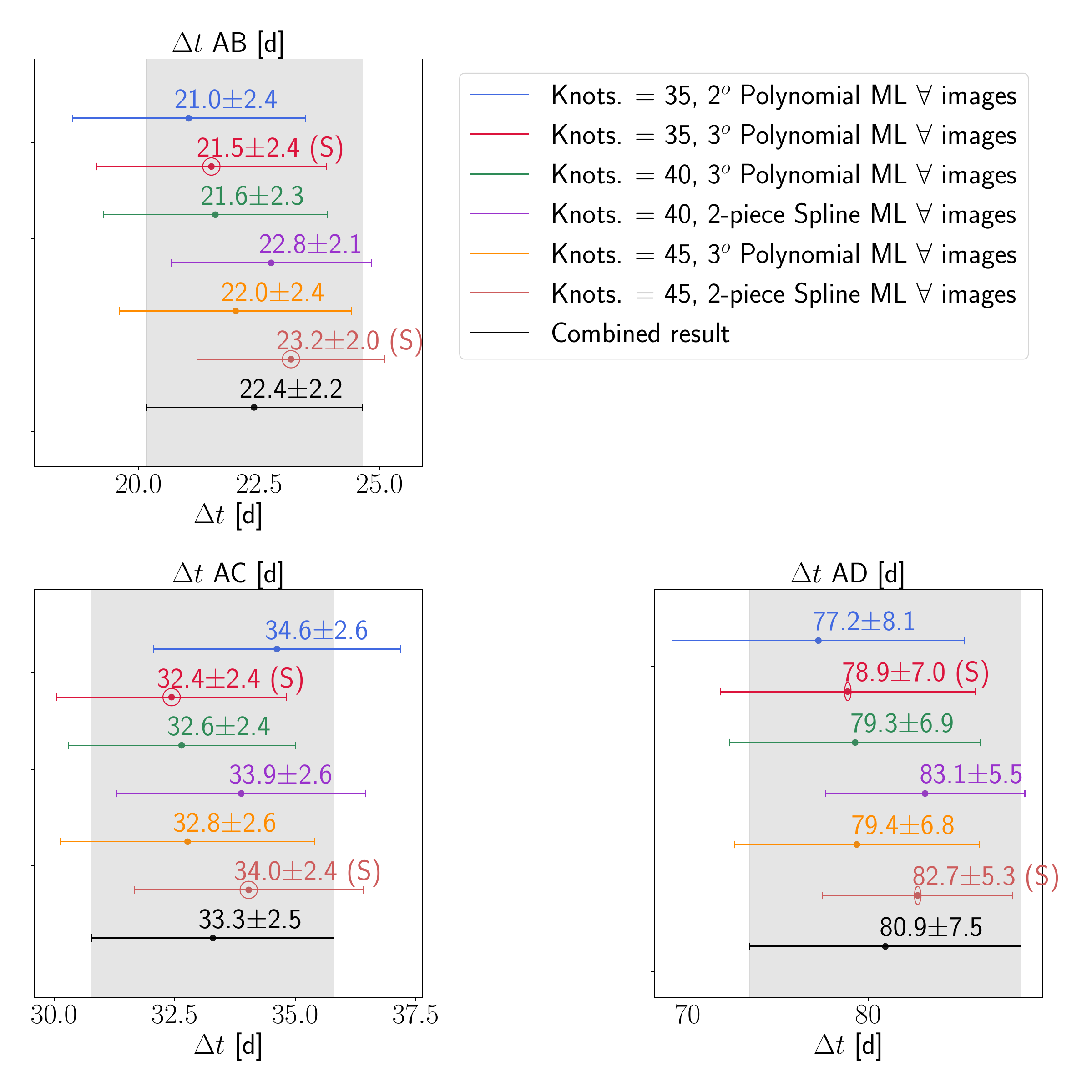}
    \caption{Time delay results for the various lightcurves couples and the combined final time delay. The (S) indicates which results are iteratively selected to be combined in the final ``Combined result'' (see Section \ref{subsec:comb_dt}).}
    \label{fig:dt_combined_res}
\end{figure}

In Figure \ref{fig:dt_combined_res} all results are shown along with their errors and the combined result. 
The combined results are also shown in Table \ref{tab:timedelay_results}. 
\begin{table}\centering
\resizebox{.8\columnwidth}{!}{
\begin{tabular}{ccc}
\hline
  $\Delta t$ AB [d] & $\Delta t$ AC [d] & $\Delta t$ AD [d]\\
\hline
  22.1 $\pm$ 2.4 & 33.3 $\pm$ 2.5& 80.9$ \pm$ 7.5 \\
\hline
\end{tabular} }

\caption{Combined results for the time delay between the various couple of images.}
\label{tab:timedelay_results}
\end{table}

\section{Flux Ratio Anomaly}\label{sec:flux_ratio}

We discuss here the flux ratio anomaly of the results, which is the difference between the flux ratio ($\text{FR}$) obtained from the observed flux in the images and the one that is expected from the magnification obtained from the lens model. 
Given a luminosity $L$ for the source, the observed luminosity of each image would be $L_i=\mu(\vec{\theta_i})\cdot L(t+\tau(\vec{\theta_i}))$, where $\mu(\vec{\theta})$ is the magnification map and  $\tau(\vec{\theta})$ being the time delay map due to the lens. If $L$ is assumed to be constant over time, i.e. $L(t)=L(t+\tau(\vec{\theta_i}))=L$, taking two images $A$ and $B$, the $\text{FR}$ should then correspond to $\text{FR}_{AB}=\frac{L_A}{L_B}=\frac{\mu(\vec{\theta_A})}{\mu(\vec{\theta_B})}$. However, this is not the case here, as the luminosity of the QSO varies over time, as observed in the lightcurves of Figure (\ref{fig:lc_init}).  
Moreover, the single image could be temporarily microlensed by an intervening massive object, as discussed in Section \ref{sec:lc_analysis}, which would further affect the $\text{FR}$, as $L_i(t) = (\mu(\vec{\theta_i}) + \delta\mu(t))\cdot L(t+\tau(\vec{\theta_i})) $.
For this reason, $\text{FR}$ cannot be measured directly from the \textit{HST} images, as it would be biased by the time delay and microlensing events.

Instead, the \textit{WST} lightcurves were used to measure the observed $\text{FR}$. Once analysed with \texttt{PyCS3}, they can be corrected for time delays between the images. Assuming an effective microlensing correction, microlensing effects should also be accounted for in this case, but only for events with time scales shorter than the observation campaign. Very long microlensing events, appearing as a constant magnification of one image over the observation campaign, would not affect the time delay analysis but would bias the measured $\text{FR}$. It is to be noted that such events are more unlikely to happen the longer the observation, as they would require the orbit(s) of the microlens(es) to be very well aligned with the line of sight of the image. A more likely scenario would be the presence of substructures (e.g. subhalos) of the lens which are not modelled by the smooth PEMD lens model.

The magnitude shift computed in \texttt{PyCS3} can then be converted in $\text{FR}$  as $\text{FR}_{Ai}=\frac{F_i}{F_A} = 10^{\frac{-2\cdot\Delta \text{mag}_{Ai} }{5}}$, where $\Delta \text{mag}_{Ai} = \text{mag}_i-\text{mag}_A$ is the magnitude shift between images $i$ and A, where $i$=B, C, D. 

The relative magnitude shift is then calculated as the average of the microlensing correction, being a polynomial or a spline, added to the initial magnitude shift. The latter is previously obtained by taking the difference of the weighted mean for the magnitude of the lightcurves, as described in Section \ref{subsec:td_analysis}.
Also note that, as previously stated, a constant microlensing and a magnitude shift are indistinguishable in the lightcurves, thus in the analysis the latter is formally defined as a polynomial microlensing of degree 0.

This magnitude shift was computed in parallel with the time delay, following the same procedure. 

\begin{figure}
    \includegraphics[width=\columnwidth]{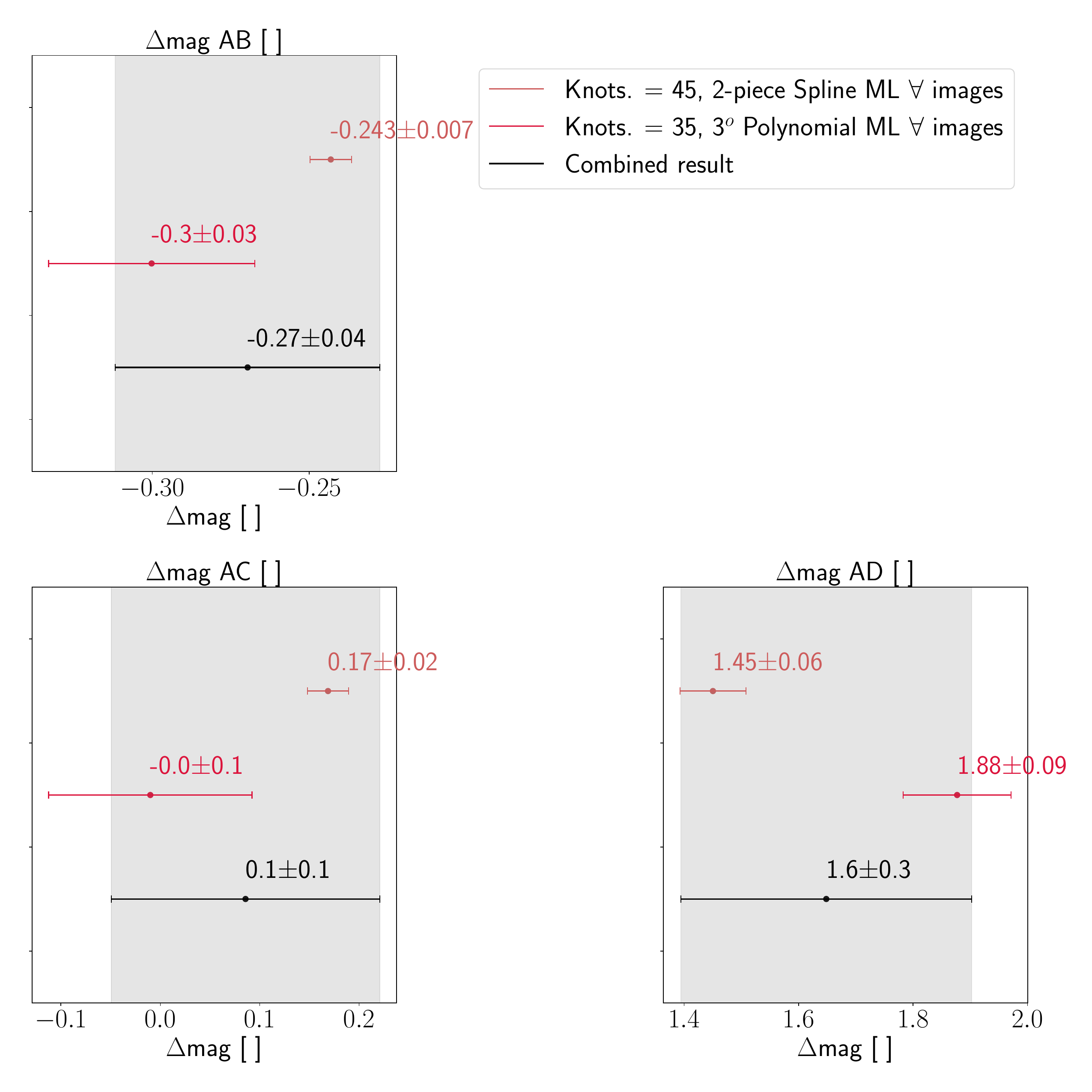}
    \caption{Results of the magnitude shift measurements from the lightcurves analysis, from which the flux ratio is computed as shown in \ref{tab:FR}. ``Knots.'' indicate the knot steps of the intrinsic spline, inversely proportional to the freedom of the model. The ML refers to microlensing, which is either fitted with a polynomial or with a 2-degree spline.}
    \label{fig:FR_pycs}
\end{figure}

The results to be combined were selected by taking only the groups that were combined for the time delay analysis.  The result is shown in Figure \ref{fig:FR_pycs}.

On the other hand, the magnification obtained from the lens model is a multiplicative factor that scales the unlensed flux at the position of the image $i$ as $F_i = |\mu_i| F_{\text{unlensed}}$, such that $\frac{F_i}{F_A} =|\frac{\mu_i}{\mu_A} |$. 

Also note that, as for $\Delta t$, the absolute value of the magnification for a single image is of little interest. Only the relative value (here chosen to be the ratio with respect to $\mu_A$) is observable.\

We calculate the posterior of the magnification ratio with the same approach as for the combined posterior of $\Delta \phi$  from the converged MCMC chains.

\begin{figure}
    \includegraphics[width=\columnwidth]{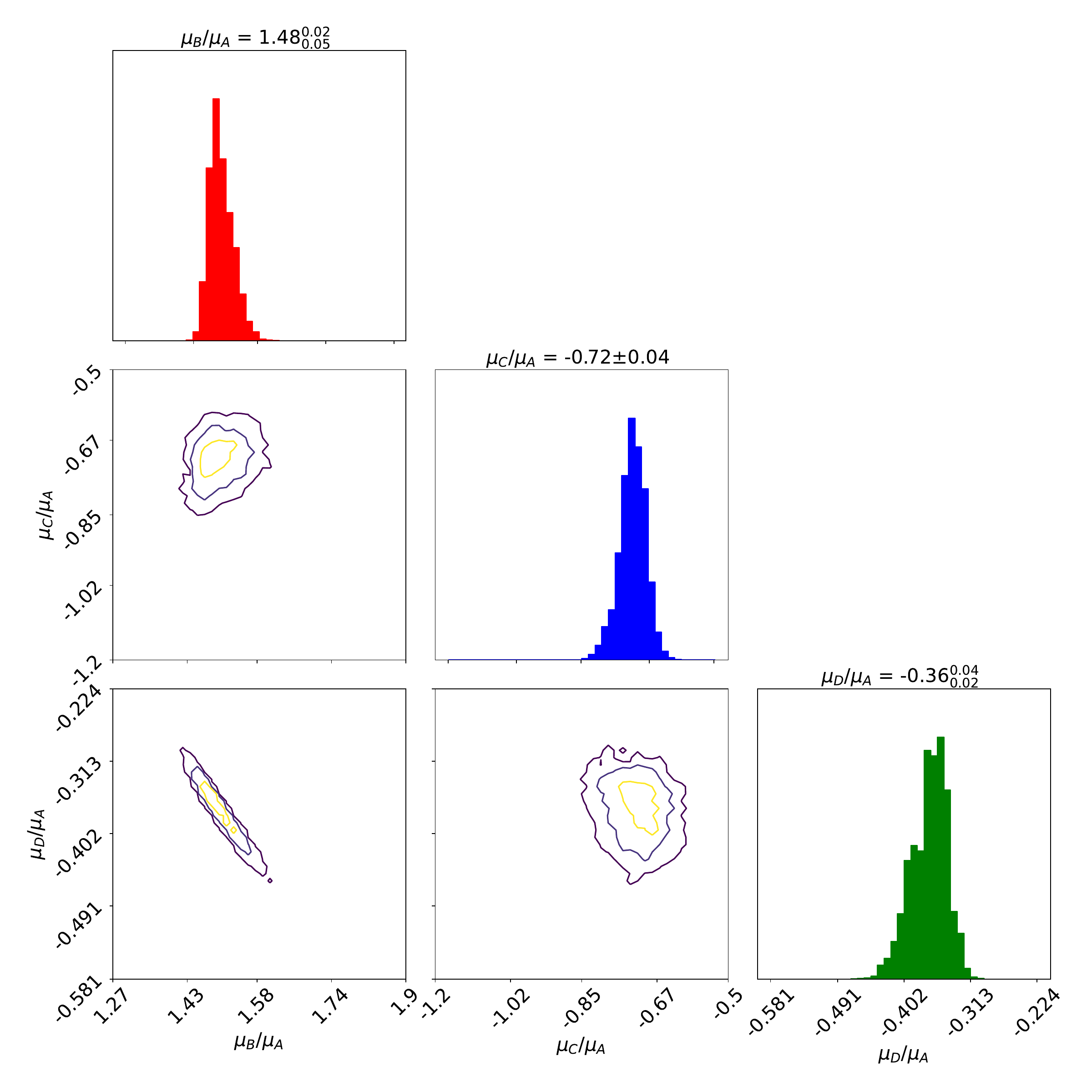}
    \label{fig:mag_post}
   \caption{ Resulting combined posterior of the magnification ratio $\mu_{\textit{i}}/\mu_A$ at the positions of the images B, C, and D from the modelling \ref{subsec:lens_modelling} with respect to image A. The contour levels indicate the 68, 95 and 99.7$\%$ confidence level.}
\end{figure}

Both results are then reported in the Table \ref{tab:FR}. 
\begin{table}\centering
\resizebox{.9\columnwidth}{!}{
\begin{tabular}{|cccc|}
\hline 
$\M{FR}$ & LCs Analysis &Lens Model & Tension\\ 
\hline
$\M{FR}_{\M{AB}}$ & $ 1.28\pm0.05 $ & $ 1.48^{+0.02}_{-0.05} $&$2.9$\\
$\M{FR}_{\M{AC}}$ & $ 0.9\pm0.1 $ & $ 0.72\pm0.04 $&$1.7$\\
$\M{FR}_{\M{AD}}$ & $ 0.22\pm0.05 $ & $ 0.36^{+0.04}_{-0.02} $&$2.6$\\

\hline
\end{tabular} }

\caption{Comparison between $\M{FR}$ obtained from lens modelling and lightcurve analysis, and their tension $\tau$ (see \ref{eq:tau}).} \label{tab:FR}
\end{table}
All images couple have some degree of tension between the expected $\M{FR}$ from the lens model and the observed one from the lightcurves analysis.
This can be explained by several possible factors: absorption along the line of sight, extremely long timescale microlensing variation (as shorter timescales would be washed out) or some additional lensing structures which are not accounted for in the lens modelling. These possible explanations have been presented in order of plausibility, as well as their effect on the final measurement of $H_0$. In the case of absorption, this could be due to gas or dust along the line of sight which could affect the flux of a single image while leaving unaltered the Fermat potential, hence without affecting the time delay measurement. Secondly, long microlensing on the timescale of years is fairly unlikely to be strong enough to be significant but would contribute to the noise in the time delay measurement and possibly hinder it. Finally, an undetected lensing perturber in proximity to the image position could be explained by dark matter substructures of the main lens unaccounted for by the mass model. This latter case could bias significantly the present measurement of $H_0$ but would be an interesting case study for the research in dark matter properties.

\section{Constraints on $H_0$}\label{sec:h0}
We review here briefly how the constraints obtained in the previous sections can be used to infer $H_0$ within the TDC method. The time delay $\Delta t$ induced by the gravitational lens depends on the `time delay distance' $D_{\Delta t}$ and on the Fermat potential difference $\Delta \phi$; $\Delta t=D_{\Delta t}\cdot \Delta\phi/c$, where $c$ is the speed of light. 
While the Fermat potential depends on the mass distribution of the lens, the time delay distance is a function of angular diameter distances defined as 
\begin{equation}
     D_{\Delta t} \equiv (1+z_{\M{l}}) \frac{D_{\M{l}} D_{\M{s}}}{D_{\M{ls}}}
\end{equation}
\citep{refsdal1964possibility,suyu2010dissecting}, where $z_{\M{l}}$ is the redshift of the lens, while $D_{\M{l}}$ is the angular diameter distance between the observer and the lens, $D_{\M{s}}$ between the observer and the source, and $D_{\M{ls}}$  between the lens and the source.
Thus $D_{\Delta t}$ can be obtained by measuring $\Delta t$ (\ref{sec:lc_analysis}) and $\Delta \phi$ (\ref{sec:lens_mod}). From its definition, $ D_{\Delta t}$ is primarily sensible to $H_0$ and can be rewritten as $D_{\Delta t} = k/H_0$, where all cosmological priors are contained in $k$. Assuming a flat $\Lambda$CDM, $k$ depends on the redshifts of the lens $z_{\M{l}}$ and of the source $z_{\M{s}}$, and only weakly on $\Omega_{\M{m}}$:
\begin{equation*}
k(\Omega_{\M{m}},z_{\M{l}},z_{\M{s}}) =\frac{\int_0^{z_{\M{s}}} \frac{dz'}{E(\Omega_{\M{m}}, z')} \int_0^{z_{\M{l}}} \frac{dz'}{E(\Omega_{\M{m}},,z')}}{\int_{z_{\M{l}}}^{z_{\M{s}}} \frac{dz'}{E(\Omega_{\M{m}},z')}}
\end{equation*}
where $E(\Omega_{\M{m}},z) = \sqrt{\Omega_{\M{m}} (1+z)^3 + (1-\Omega_{\M{m}})}$.

The linear relation between $H_0$ and the observed parameter is then
\ref{eq:h0}:
\begin{equation}
    H_0(\Delta \phi_{\text{ij}},\Delta t_{\text{ij}}, \Omega_{\M{m}},z_{\M{l}},z_{\M{s}}) = k(\Omega_{\M{m}},z_{\M{l}},z_{\M{s}}) \frac{\Delta \phi_{\text{ij}}}{\Delta t_{\text{ij}}}
    \label{eq:h0}
\end{equation}
where i and j indicate two different lensed images. In this study, we further assume a fixed value for $\Omega_{{\M{m}}}=0.3$. Thus, given the redshift of the source and the lens, the function $k(\Omega_{\M{m}},z_{\M{l}},z_{\M{s}})$ is effectively a constant $k$ for the system under consideration.

A known systematic of such methodology is the so-called ``Mass-sheet Degeneracy'' \citep[MSD,][]{falco1985model}, where the projected mass distribution $\kappa(\theta)$ can be remapped into $\kappa_\lambda(\theta)$ without affecting the fit of the lensed light apart from absolute magnification and relative time delay of the images:
\begin{equation*}
    \kappa_\lambda(\theta) = \lambda\kappa(\theta) + (1-\lambda),
\end{equation*}
combined with an isotropic rescaling the source coordinates $\beta\rightarrow \lambda\beta$ in the source plane.
Both time delays and magnification can not be used in this study to constrain the MSD as the intrinsic luminosity of the QSO is unknown while the time delay is used to infer $H_0$. Conversely, the Fermat potential differences would be affected by a factor $(1-\lambda)$, thus the measured Hubble parameter would be biased as $H_0^{\M{true}}=(1-\lambda)H_0^{\M{meas}}$ \citep{suyu2010dissecting}.

The MSD has been thoroughly investigated over the years \citep{MSD_Suyu_2014,wertz2018ambiguities} and recognised to be a special case of the more general Source Position Transformation \citep{schneider2013mass}. It is commonly divided between internal ($\kappa_{\text{int}}$) and external ($\lambda_{\text{ext}}$) MSD, depending on whether it is due to degeneracy of the lens mass model considered to fit the data or line-of-sight structures and environment under/over-densities, respectively. This division is based on the observables used to break such degeneracy. 

$\kappa_{\text{int}}$ can be estimated from 2D spectroscopically derived stellar velocity dispersion of the lens correlated with the imaging-based mass model \citep{Treu02MSD,shajib2018improving}. On the other hand, $\lambda_{\text{ext}}$ is constrained on the weighted number counts of surrounding galaxies along with weak gravitational lensing information \citep{suyu2010dissecting,suyu2013_MSD_ext}. 

In this paper, neither of such approaches is implemented due to limitations on the available data. We can observe that the lens system is not located in a group or cluster, and thus $\lambda_{\text{ext}}$ should have a limited effect. 
They are therefore purposely left out of the scope of this research and should be later faced in a future study. 
In order to constrain $H_0$ given the obtained posteriors of $\Delta t$ and $\Delta \phi$ a Bayesian approach was followed.  Using equation \ref{eq:h0} the posterior distribution of $\Delta t$ can be converted into a posterior distribution on the $\Delta \phi$ parametric space given a certain value of $H_0$. This ``mapped'' posterior $P_{\Delta t}^{\text{map}}(\Delta\phi | H_0)$, once opportunely normalised, can be multiplied with the posterior of the measured Fermat potential, $ P_{\Delta \phi}^{\text{meas}}(\Delta\phi)$, in the Fermat potential parametric space. Marginalising over all possible values of $\Delta \phi$ and multiplying by the prior of $H_0$, the combined posterior of $H_0$ was obtained with respect to the time delay data and the Fermat potential data. This is described in equation \ref{eq:comb_H0}:
\begin{dmath}
\label{eq:comb_H0}
 \large P (H_0 \mid \Delta t_{\text{ij}},\Delta \phi_{\text{ij}}) = \int d\Delta \phi_{\text{ij}} P_{\Delta \phi}^{\text{mea.}}(\Delta \phi_{\text{ij}})  \frac{P_{\Delta t}^{\text{map.}} (\Delta \phi_{\text{ij}}\mid H_0) \, \text{Prior}(H_0)}{\int dH_0 ' P_{\Delta t}^{\text{map}}(\Delta \phi_{\text{ij}} \mid  H_0') }.
\end{dmath}
Here only the image combination $ij$ was shown, but it is trivial to generalise over multiple combinations by integrating over the multidimensional $\Delta \phi$ parametric space, considering a multi-dimensional posterior for both $P_{\Delta \phi}^{\text{meas}}$ and  $P_{\Delta t}^{\text{map}}$. Given the 4 images, hence 3 independent differences, the integral results to be in 3 dimensions. 

Three main elements enter this equation: the posterior probability of the measured Fermat potential $P_{\Delta \phi}^{\text{meas}}(\Delta \phi)$, the prior of $H_0$ Prior$(H_0)$,  and the posterior probability of the time delay mapped on the Fermat potential domain given $H_0$ $P_{\Delta t}^{\text{map}}(\Delta \phi|H_0)$. The first element is obtained directly from Section \ref{subsec:compare_&_combine}, as the method employed to combine the posteriors for the various filters yielded a binned posterior, which binning is here considered for the integration $d\Delta \phi$ in the Fermat potential domain.

Secondly, a uniform sample of test values of $H_0$ is taken, ranging from $50$ to $100 \frac{\M{km}}{\M{Mpc\;s}}$ evenly separated by $0.1\frac{\M{km}}{\M{Mpc\;s}}$. By construction, this implied a top-hat prior Prior$(H_0)$ with the above-defined boundaries, which is then ignored in the following steps as it only affects the normalisation.

The time delay posterior was considered to be defined as a multivariate Gaussian centred around the measured time delay and which covariance is given by the covariance matrix described in \ref{subsec:td_analysis}. Given the linearity of the transformation $\Delta t \rightarrow \Delta \phi$, it is straightforward to obtain the equivalent distribution in the time delay domain:
\begin{align}
\Delta \phi(\Delta t,H_0) &= \Delta t \frac{H_0}{k} \\
\M{cov}(\Delta \phi (\Delta t,H_0)) &= \M{cov}(\Delta t) \left(\frac{H_0}{k}\right)^2,
\end{align}
where $H_0$ is given from the sampled prior. It is necessary to explicitly normalise such mapped posterior by marginalising over all $H_0$, as this normalisation is not constant by default.

\begin{figure}
    \includegraphics[width=\columnwidth]{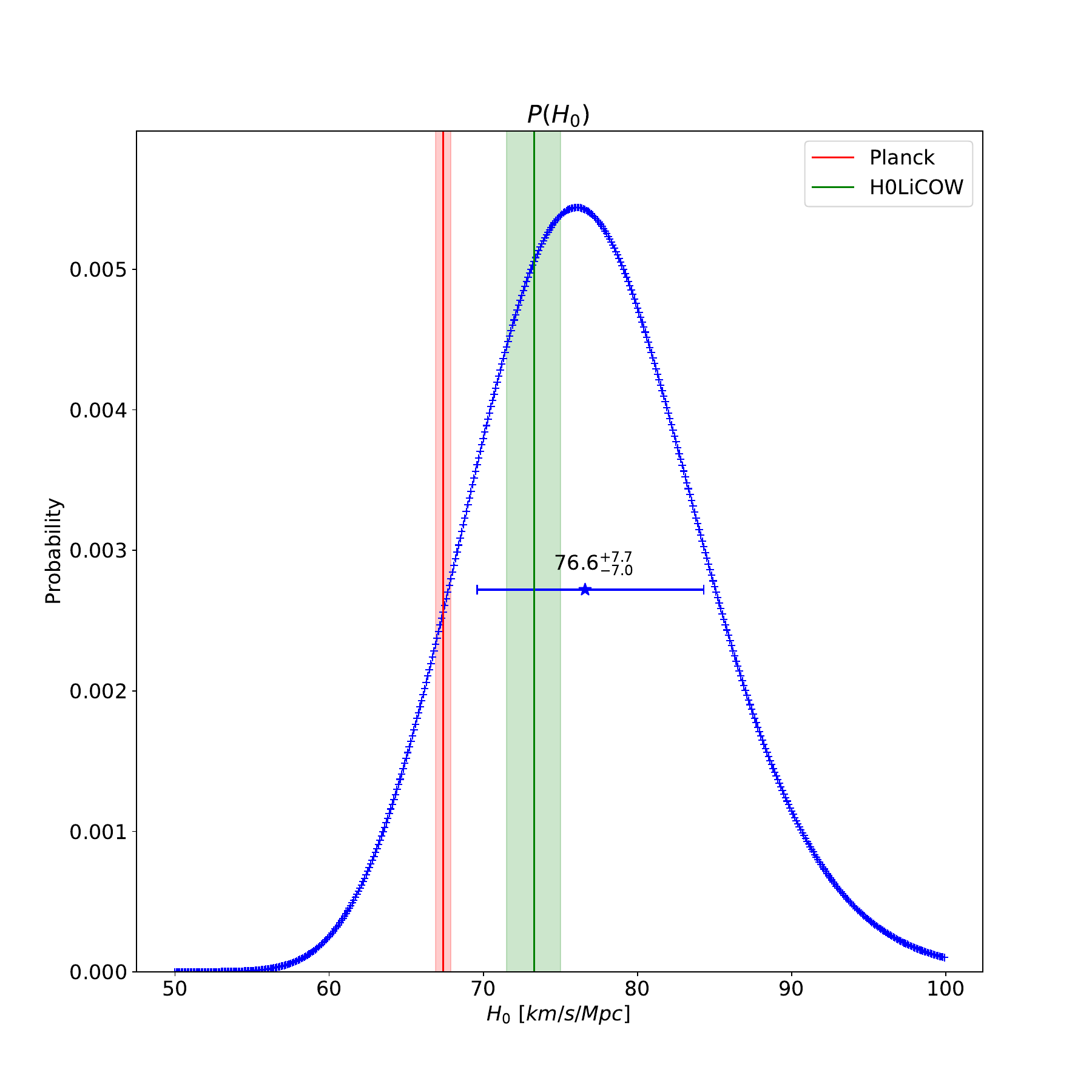}
    \caption{Posterior probability for $H_0$ constrained by lightcurves A, B, C, and D of SDSSJ133 given a PEMD mass profile compared with results from the Planck Collaboration \citet{aghanim2020planck} and H0LiCOW Collaboration \citet{h0licow_XIII}}
    \label{fig:res_H0}
\end{figure}

The result is shown in Figure \ref{fig:res_H0} where two characteristics results of the Hubble tension are reported for illustrative purposes: the Planck collaboration's 
\citep[``Planck'': $67.4\pm0.5\frac{\M{km}}{\M{Mpc\;s}} $, ][]{aghanim2020planck} and the H0LiCOW collaboration's 
\citep[``H0LiCOW'': $73.3_{-1.8}^{+1.7} \frac{\M{km}}{\M{Mpc\;s},} $][]{h0licow_XIII} results.

The resulting value is 
\begin{equation}
 H_0=76.6^{+7.7}_{-7.0}\frac{\M{km}}{\M{Mpc\;s}}.
\end{equation} 
The total error budget of  $\sim 9.6\%$ is dominated by the time delay uncertainty, whose relative error contribution is $11.2\%$, which in turn is divided between the image couples depending on their accuracy (ranging between $8\%$ for AC and $15\%$ for AD). The difference of Fermat potential relative error contribution is $\sim6.3\%$, and is instead divided evenly to $6\%$ for every image couple. 
Note that the combined error of $H_0$ is based on the combination of three measurements, given the three combinations of images, which are considered independent. We remind that this analysis does consider explicitly the contribution of random errors introduced by the time and Fermat potential measurements but does not take into account the possible systematics, such as the MSD.

\section{Conclusion}\label{sec:conclusion}
In this work, we have presented the first integral analysis of J1433 in the TDC framework. We have combined the results from HST data lens modelling with time delay analyses based on a 3-year monitoring campaign with the 2.1m Fraunhofer telescope on the Wendelstein Observatory in the $\textit{g'}$-band. The quality of the Wendelstein site allows for taking data with a cadence of $\sim 4$ days and a median PSF of $1.16''$ throughout the year and lunar phases. The relative photometric errors for the QSO fluxes range between 1.6$\%$ to 5.6$\%$. This demonstrates that this site and instrumentation qualify for light curve monitoring of multiple imaged QSO-galaxy lensing systems for time delay cosmography. 

Regarding the lens modelling, we have closely followed the approaches of the H0LICOW and TDCOSMO collaborations and we have used the public \texttt{lenstronomy} modelling tool, which ensures that our analysis is reproducible by other researchers. When we compare our lens modelling results with previous ones, that have been obtained with data of different amounts and quality and within a different scope, we find that some parameters are in tension with ours within 1.7 to 3.8 sigma (see Table \ref{tab:comp_s22}). These discrepancies can be explained by the presence of a ``contaminant'' source of light in the F160W exposures, which is either not related to the lensed source, or points to a more complex source in this band than can be modelled with our approach and the available data.

This further highlights the critical importance of high-resolution data in multiple bands for cosmography-grade lens modelling. As shown in this work low-resolution images and the usage of low signal-to-noise regions for lens modelling can lead to modelling light features in the data which could potentially bias the modelling procedure. 

This resolution and noise aspect is especially present for near-infrared and infrared images HST images, which traditionally carry more constraints given by the presence of brighter lensed arcs (usually originating from the lensed QSO host galaxies). Therefore, we recommend independent modelling for each filter to probe for consistency in both the data and the modelling procedure. We further note that, in a hypothetical scenario in which only F160W images would have been available for a similar lensing system, no tension between different models would have pointed us to the discovery of the contaminant. The results then would have been strongly biased. The obtained posterior of $\Delta\phi$ would have been affected on the order of 1.7$\sigma$ (see Figure  \ref{fig:Df_f160w_wo_mask}), and given its linear relation to the $H_0$ posterior, a similar bias would have entered our cosmological inference. 

We finally point out the limitations of our analysis, as it has not been blinded to avoid confirmation biases in the cosmological parameters. Furthermore, while the random error has been fully accounted for, systematic effects such as internal and external mass sheet degeneracy remain to be addressed for this system with additional studies.
Once we combine our posteriors for the Fermat potentials and time delays for all images we obtain a posterior for the Hubble constant of 
$H_0=76.6^{+7.7}_{-7.0}\frac{\M{km}}{\M{Mpc\;s}}$ ($9.6\%$ precision). Our result might be on the higher end of the Hubble tension, but one has to bear in mind, that the posteriors for $H_0$ from the analysis of individual lens systems have a fairly large scatter \citep[see, for example, Figure 2 of][]{h0licow_XIII}.

Furthermore, our result is in agreement within 1-$\sigma$ with previous TDC measurements 
although with a comparatively large uncertainty due to the underconstrained $\Delta t_{\M{AD}}$
 \citep[see Figure \ref{fig:dt_combined_res} and considering single strong lensed QSOs systems, e.g. ][]{chen2019sharp}. 
 
Therefore, our study of J1433 adds an independent TDC measurement of $H_0$ for a strong lensing system that was not yet analysed.

\section*{Acknowledgements}
The authors gratefully acknowledge support from the Excellence Cluster ORIGINS which is funded by the Deutsche Forschungsgemeinschaft (DFG, German Research Foundation) under Germany´s Excellence Strategy – EXC-2094 – 390783311.
The Wendelstein 2.1m telescope project was funded by the Bavarian government and by the German Federal government through a common funding process. Furthermore, part of the 2.1m instrumentation including some of the upgrades for the infrastructure was funded by the Excellence Cluster ORIGINS of the DFG.
We furthermore thank the fruitful input of Martin Millon for the discussions over the implementation of \texttt{PyCS3}, Simon Birrer for the advice and help in \texttt{lenstronomy} and Sherry Suyu and Roberto Saglia for multiple and useful discussions during the project.

The figures in this work were created with \texttt{matplotlib} \citep{matplotlib}. We further make use of the following software packages: \texttt{numpy} \citep{numpy}, \texttt{scipy} \citep{scipy}, \texttt{astropy} \citep{astropy}, \texttt{corner} \citep{corner}, \texttt{scikit-learn} \citep{scikit-learn}, \texttt{emcee} \citep{emcee}, \texttt{pathos} and \texttt{dill} \citep{mckerns2010pathos,mckerns2012building}.
This research also makes use of the ``K-corrections calculator'' service available at http://kcor.sai.msu.ru/.

\textit{Author contributions}: 
The authorship list is divided into two groups: the lead authors (GQ, SS, AR, MK), followed by an alphabetical group including those who made a significant contribution either to the observations and data collection or to the building and maintenance of the telescope. The implementation of the analysis and most of the writing of the paper have been carried out by GQ. SS supervised and contributed to all steps of the project. AR contributed to the modelling process as well as providing extensive knowledge and support in the \textit{WST} data analysis. MK was responsible for the implementation of the isophotal lens light model (see Section \ref{subsec:light_prof&llm}) as well as for the development of the lightcurve collection code described in Section \ref{subsec:WST_data} and the pipeline for the \textit{WWFI} data reduction. The observations have been carried out by (in parenthesis the number of observed nights): CR (195), MS (177), CG (17), AR (9), RZ (8), MK (6). Furthermore, the telescope builders are CG, UH and RB. Finally, RZ also contributed to the colour measurement (Section \ref{subsubsec:main_lens_colour}) and the data reduction (Section \ref{subsec:WST_data})
All co-authors have read and commented on the present paper previous to submission.
\section*{Data Availability}

 The data that support the findings of this study are available from the corresponding author, GQ,  upon reasonable request.



\bibliographystyle{mnras}
\bibliography{biblio}

\begin{thebibliography}{}
\makeatletter
\relax
\def\mn@urlcharsother{\let\do\@makeother \do\$\do\&\do\#\do\^\do\_\do\%\do\~}
\def\mn@doi{\begingroup\mn@urlcharsother \@ifnextchar [ {\mn@doi@}
  {\mn@doi@[]}}
\def\mn@doi@[#1]#2{\def\@tempa{#1}\ifx\@tempa\@empty \href
  {http://dx.doi.org/#2} {doi:#2}\else \href {http://dx.doi.org/#2} {#1}\fi
  \endgroup}
\def\mn@eprint#1#2{\mn@eprint@#1:#2::\@nil}
\def\mn@eprint@arXiv#1{\href {http://arxiv.org/abs/#1} {{\tt arXiv:#1}}}
\def\mn@eprint@dblp#1{\href {http://dblp.uni-trier.de/rec/bibtex/#1.xml}
  {dblp:#1}}
\def\mn@eprint@#1:#2:#3:#4\@nil{\def\@tempa {#1}\def\@tempb {#2}\def\@tempc
  {#3}\ifx \@tempc \@empty \let \@tempc \@tempb \let \@tempb \@tempa \fi \ifx
  \@tempb \@empty \def\@tempb {arXiv}\fi \@ifundefined
  {mn@eprint@\@tempb}{\@tempb:\@tempc}{\expandafter \expandafter \csname
  mn@eprint@\@tempb\endcsname \expandafter{\@tempc}}}

\bibitem[\protect\citeauthoryear{Ade et~al.,}{Ade et~al.}{2016}]{ade2016planck}
Ade P.~A.,  et~al., 2016, Astronomy \& Astrophysics, 594, A13

\bibitem[\protect\citeauthoryear{Aghanim et~al.,}{Aghanim
  et~al.}{2020}]{aghanim2020planck}
Aghanim N.,  et~al., 2020, Astronomy \& Astrophysics, 641, A6

\bibitem[\protect\citeauthoryear{Agnello}{Agnello}{2018}]{agnello_SDSSJ1433}
Agnello G. e.~a.,  2018, Monthly Notices of the Royal Astronomical Society,
  474, 3391

\bibitem[\protect\citeauthoryear{Bertin \& Arnouts}{Bertin \&
  Arnouts}{1996}]{sextractor}
Bertin E.,  Arnouts S.,  1996, Astronomy and astrophysics supplement series,
  117, 393

\bibitem[\protect\citeauthoryear{Birrer \& Amara}{Birrer \&
  Amara}{2018}]{BIRRER2018}
Birrer S.,  Amara A.,  2018, \mn@doi [Physics of the Dark Universe]
  {https://doi.org/10.1016/j.dark.2018.11.002}, 22, 189

\bibitem[\protect\citeauthoryear{Birrer, Amara  \& Refregier}{Birrer
  et~al.}{2015}]{Birrer_2015}
Birrer S.,  Amara A.,   Refregier A.,  2015, \mn@doi [The Astrophysical
  Journal] {10.1088/0004-637x/813/2/102}, 813, 102

\bibitem[\protect\citeauthoryear{Birrer et~al.,}{Birrer
  et~al.}{2020}]{birrer2020tdcosmo}
Birrer S.,  et~al., 2020, Astronomy \& Astrophysics, 643, A165

\bibitem[\protect\citeauthoryear{Bradley et~al.,}{Bradley
  et~al.}{2022}]{photutils}
Bradley L.,  et~al., 2022, astropy/photutils: 1.5.0,
  \mn@doi{10.5281/zenodo.6825092}, \url
  {https://doi.org/10.5281/zenodo.6825092}

\bibitem[\protect\citeauthoryear{C.~Lemon/University~of
  Cambridge}{C.~Lemon/University~of Cambridge}{2019}]{GLQ_site}
C.~Lemon/University~of Cambridge I. o.~A.,  12/2019, {Gravitationally Lensed
  Quasar Database},
  \url{https://research.ast.cam.ac.uk/lensedquasars/index.html}

\bibitem[\protect\citeauthoryear{Chen et~al.,}{Chen
  et~al.}{2019}]{chen2019sharp}
Chen G.~C.,  et~al., 2019, Monthly Notices of the Royal Astronomical Society,
  490, 1743

\bibitem[\protect\citeauthoryear{{Chilingarian} \& {Zolotukhin}}{{Chilingarian}
  \& {Zolotukhin}}{2012}]{2012K_corr}
{Chilingarian} I.~V.,  {Zolotukhin} I.~Y.,  2012, \mn@doi [\mnras]
  {10.1111/j.1365-2966.2011.19837.x}, \href
  {https://ui.adsabs.harvard.edu/abs/2012MNRAS.419.1727C} {419, 1727}

\bibitem[\protect\citeauthoryear{{Chilingarian}, {Melchior}  \&
  {Zolotukhin}}{{Chilingarian} et~al.}{2010}]{2010K_corr}
{Chilingarian} I.~V.,  {Melchior} A.-L.,   {Zolotukhin} I.~Y.,  2010, \mn@doi
  [\mnras] {10.1111/j.1365-2966.2010.16506.x}, \href
  {https://ui.adsabs.harvard.edu/abs/2010MNRAS.405.1409C} {405, 1409}

\bibitem[\protect\citeauthoryear{Collaboration et~al.,}{Collaboration
  et~al.}{2014}]{collaboration2014planck}
Collaboration P.,  et~al., 2014, A\&A, 571, A16

\bibitem[\protect\citeauthoryear{Collett}{Collett}{2015}]{collett2015population}
Collett T.~E.,  2015, The Astrophysical Journal, 811, 20

\bibitem[\protect\citeauthoryear{Di~Valentino et~al.,}{Di~Valentino
  et~al.}{2021}]{di2021realm}
Di~Valentino E.,  et~al., 2021, Classical and Quantum Gravity, 38, 153001

\bibitem[\protect\citeauthoryear{Ertl, Schuldt, Suyu, Schmidt, Treu, Birrer,
  Shajib  \& Sluse}{Ertl et~al.}{2023}]{ertl23}
Ertl S.,  Schuldt S.,  Suyu S.,  Schmidt T.,  Treu T.,  Birrer S.,  Shajib A.,
   Sluse D.,  2023, Astronomy and Astrophysics, 672

\bibitem[\protect\citeauthoryear{Falco, Gorenstein  \& Shapiro}{Falco
  et~al.}{1985}]{falco1985model}
Falco E.,  Gorenstein M.,   Shapiro I.,  1985, The Astrophysical Journal, 289,
  L1

\bibitem[\protect\citeauthoryear{Flewelling et~al.,}{Flewelling
  et~al.}{2016}]{panstarrs}
Flewelling H.,  et~al., 2016, arXiv preprint arXiv:1612.05243

\bibitem[\protect\citeauthoryear{Foreman-Mackey}{Foreman-Mackey}{2016}]{corner}
Foreman-Mackey D.,  2016, \mn@doi [The Journal of Open Source Software]
  {10.21105/joss.00024}, 1, 24

\bibitem[\protect\citeauthoryear{Foreman-Mackey, Hogg, Lang  \&
  Goodman}{Foreman-Mackey et~al.}{2013}]{emcee}
Foreman-Mackey D.,  Hogg D.~W.,  Lang D.,   Goodman J.,  2013, Publications of
  the Astronomical Society of the Pacific, 125, 306

\bibitem[\protect\citeauthoryear{Freedman et~al.,}{Freedman
  et~al.}{2020}]{freedman2020calibration}
Freedman W.~L.,  et~al., 2020, The Astrophysical Journal, 891, 57

\bibitem[\protect\citeauthoryear{Gong et~al.,}{Gong
  et~al.}{2019}]{gong2019cosmology}
Gong Y.,  et~al., 2019, The Astrophysical Journal, 883, 203

\bibitem[\protect\citeauthoryear{Grillo, Rosati, Suyu, Caminha, Mercurio  \&
  Halkola}{Grillo et~al.}{2020}]{grillo2020accuracy}
Grillo C.,  Rosati P.,  Suyu S.,  Caminha G.,  Mercurio A.,   Halkola A.,
  2020, The Astrophysical Journal, 898, 87

\bibitem[\protect\citeauthoryear{Harris et~al.,}{Harris et~al.}{2020}]{numpy}
Harris C.~R.,  et~al., 2020, \mn@doi [Nature] {10.1038/s41586-020-2649-2}, 585,
  357

\bibitem[\protect\citeauthoryear{Holloway, Verma, Marshall, More  \&
  Tecza}{Holloway et~al.}{2023}]{holloway_23}
Holloway P.,  Verma A.,  Marshall P.~J.,  More A.,   Tecza M.,  2023, \mn@doi
  [Monthly Notices of the Royal Astronomical Society] {10.1093/mnras/stad2371},
  525, 2341

\bibitem[\protect\citeauthoryear{{Hopp}, {Bender}, {Grupp}, {Goessl},
  {Lang-Bardl}, {Mitsch}, {Riffeser}  \& {Ageorges}}{{Hopp}
  et~al.}{2014}]{wwfi_2014}
{Hopp} U.,  {Bender} R.,  {Grupp} F.,  {Goessl} C.,  {Lang-Bardl} F.,  {Mitsch}
  W.,  {Riffeser} A.,   {Ageorges} N.,  2014, in Ground-based and Airborne
  Telescopes V. p. 91452D, \mn@doi{10.1117/12.2054498}

\bibitem[\protect\citeauthoryear{Hunter}{Hunter}{2007}]{matplotlib}
Hunter J.~D.,  2007, \mn@doi [Computing in Science \& Engineering]
  {10.1109/MCSE.2007.55}, 9, 90

\bibitem[\protect\citeauthoryear{Kelly et~al.,}{Kelly
  et~al.}{2015}]{kelly2015multiple}
Kelly P.~L.,  et~al., 2015, Science, 347, 1123

\bibitem[\protect\citeauthoryear{Kennedy \& Eberhart}{Kennedy \&
  Eberhart}{1995}]{pso}
Kennedy J.,  Eberhart R.,  1995, in Proceedings of ICNN'95 - International
  Conference on Neural Networks. pp 1942--1948 vol.4,
  \mn@doi{10.1109/ICNN.1995.488968}

\bibitem[\protect\citeauthoryear{{Kluge}}{{Kluge}}{2020}]{Kluge_diss}
{Kluge} M.,  2020, PhD thesis, Ludwig-Maximilians University of Munich, Germany

\bibitem[\protect\citeauthoryear{{Kluge} \& {Bender}}{{Kluge} \&
  {Bender}}{2023}]{klugeisophotespy}
{Kluge} M.,  {Bender} R.,  2023, \mn@doi [\apjs] {10.3847/1538-4365/ace052},
  \href {https://ui.adsabs.harvard.edu/abs/2023ApJS..267...41K} {267, 41}

\bibitem[\protect\citeauthoryear{Kluge et~al.,}{Kluge
  et~al.}{2020}]{kluge2020structure}
Kluge M.,  et~al., 2020, The Astrophysical Journal Supplement Series, 247, 43

\bibitem[\protect\citeauthoryear{{Kluge}, {Remus}, {Babyk}, {Forbes}  \&
  {Dolfi}}{{Kluge} et~al.}{2023}]{Kluge2023rhea}
{Kluge} M.,  {Remus} R.-S.,  {Babyk} I.~V.,  {Forbes} D.~A.,   {Dolfi} A.,
  2023, \mn@doi [\mnras] {10.1093/mnras/stad882}, \href
  {https://ui.adsabs.harvard.edu/abs/2023MNRAS.521.4852K} {521, 4852}

\bibitem[\protect\citeauthoryear{Kosyra}{Kosyra}{2014}]{WWFI_article}
Kosyra R. e.~a.,  2014, Experimental Astronomy, 38, 213

\bibitem[\protect\citeauthoryear{Lee, Freedman  \& Madore}{Lee
  et~al.}{1993}]{lee1993tip}
Lee M.~G.,  Freedman W.~L.,   Madore B.~F.,  1993, Astrophysical Journal v.
  417, p. 553, 417, 553

\bibitem[\protect\citeauthoryear{MAST}{MAST}{2023}]{HST_archive}
MAST 09/2023, Search MAST for Hubble,
  \url{https://archive.stsci.edu/hst/search.php}

\bibitem[\protect\citeauthoryear{McKerns \& Aivazis}{McKerns \&
  Aivazis}{2010}]{mckerns2010pathos}
McKerns M.,  Aivazis M.,  2010, URL http://trac. mystic. cacr. caltech.
  edu/project/pathos

\bibitem[\protect\citeauthoryear{McKerns, Strand, Sullivan, Fang  \&
  Aivazis}{McKerns et~al.}{2012}]{mckerns2012building}
McKerns M.~M.,  Strand L.,  Sullivan T.,  Fang A.,   Aivazis M.~A.,  2012,
  arXiv preprint arXiv:1202.1056

\bibitem[\protect\citeauthoryear{Meylan, Jetzer, North, Schneider, Kochanek  \&
  Wambsganss}{Meylan et~al.}{2006}]{meylan2006gravitational}
Meylan G.,  Jetzer P.,  North P.,  Schneider P.,  Kochanek C.~S.,   Wambsganss
  J.,  2006, Saas-Fee Advanced Course 33: Gravitational Lensing: Strong, Weak
  and Micro

\bibitem[\protect\citeauthoryear{Millon et~al.,}{Millon
  et~al.}{2020a}]{millon2020tdcosmo}
Millon M.,  et~al., 2020a, Astronomy \& Astrophysics, 639, A101

\bibitem[\protect\citeauthoryear{Millon et~al.,}{Millon
  et~al.}{2020b}]{cosmog_pycs_XIX}
Millon M.,  et~al., 2020b, Astronomy \& Astrophysics, 640, A105

\bibitem[\protect\citeauthoryear{NASA/IPAC}{NASA/IPAC}{2023}]{NED_ext}
NASA/IPAC 2023, NED - NASA/IPAC Extragalactic Database,
  \url{https://ned.ipac.caltech.edu/extinction_calculator?in_csys=Equatorial&in_equinox=J2000.0&obs_epoch=2000.0&ra=14%3A33%3A26.5602&dec=%2B60%3A07%3A42.858}

\bibitem[\protect\citeauthoryear{Oguri \& Marshall}{Oguri \&
  Marshall}{2010}]{Oguri_2010}
Oguri M.,  Marshall P.~J.,  2010, \mn@doi [Monthly Notices of the Royal
  Astronomical Society] {10.1111/j.1365-2966.2010.16639.x}, 405, 2579

\bibitem[\protect\citeauthoryear{Pedregosa et~al.,}{Pedregosa
  et~al.}{2011}]{scikit-learn}
Pedregosa F.,  et~al., 2011, Journal of Machine Learning Research, 12, 2825

\bibitem[\protect\citeauthoryear{Price-Whelan et~al.,}{Price-Whelan
  et~al.}{2022}]{astropy}
Price-Whelan A.~M.,  et~al., 2022, The Astrophysical Journal, 935, 167

\bibitem[\protect\citeauthoryear{Refsdal}{Refsdal}{1964}]{refsdal1964possibility}
Refsdal S.,  1964, Monthly Notices of the Royal Astronomical Society, 128, 307

\bibitem[\protect\citeauthoryear{Riess et~al.,}{Riess
  et~al.}{1998}]{riess1998SNeIa}
Riess A.~G.,  et~al., 1998, The astronomical journal, 116, 1009

\bibitem[\protect\citeauthoryear{Riess et~al.,}{Riess
  et~al.}{2016}]{Riess_2016}
Riess A.~G.,  et~al., 2016, \mn@doi [The Astrophysical Journal]
  {10.3847/0004-637X/826/1/56}, 826, 56

\bibitem[\protect\citeauthoryear{{Riffeser}}{{Riffeser}}{2006}]{Rieffeser_psf}
{Riffeser} A.,  2006, PhD thesis, Ludwig-Maximilians University of Munich,
  Germany

\bibitem[\protect\citeauthoryear{STSI}{STSI}{2023}]{STSCI_EE}
STSI S. T. S.~I.,  07/10/2023, HST Data Search,
  \url{https://www.stsci.edu/hst/instrumentation/wfc3/data-analysis/photometric-calibration/}

\bibitem[\protect\citeauthoryear{Saglia, Maraston, Greggio, Bender  \&
  Ziegler}{Saglia et~al.}{2000}]{saglia2000evolution}
Saglia R.,  Maraston C.,  Greggio L.,  Bender R.,   Ziegler B.,  2000, arXiv
  preprint astro-ph/0007038

\bibitem[\protect\citeauthoryear{Schmidt et~al.,}{Schmidt
  et~al.}{2022}]{schmidt22_STRIDES}
Schmidt T.,  et~al., 2022, Monthly Notices of the Royal Astronomical Society,
  518, 1260

\bibitem[\protect\citeauthoryear{Schneider \& Sluse}{Schneider \&
  Sluse}{2013}]{schneider2013mass}
Schneider P.,  Sluse D.,  2013, Astronomy \& Astrophysics, 559, A37

\bibitem[\protect\citeauthoryear{Shajib, Treu  \& Agnello}{Shajib
  et~al.}{2018}]{shajib2018improving}
Shajib A.~J.,  Treu T.,   Agnello A.,  2018, Monthly Notices of the Royal
  Astronomical Society, 473, 210

\bibitem[\protect\citeauthoryear{Shajib et~al.,}{Shajib
  et~al.}{2019}]{shajib_SDSSJ1433}
Shajib A.~J.,  et~al., 2019, Monthly Notices of the Royal Astronomical Society,
  483, 5649

\bibitem[\protect\citeauthoryear{Shajib et~al.,}{Shajib
  et~al.}{2022}]{shajib2022tdcosmo}
Shajib A.,  et~al., 2022, Astronomy \& Astrophysics, 667, A123

\bibitem[\protect\citeauthoryear{Shapiro}{Shapiro}{1964}]{shapiro1964fourth}
Shapiro I.~I.,  1964, Physical Review Letters, 13, 789

\bibitem[\protect\citeauthoryear{Smith, Robertson, Bianconi  \& Jauzac}{Smith
  et~al.}{2019}]{smith2019discovery}
Smith G.~P.,  Robertson A.,  Bianconi M.,   Jauzac M.,  2019, arXiv preprint
  arXiv:1902.05140

\bibitem[\protect\citeauthoryear{Suyu, Marshall, Auger, Hilbert, Blandford,
  Koopmans, Fassnacht  \& Treu}{Suyu et~al.}{2010}]{suyu2010dissecting}
Suyu S.,  Marshall P.,  Auger M.,  Hilbert S.,  Blandford R.,  Koopmans L.,
  Fassnacht C.,   Treu T.,  2010, The Astrophysical Journal, 711, 201

\bibitem[\protect\citeauthoryear{Suyu et~al.,}{Suyu
  et~al.}{2013}]{suyu2013_MSD_ext}
Suyu S.,  et~al., 2013, The Astrophysical Journal, 766, 70

\bibitem[\protect\citeauthoryear{Suyu et~al.,}{Suyu
  et~al.}{2014}]{MSD_Suyu_2014}
Suyu S.~H.,  et~al., 2014, \mn@doi [The Astrophysical Journal Letters]
  {10.1088/2041-8205/788/2/L35}, 788, L35

\bibitem[\protect\citeauthoryear{Suyu et~al.,}{Suyu
  et~al.}{2017}]{suyu2017h0licow}
Suyu S.~H.,  et~al., 2017, Monthly Notices of the Royal Astronomical Society,
  468, 2590

\bibitem[\protect\citeauthoryear{Suyu et~al.,}{Suyu
  et~al.}{2020}]{suyu2020holismokes}
Suyu S.,  et~al., 2020, Astronomy \& Astrophysics, 644, A162

\bibitem[\protect\citeauthoryear{Tewes, Courbin  \& Meylan}{Tewes
  et~al.}{2013a}]{cosmog_pycs_XI}
Tewes M.,  Courbin F.,   Meylan G.,  2013a, Astronomy \& Astrophysics, 553,
  A120

\bibitem[\protect\citeauthoryear{Tewes, Courbin  \& Meylan}{Tewes
  et~al.}{2013b}]{cosmograil_XI}
Tewes M.,  Courbin F.,   Meylan G.,  2013b, Astronomy \& Astrophysics, 553,
  A120

\bibitem[\protect\citeauthoryear{Treu \& Koopmans}{Treu \&
  Koopmans}{2002}]{Treu02MSD}
Treu T.,  Koopmans L. V.~E.,  2002, \mn@doi [Monthly Notices of the Royal
  Astronomical Society] {10.1046/j.1365-8711.2002.06107.x}, 337, L6

\bibitem[\protect\citeauthoryear{Treu, Suyu  \& Marshall}{Treu
  et~al.}{2022}]{treu2022strong}
Treu T.,  Suyu S.~H.,   Marshall P.~J.,  2022, The Astronomy and Astrophysics
  Review, 30, 8

\bibitem[\protect\citeauthoryear{Virtanen et~al.,}{Virtanen
  et~al.}{2020}]{scipy}
Virtanen P.,  et~al., 2020, \mn@doi [Nature Methods]
  {10.1038/s41592-019-0686-2}, \href {https://rdcu.be/b08Wh} {17, 261}

\bibitem[\protect\citeauthoryear{Weiner, Serjeant  \& Sedgwick}{Weiner
  et~al.}{2020}]{weiner2020predictions}
Weiner C.,  Serjeant S.,   Sedgwick C.,  2020, Research Notes of the AAS, 4,
  190

\bibitem[\protect\citeauthoryear{Wertz, Orthen  \& Schneider}{Wertz
  et~al.}{2018}]{wertz2018ambiguities}
Wertz O.,  Orthen B.,   Schneider P.,  2018, Astronomy \& Astrophysics, 617,
  A140

\bibitem[\protect\citeauthoryear{Wong et~al.,}{Wong
  et~al.}{2020}]{h0licow_XIII}
Wong K.~C.,  et~al., 2020, Monthly Notices of the Royal Astronomical Society,
  498, 1420

\bibitem[\protect\citeauthoryear{{Z{\"o}ller}, {Kluge}, {Staiger}  \&
  {Bender}}{{Z{\"o}ller} et~al.}{2023}]{zoeller23}
{Z{\"o}ller} R.,  {Kluge} M.,  {Staiger} B.,   {Bender} R.,  2023, \mn@doi
  [arXiv e-prints] {10.48550/arXiv.2310.09330}, \href
  {https://ui.adsabs.harvard.edu/abs/2023arXiv231009330Z} {p. arXiv:2310.09330}

\makeatother
\end{thebibliography}



\appendix

\section{Appendix}
\subsection{Encircle Energy}
\begin{figure}

 \begin{multicols}{2}
 \includegraphics[width=1.1\columnwidth]{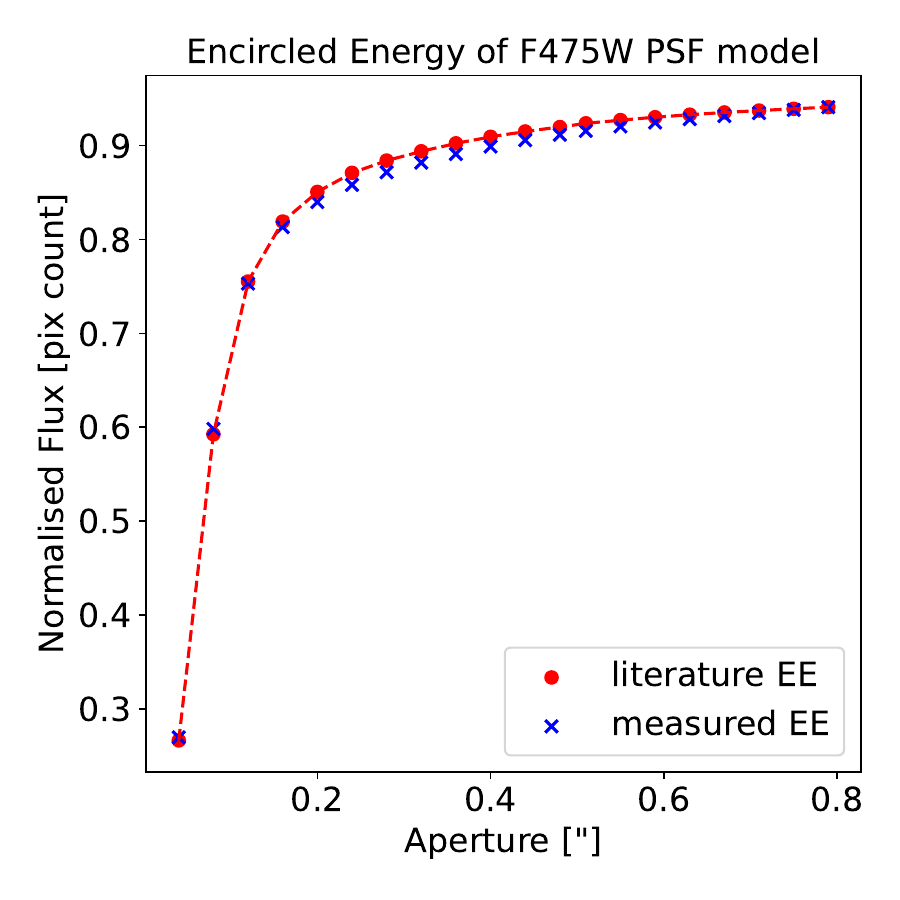}\par
\includegraphics[width=1.1\columnwidth]{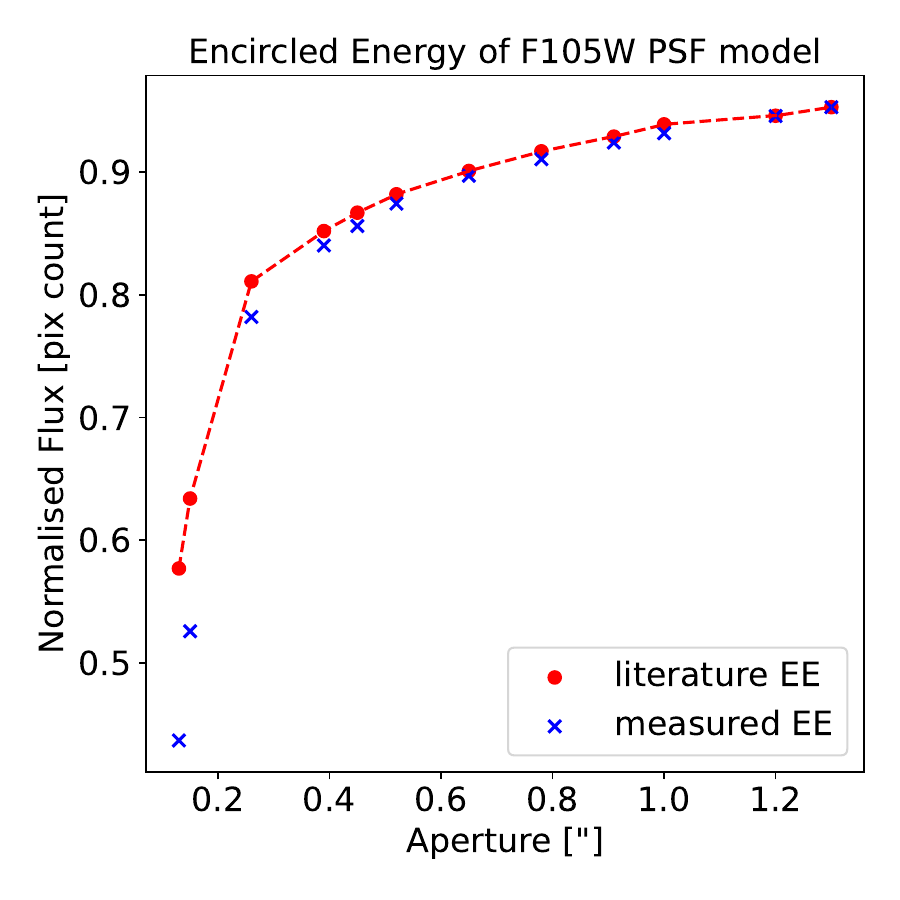} 
\end{multicols}
 \begin{multicols}{2}
\includegraphics[width=1.1\columnwidth]{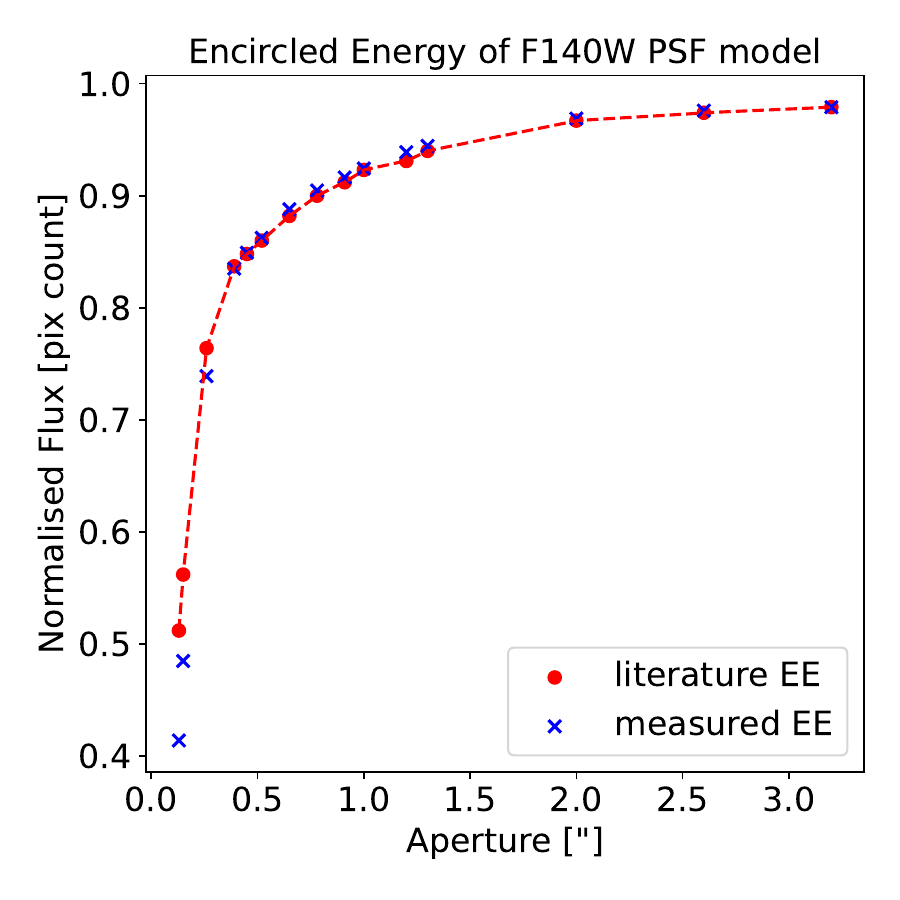} \par
\includegraphics[width=1.1\columnwidth]{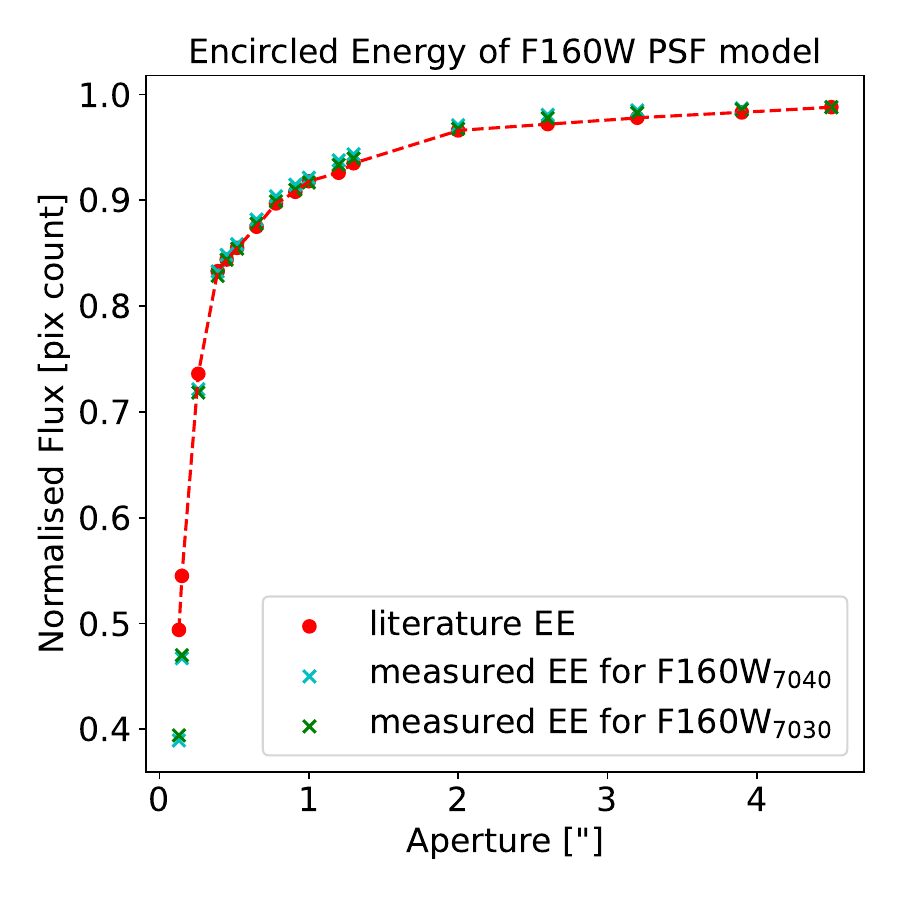} 
\end{multicols}
\caption{Encircled energy of the PSF model at increasing radii compared to the literature values \citep{STSCI_EE} for filter F475X, F105W, F140W and F160W (F160W$_{\M{7030}}$ and F160W$_{\M{7040}}$ are shown in the same plot, at the bottom right).}
    \label{fig:app_EE}
\end{figure}

We compared the encircled energy (EE) trend of the PSF model with respect to the radius (i.e. the growth curves) with the values available in the literature \citep{STSCI_EE}. This step was presented in Section \ref{subsubsec:psf} and the equivalent plot for F814W is shown in Figure (\ref{fig:psf_EE_f814}). We report in Figure (\ref{fig:app_EE}) the equivalent plots for all remaining filters.


\bsp	
\label{lastpage}

\end{document}